\documentclass[12pt]{article} %,draft,openright]
\usepackage{axodraw}
\usepackage{amsmath}
\usepackage{graphicx}
\usepackage{latexsym,amssymb,amsmath}
\usepackage{sidecap}
\usepackage[section] {placeins}
\usepackage[all]{xy}
\usepackage{slashed}
\usepackage{a4wide}
\usepackage{subfigure}

\newcommand{\rref}[1]{{(\ref{#1})}}
\newcommand{\tref}[1]{{\ref{#1}}}
\newcommand{\rlabel}[1]{{\label{#1}}}

\newcommand{\be}{\begin{equation}}
\newcommand{\ee}{\end{equation}}
\newcommand{\ba}{\begin{eqnarray}}
\newcommand{\ea}{\end{eqnarray}}

\newcommand{\fourp}{\Pi^{\rho\nu\alpha\beta}(p_1,p_2,p_3)}
\newcommand{\mathrom}[1]{{\rm #1}}
\newcommand{\tr}{\mathrom{tr}}

\begin{document}

%%%%%%%%%%%%%%%%%%%%%%%%titlepage%%%%%%%%%%%%%%%%%%%%%%%%%%%%%%%%%%%%%%%%
\begin{titlepage}

\begin{flushright}
LU TP 12-26\\
20 June 2012
\end{flushright}

\hfill

\begin{center}
{\Large\bf The Anatomy of the Pion Loop Hadronic Light by Light
Scattering Contribution to the Muon Magnetic Anomaly}\\[2cm]

{\large {\bf
Mehran Zahiri Abyaneh}\\[0.5cm]
\textit{Thesis advisor}: Johan Bijnens\\[1cm]
\vspace{0.5cm}
Department of Theoretical Physics, Lund University\\[0.3cm]
S\"olvegatan 14A, S22362 Lund, Sweden
}

\end{center}

\hfill

\begin{abstract}

This thesis investigates the Hadronic Light by Light (HLL)
scattering contribution to the muon $g-2$, which is one of the
most important low energy hadronic effects and consists mainly of
the quark loop, the pion pole and the charged pion and kaon loops.
In this work the charged pion loop has been investigated more
closely. After reviewing the subject a preliminary introduction
to Chiral Perturbation Theory
(ChPT), Hidden Local Symmetry (HLS) model and the full Vector
Meson Dominance (VMD) model is given, and they
are used to calculate the pion loop HLL scattering contribution to
the muon anomalous magnetic moment. The momentum regions where the
contributions of the bare pion loop, the VMD model, and the HLS
come from, have been studied, to understand why different
models give very different results. The effects of pion
polarizability and charge radius on the HLL scattering,
which appear at order $p^4$ in ChPT, from $ L_9$ and $ L_{10}$
Lagrangian terms and their momentum regions have been
studied.

\end{abstract}

\begin{center}
\large
Master of Science Thesis
\end{center}

\hfill

\begin{center}
\end{center}
\hfill
\end{titlepage}

%if you want a quote
%\begin{center}
% \begin{quote}
%\it
%  %There is no excellent beauty that hath not some strangeness in the proportion.
% \end{quote}
%
%\end{center}
%\hfill{\small Francis Bacon}

\tableofcontents

\newpage

%sets up headers for lefthand and righthand pages. To alter, edit
%these lines and the chaptermark/sectionmark lines above
%NOTE some of these lines are redundant, but I haven't had time to
%optimize the size of this file, and anyway, IT WORKS as is.
%\addtolength{\headheight}{3pt}
%\fancyhead{}
%\fancyhead[LE]{\sl\leftmark}
%\fancyhead[LO,RE]{\rm\thepage}
%\fancyhead[RO]{\sl\rightmark}
%\fancyfoot[C,L,E]{}
%\renewcommand{\headrulewidth}{0.5pt}
%\pagenumbering{arabic}
%\setcounter{page}{1} %\pagestyle{headings}

%\singlespacing  % you must use setspace.sty to get this. setspace also
                % defines the below two spacing options. It's magic.
%\doublespacing
%\onehalfspacing

\section{Introduction}
\label{chap:intro}
\setcounter{equation}{0}
\subsection{Theory}\rlabel{th}
Elementary particles have some inherent properties including
charge, mass, spin and lifetime. As important as these quantities,
are the magnetic and electric
 dipole moments  which are typical for charged particles with spin. Classically, an orbiting
 particle with electric charge $e$ carrying mass $m$ entails a magnetic dipole moment given by
\begin{equation}\label{1-1}
\mbox{\boldmath$\mu$}=\frac{e}{2m}\mathbf{L}\, \ ,
\end{equation}
where $\mbox{\boldmath$L$}$ is the angular momentum of the particle. Magnetic and electric
 moments interact with external magnetic and electric fields via the Hamiltonian
\begin{equation}\label{2}
H=-\mbox{\boldmath$\mu$}\cdot\mathbf{B}-\mathbf{d}\cdot\mathbf{E}\, \ ,
\end{equation}
 where $\mathbf{B}$ and $\mathbf{E}$ are the magnetic and electric field strengths and $\mbox{\boldmath$\mu$}$ and $\mathbf{d}$
 the magnetic and electric dipole moment operators.
 The magnetic moment is often measured in units of the Bohr
magneton $\mbox{\boldmath$\mu$}_B$ which is defined as
\begin{equation}\label{3}
\mu_B=\frac{e}{2m_e}=5.788381804(39)\times 10^{-11}
\mathrm{MeVT^{-1}}\, \ ,
\end{equation}
where T stands for Tesla. When it comes to spinning particles, the
angular momentum operator in~(\ref{1-1}) should be replaced by
the spin operator~\cite{Jegerlehner:2009ry}.

For a charged elementary particle with intrinsic spin and charge $q$, the magnetic moment is written
\begin{equation}\label{4}
\mbox{\boldmath$\mu$}=g_s\frac{q}{2m}\mathbf{S}\, \ ,
\end{equation}
where, $\mathbf{S}$ is the spin operator. The constant $g_s$ is
the Lande g-factor. Although the Dirac equation predicts that
$g_s=2$ for electron-like particles, it is slightly greater than
$2$, and theoretically it is useful to break the magnetic moment
into two pieces
\begin{equation}\label{5}
\mu=(1+a)\frac{q\hbar}{m}\, \ ,
\end{equation}
where $a=\frac{g-2}{2}$. The first piece, called the Dirac moment,
is $2$ in units of the Bohr magnetic moment. The second piece is
called the anomalous (Pauli) moment, and $a$ is a dimensionless
quantity referred to as the anomaly.

\begin{figure}
\begin{center}
\includegraphics[width=9cm,height=3.5cm]{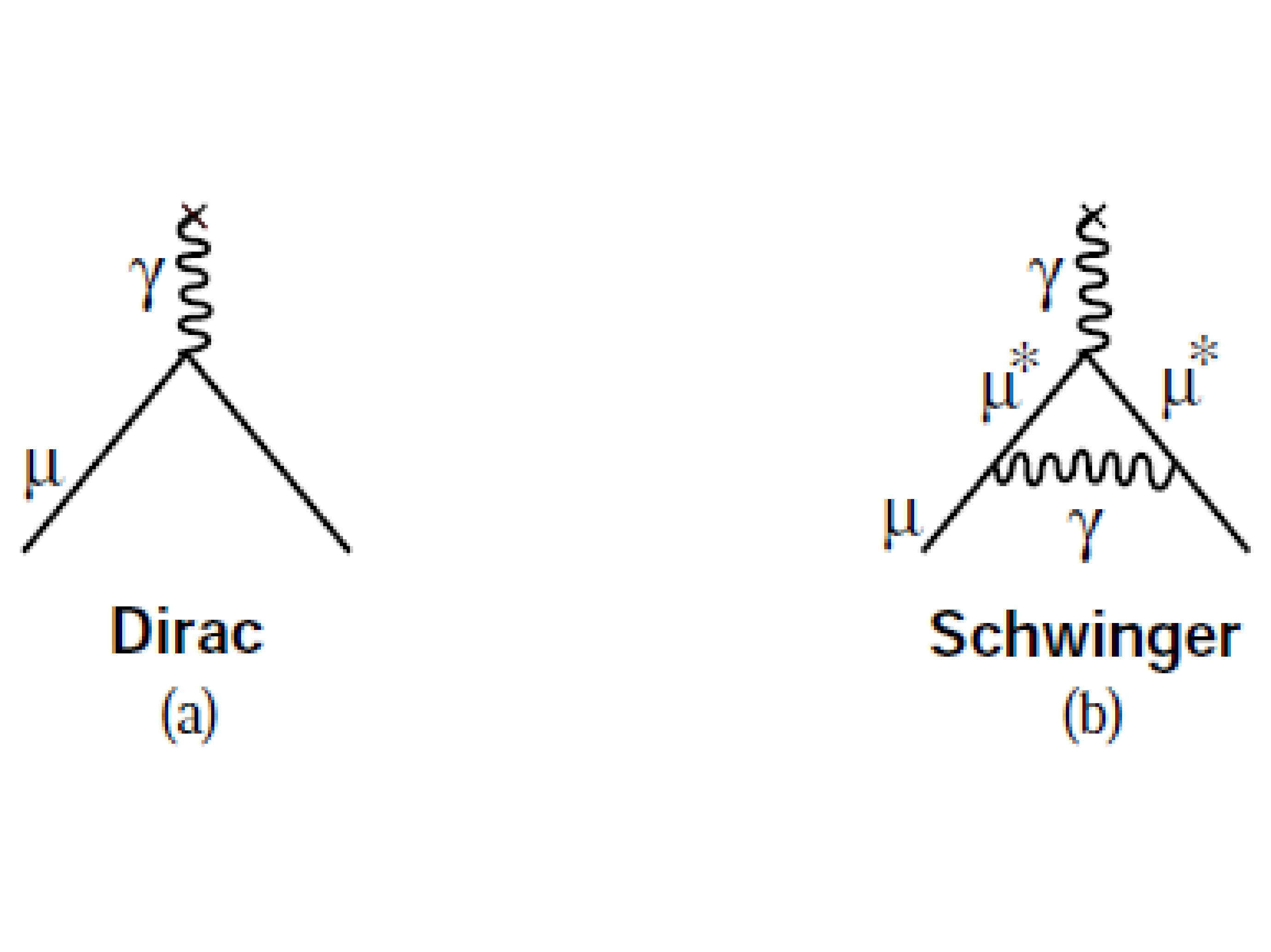}
\end{center}
\caption{The Feynman graphs for: (a) g = 2; (b) the lowest-order radiative correction
first calculated by Schwinger. Figure from~\cite{Miller:2007kk}.}
\label{f1}
\end{figure}

\begin{figure}
\begin{center}
\includegraphics[width=10cm,height=8cm]{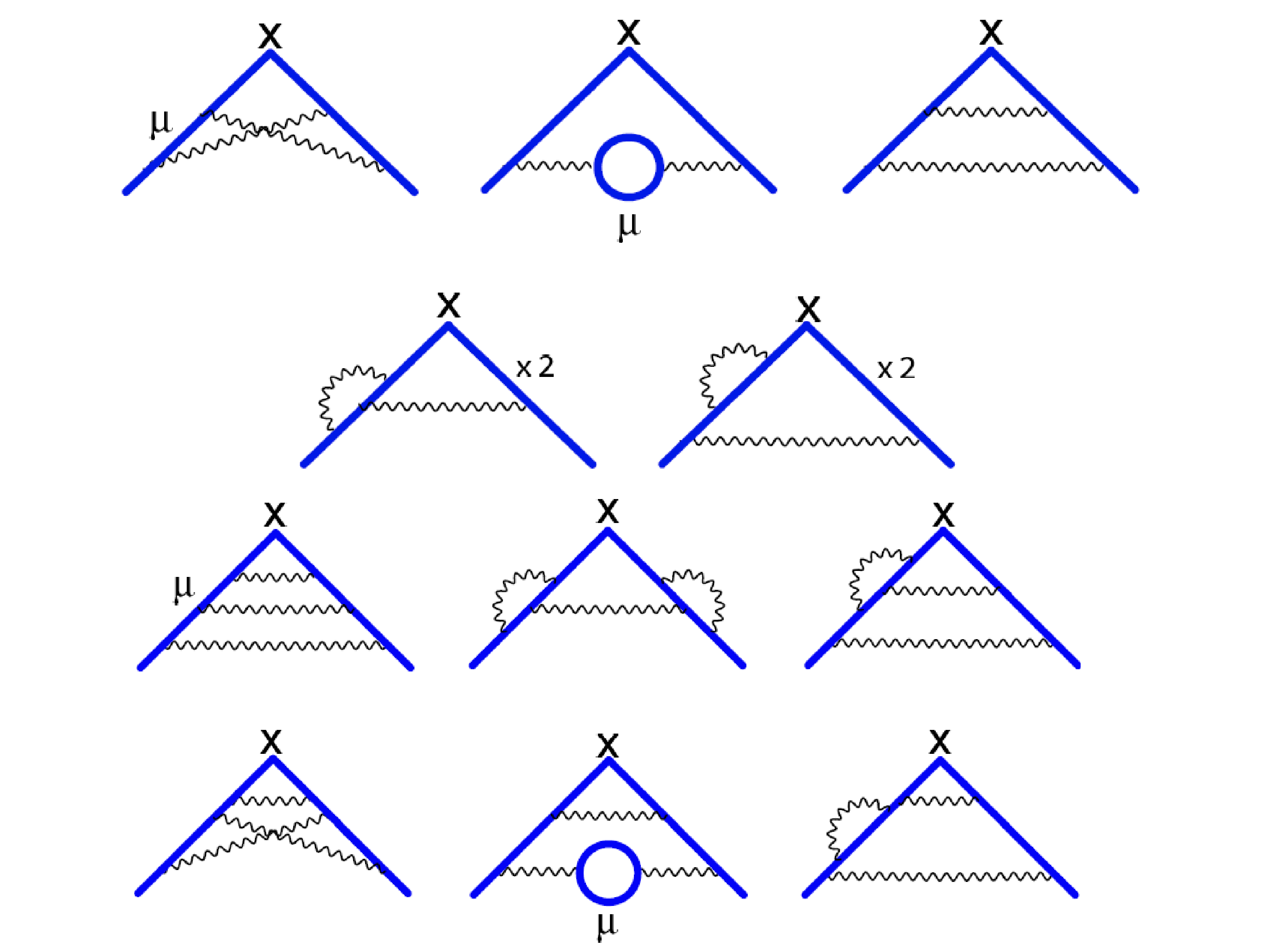}
\end{center}
\caption{Typical second and third order QED loop corrections. Figure from~\cite{Miller:2007kk}.}
\label{QED}
\end{figure}

In 1947, Schwinger, having managed to eliminate divergencies
arising in the calculation in loop corrections in
QED, showed that the deviation of $g_s$ from $2$ can be ascribed
to radiative corrections. The first order correction known as the
one-loop correction to g = 2, is shown diagrammatically in
Figure~\tref{f1}.
More generally, the Standard-Model corrections to the electron,
muon or tau anomaly, $a(SM)$, arise from virtual leptons,
hadrons, gauge bosons and the Higgs boson. This includes the dominant QED terms,
which contain only leptons and photons; terms which involve
hadrons including hadronic vacuum polarization and hadronic light
by light (HLL) corrections,
 and electroweak terms, which contain the Higgs, W and Z. That is,
 the anomaly for lepton $l$ is calculated as
\begin{equation}\label{6}
a_l=a_l^{QED}+a_l^{Had}+a_l^{Weak}\, \ .
\end{equation}
An introduction to the theory can be found in~\cite{Knecht:2003kc}.
 It should be mentioned that, in the Standard Model calculations of
$a_l$, all contributions coming from the mass scale $m_l\gg M$ in
loops are suppressed by powers of $m_l/M$, and all with in the
range $M\gg m_l$ are enhanced by powers of $ln(m_l/M)$. Therefore,
for the electron, the most important parts come from the QED part
where the mediator is the massless photon~\cite{Jegerlehner:2009ry} and the
 sensitivity to hadronic and weak effects as well as the
 sensitivity to physics beyond the SM is very small.
 Typical Feynman
diagrams which contribute to the electron magnetic anomaly are
shown in Figure~\tref{QED}.
This allows for a
 very precise and model independent prediction of $a_e$
and hence to determine the fine
 structure constant $\alpha$ with the highest accuracy, which is
needed as an input to be able to make precise predictions for
 other observables like $a_{\mu}$.
 This could be done, matching the predicted value of
$a_e^{SM}$~\cite{Knecht:2003kc}
\begin{equation}\label{7}
a_e^{SM}= 0.5\frac{\alpha}{\pi}-0.32847844400(\frac{\alpha}{\pi})^2+
1.181234017(\frac{\alpha}{\pi})^3-1.7502(384)(\frac{\alpha}{\pi})^4+1.70(3)\times 10^{-12}\, \ ,
\end{equation}
where the hadronic and weak contributions are also accounted for,
with the observed value
$a_e^{exp}=0.0011596521883(42)$ to find~\cite{Knecht:2003kc}
\begin{equation}\label{8}
\alpha^{-1}(a_e)=137.03599875(52)\, \ .
\end{equation}
This value is six times more accurate than the other best
assessment via the quantum Hall effect, which returns
\begin{equation}\label{9}
\alpha^{-1}(qH)=137.03600300(270)\, \ .
\end{equation}

As discussed above, the QED contributions to $a_\mu$ are the same
as for the electron however, the heavy leptons are also allowed
inside the loop this time. The overall QED contribution to $a_\mu$
then reads~\cite{Passera007}
\begin{equation}\label{10}
a_\mu^{QED}=11658471.809(0.016)\times 10^{-10}\, \ .
\end{equation}
 On the other hand, $a_{\mu}$
is much more sensitive to all three types of effects accounted
above, and even to physics beyond the Standard Model due to the
higher mass of the muon~\cite{Jegerlehner:2009ry, Miller:2007kk}.

\begin{figure}
\begin{center}
\includegraphics[width=10cm,height=5cm]{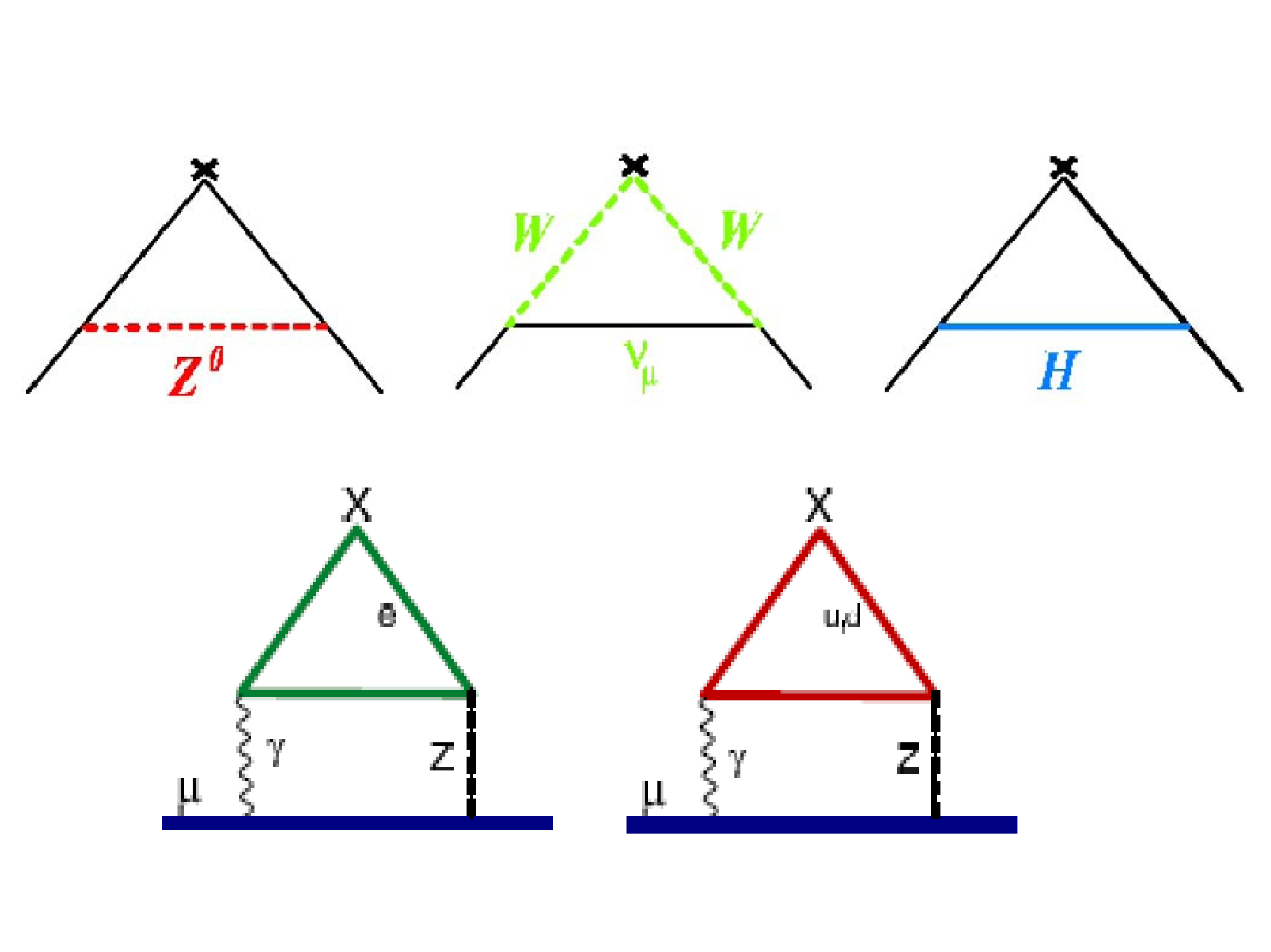}
\end{center}
\caption{Electroweak one loop and two loop contributions to $a_\mu$. Figure from~\cite{Miller:2007kk}.}
\label{weak}
\end{figure}

The Electroweak contribution to $a_\mu$ is divided into two parts,
one and two--loop contributions as shown in Figure~\tref{weak}, so
that

\begin{equation}\label{11}
a_\mu^{EW} = a_\mu^{W(1)}+a_\mu^{W(2)}\, \ ,
\end{equation}
which results in
\begin{eqnarray}\label{111}
a_\mu^{EW(1)} &=& 19.48\times 10^{-10}\nonumber\\
a_\mu^{EW(2)} &=& -4.07(0.1)(0.18)\times 10^{-10}\nonumber\\
a_\mu^{EW} &=& 15.4(0.1)(0.2)\times 10^{-10}\, \ .
\end{eqnarray}

Both the QED and electroweak contributions can be calculated to
high precision. In contrast, the hadronic contribution to
$a_{\mu}$ cannot be accurately evaluated from low-energy quantum
chromodynamics (QCD), and leads to the dominant theoretical
uncertainty on the Standard-Model prediction~\cite{Miller:2007kk}.
In fact, since effects of the energies higher than the muon mass
are suppressed by powers of $(m_\mu/M)$, the relevant QCD
contributions to $a_\mu$ are in the non perturbative regime.
Nevertheless, there exists a consistent theory to
 control strong interaction dynamics at very low energies, which is called chiral
 perturbation theory (ChPT)~\cite{Scherer} and will
 be discussed in Sec.~\ref{Chpt}.

\begin{figure}
\begin{center}
\includegraphics[width=8cm,height=4cm]{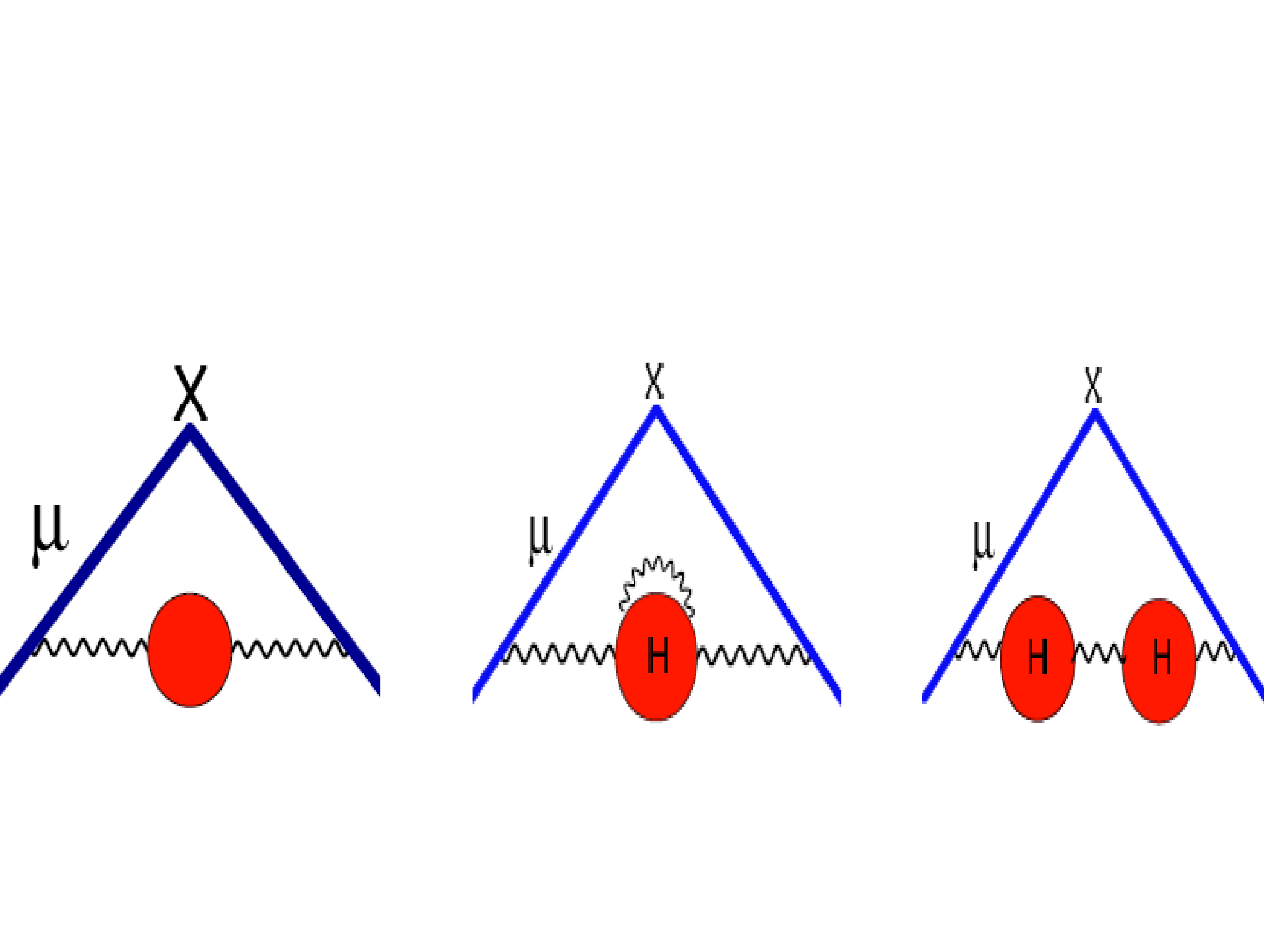}
\end{center}
\caption{The hadronic vacuum polarization contribution, lowest and
higher orders. Figure from~\cite{Miller:2007kk}}
\label{hadpol}.
\end{figure}

\begin{figure}
\begin{center}
\includegraphics[width=7.5cm,height=5cm]{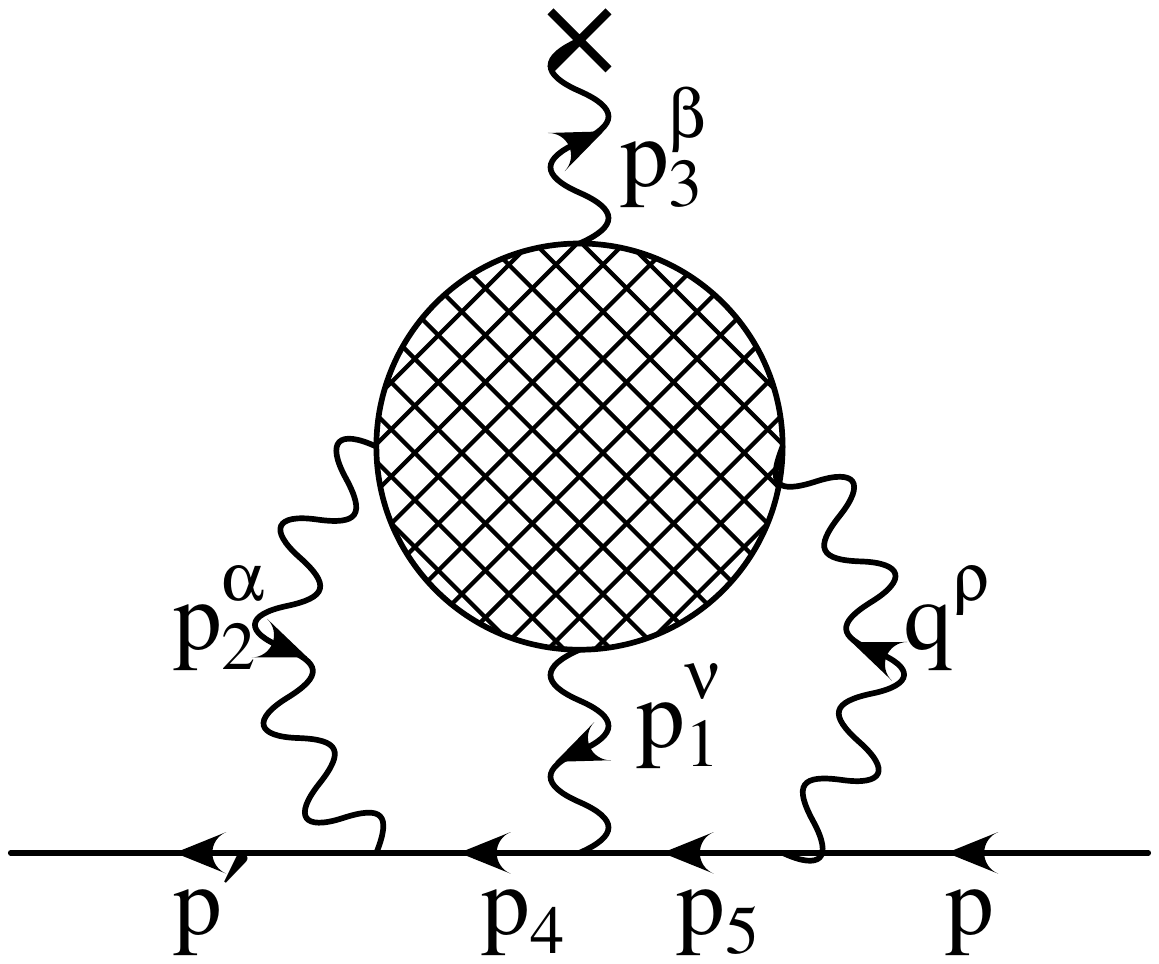}
\end{center}
\caption{Hadronic light by light contribution. Figure from~\cite{Bijnens:2007pz}.}
\label{HLL}
\end{figure}

 The hadronic contribution is divided in two
parts: the hadronic vacuum polarization contribution Figure~
\tref{hadpol}, and the HLL, Figure~\tref{HLL}, that is

\be\label{1111} a_\mu^{had}=a_\mu^{(hvp)}+a_\mu^{(HLL)}\, \ . \ee
 The
vacuum polarization is divided into the leading order and
next-to-leading order, whose contributions
are~\cite{Dorokhov:2008pw}

\be\label{11111}
 a_\mu^{Had,LO}=690.9(4.4)\times 10^{-10} \ee and \be
a_\mu^{Had,HO}=-9.8(0.1)\times 10^{-10}\, \ . \ee The part we are
interested in in this work, is the hadronic light by light
scattering, which, contrary to the vacuum polarization part,
 can not be expressed fully in terms of any experimental
data and should be dealt with only theoretically and hence, it can
be a source of more serious errors~\cite{Jegerlehner:2009ry} and
makes the result model dependent. It consists of three
contributions, the quark loop, the pion exchange and the charged
pion (Kaon) loop~\cite{Miller:2007kk}. Due to considerations of
the Ref.~\cite{Prades:2009tw}, the estimation of the HLL
contribution to the muon $g-2$ is \be a_\mu^{(h.L\times
L)}=(10.5\pm 2.6)\times 10^{-10}\, \ , \ee which is suffering from
a large error, as discussed above.

Calculating the HLL part is the trickiest. Although, ChPT is a
reliable theory of hadrons at low energies, its usage for the pion
exchange brings about divergences and one should resort to certain
models to get rid of them. One can just introduce some cut off
energy, but, the way to do it systematically is to cover the
photon legs with vector mesons. These vector mesons cure the
infinities similar to what the Pauli Villars method does in QFT,
although the Pauli Villars is a pure mathematical manipulation,
while vector mesons are observable physical entities. There are
certain models to do the job~(below). Historically, after
Ref.~\cite{Kinoshita:1984} calculated the HLL part via the naive
VMD approach, which does not obviously respect the electromagnetic
Ward identities~\cite{Hayakawa:1996ki}, the first thorough
consideration, compatible with the Ward identities, was by
Bijnens, Pallante and Prades ~\cite{Bijnens:1995cc,Bijnens:1995xf}
via the Extended Nambu--Jona--Lasinio approach, assuming full VMD.
The other was by Hayakawa, Kinoshita and
Sanda~\cite{Hayakawa:1995ps} using the HLS model. Then,
 Knecht--Nyffeler recalculated the $\pi_0, \eta, \eta\prime$ exchange
 contribution via
the quark--hadron duality in the large $N_c$ limit of
QCD~\cite{Nyffeler}, and found a sign difference with the previous
results. Subsequently authors of both previous works found a sign
 mistake which was corrected~\cite{Bijnens:2001cq}. Meanwhile,
 afterwards, matching between the short and the long distance
behavior of the light-by-light scattering amplitude,
Melnikov and Vainshtein found some corrections~\cite{Melnikov:2003xd}.

 However, as mentioned above, the HLL contribution consists of three
 parts among which, we are interested in the charged pion loop correction in this
 work. The reason is, as can be seen from Table~\tref{table1}, different
 approaches to this part led to very different results.
\begin{table}[b]
\begin{center}
\begin{tabular}{|c|c|c|}
\hline
 Charged pion and Kaon Loop Contributions
&$a_\mu$ $\times$  $10^{10}$  \\
\hline
Bijnens, Pallante and Prades(Full VMD) &$-1.9\pm 0.5$\\
Hayakawa and Kinoshita (HGS) &$-0.45\pm0.85$\\
Kinoshita, Nizic and Okamoto(Naive VMD) &$-1.56\pm 0.23$\\
Kinoshita, Nizic and Okamoto(Scalar QED) &$-5.47\pm 4.6$\\
\hline
\end{tabular}
\end{center}
\caption{Results of different approaches to the charged pion loop
HLL contribution to
$a_\mu$~\cite{Bijnens:2007pz,Kinoshita:1984,Hayakawa:1995ps}\, \
.} \label{table1}
\end{table}
In fact, when the vector mesons are introduced into the
calculation, one expects that results are heavily suppressed,
compared to the bare pion loop case. However, as both VMD and HLS
models use this mechanism, one might wonder, why the full VMD
result is about three times larger than the one from the HLS one.
This is the main question which is tried to be answered in this
work.

\subsection{Experiment}
\rlabel{exp}

\begin{figure}
\begin{center}
\includegraphics[width=10cm,height=6cm]{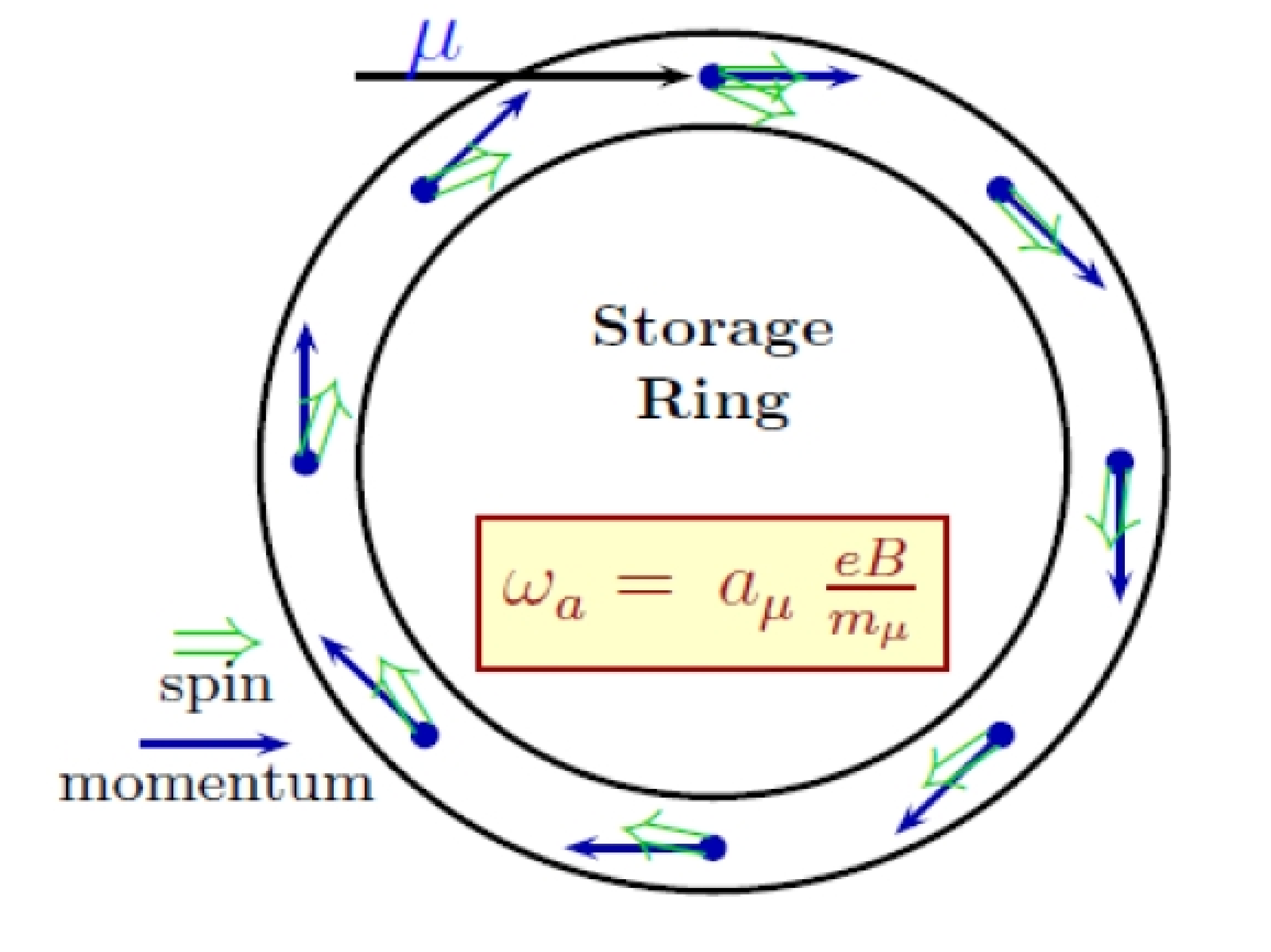}
\end{center}
\caption{Spin precession in the $g-2$ ring. Figure from~\cite{Jegerlehner:2009ry}.}
\label{lab1}
\end{figure}

\begin{figure}
\begin{center}
\includegraphics[width=10cm,height=6cm]{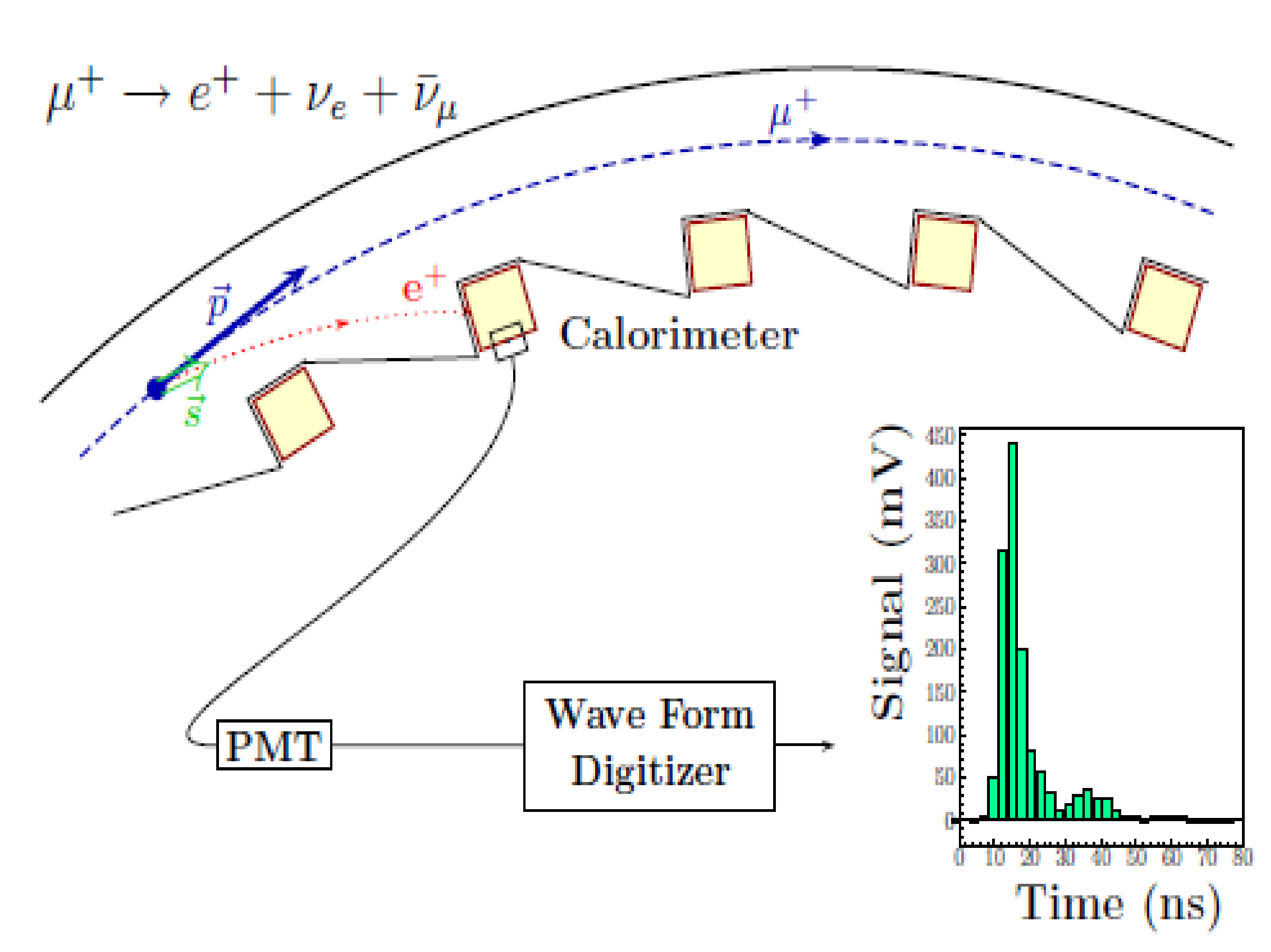}
\end{center}
\caption{Decay of $\mu^+$ and detection of the emitted $e^+$. Figure from~\cite{Jegerlehner:2009ry}.}
\label{lab2}
\end{figure}

\begin{figure}
\begin{center}
\includegraphics[width=10cm,height=6cm]{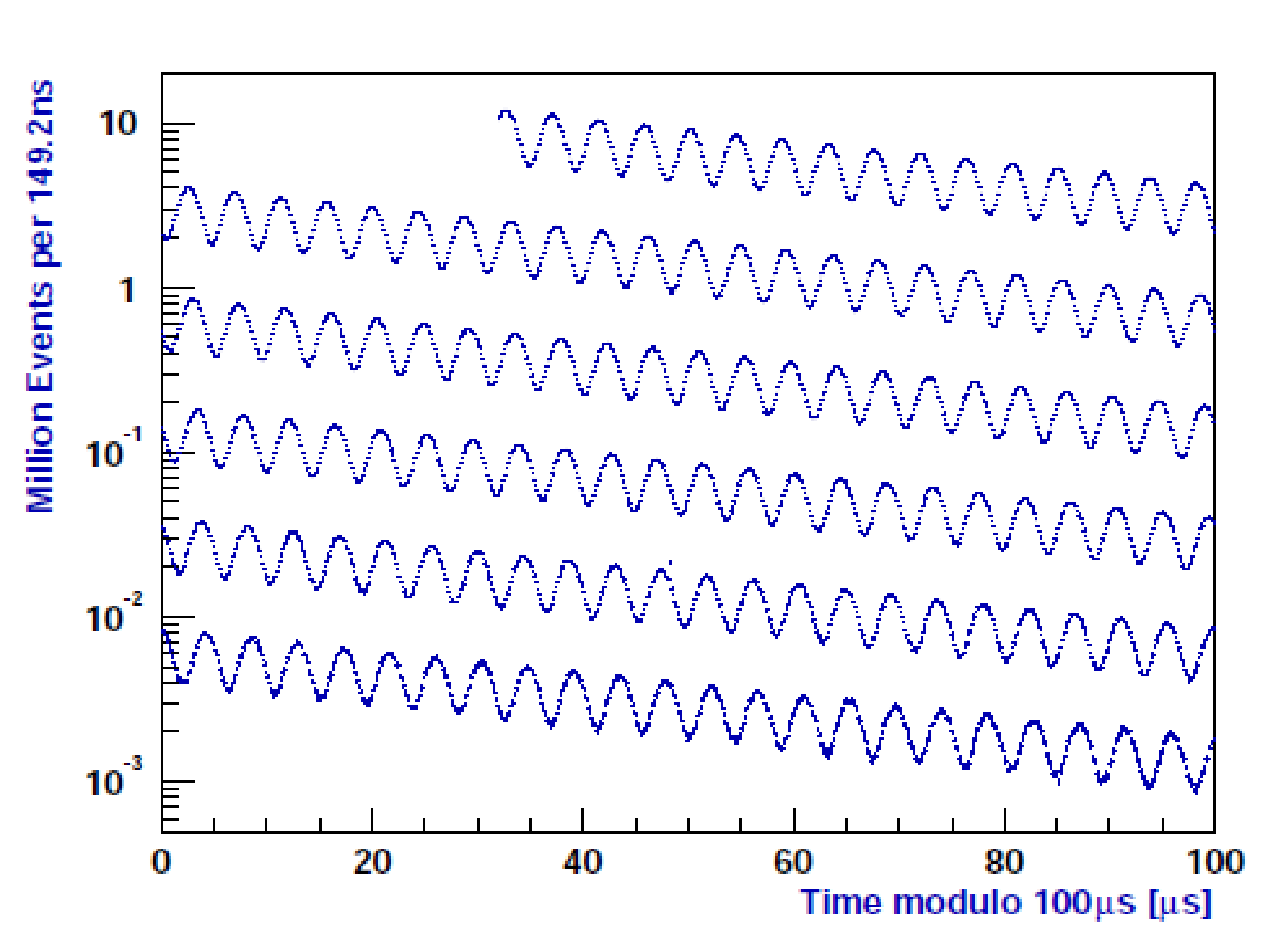}
\end{center}
\caption{Distribution of counts versus time. Figure from~\cite{Jegerlehner:2009ry}.}
\label{lab4}
\end{figure}

A diagrammatic scheme of the $a_\mu$ measurement is shown in
Figure~\tref{lab1}~\cite{Jegerlehner:2009ry}.
To measure the magnetic anomaly an electric field $\mathbf{E}$
and/or a magnetic field $\mathbf{B}$ must be applied. The general
formula, derived by Michel and Telegdi~\cite{Knecht:2003kc} in
1959 for this purpose, reads
\begin{equation}\label{12}
\mbox{\boldmath$\omega_a$}=\mbox{\boldmath$\omega_s$}-\mbox{\boldmath$\omega_c$}=
-\frac{e}{m_\mu c} \left\{a_\mu
\mathbf{B}-\left[a_\mu+\frac{1}{1-\gamma^2}\right]\mbox{\boldmath$\beta$}\times\mathbf{E}
\right\} -\frac{2d_\mu}{\hbar}\{\mbox{\boldmath$\beta$}\times\mathbf{B}+\mathbf{E}\}\, \ ,
\end{equation}
where $\omega_c= eB/m_{\mu}\gamma$ is the cyclotron frequency,
$\omega_{s}=eB/m_\mu\gamma+a_\mu eB/m_\mu$,
$\gamma=1/\sqrt{1-v^2}$ and $v$ the muon speed. If one forgets
about the electric dipole moment of the muon,$d_\mu$, so that
$\omega_a$ is independent of $d_\mu$, and chooses $\gamma$ such
that $a_{\mu}-1/(\gamma^2-1)=0$, which corresponds to the energy
$3.1$ GeV, called the magical energy, the measurement of $a_\mu$
reduces to measuring the magnetic field and the value of
$\omega_a$. As for $\omega_a$, one should note that the direction
of the muon spin is determined by detecting the electrons
resulting from the decay $\mu^-\rightarrow e^-
+\nu_e+\bar{\nu}_\mu$, or positrons from the decay of $\mu^+$ as
shown in Figure~\tref{lab2}.
 The number of electrons detected with
an energy above some threshold $E_t$, decreases exponentially with
time as shown in Figure~\tref{lab4}, according to the formula
\begin{equation}\label{13}
N_e(t)=N_0(E_t)e^{-t/\gamma\tau_\mu}\left\{1+A(E_t)\cos[(\omega_a)t+\Phi(E_t)]\right\}\
,
\end{equation}
where $\tau_\mu$ is the muon's lifetime in the laboratory frame.
This allows one to extract $\omega_a$. Then, one uses the relation
\begin{equation}\label{14}
B=\frac{\omega_p}{2\mu_p}\, \ ,
\end{equation}
between the Larmor spin precession angular velocity of the proton,
$\omega_p$, the proton Bohr magneton, $\mu_p$, and the magnetic
field $B$, to obtain
\begin{equation}\label{abrelation}
a_\mu=\frac{R}{\lambda-R}\, \ ,
\end{equation}
where $R=\omega_a/\omega_p$ and $\lambda=\mu_\mu/\mu_p$ with
$\mu_\mu$ the muon Bohr magneton. The value of $\lambda$ is
measured separately and is used by the experiment to obtain
$a_\mu$ via the relation~(\ref{abrelation}).

\begin{figure}
\begin{center}
\includegraphics[width=10cm,height=7cm]{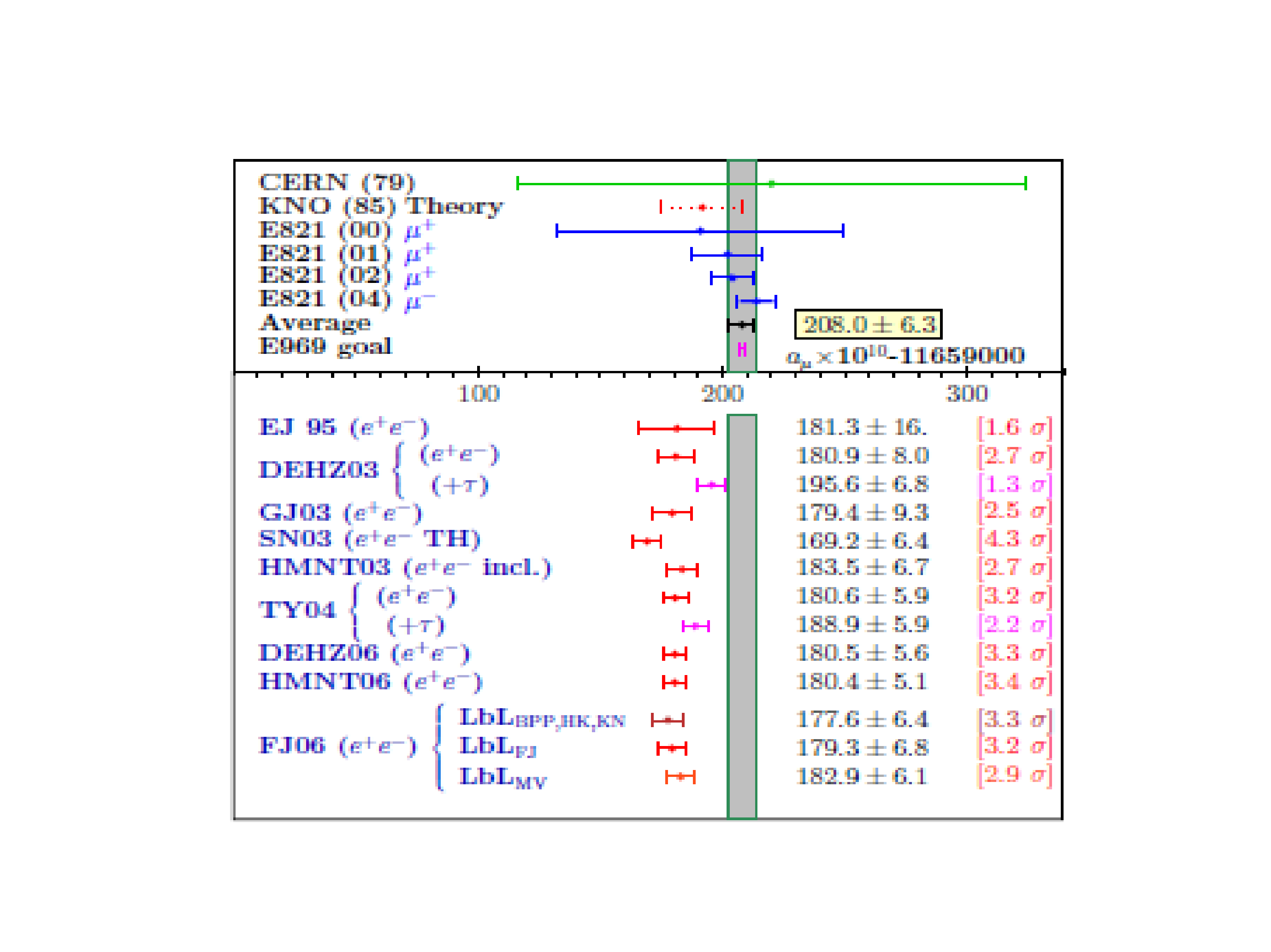}
\end{center}
\caption{Comparison between the theoretical and experimental
values of the $a_\mu$, experimental results in top and theoretical
values in below. Table from~\cite{Jegerlehner:2007}.}
\label{lab3}
\end{figure}

 Before the E821 experiment at Brookhaven national
laboratory between 2001 and 2004~\cite{Jegerlehner:2009ry}, results of a series of
measurements accomplished in the Muon Storage Ring at CERN were in
good agreement with theoretical predictions of the Standard Model
of particle physics, that is
\begin{equation}\label{15}
a_{\mu}^{exp}=1165924.0(8.5)\times10^{-9}\qquad
a_{\mu}^{th}=1165921(8.3)\times10^{-9}\, \ .
\end{equation}
The BNL experiment managed to improve the CERN experiment $14$
fold. The BNL average value is~\cite{Bennett006}
\begin{equation}\label{16}
a_{\mu}=11659208.0(3.3)[6.3]\times10^{-10}\, \ ,
\end{equation}
where the uncertainties are statistical and systematic. The
comparison between the experimental and theoretical values has
been done in Figure~\tref{lab3}.
As can be seen, judging by the experimental accuracy achieved in
the past decade at BNL, a small discrepancy at the $2$ to $3$
$\sigma$ level has persisted with the theoretical predictions.
 This discrepancy is still debated and many conjectures
have been made to link it with physics beyond the standard model.

\subsection{Overview of this work}\rlabel{Ove}
However, as mentioned above, the theoretical predictions in the
realm of the SM are still obscured by the hadronic calculations.
This work will try to address the charged pion loop, as a part of
the HLL scattering contribution to $a_\mu$. The structure of this
work is as follows. In Sec.~\ref{chap:QCD}, QCD and its chiral symmetry
will be discussed to give an introduction to ChPT. Sec.~\ref{chap:HLS} is
devoted to the Hidden Local Symmetry (HLS) model, as an extension
of the ChPT. Sec.~\ref{Had}
 will have a closer look
into generalized Feynman vertices for different models and some
short distance constraints. In Sec.~\ref{chap:Muon} the main body
of calculation of $a_\mu$ via different models is discussed and
Sec.~\ref{chp:momentum}
 deals with the role of different momentum regions contribution
to $a_\mu$. Finally,
 in Sec.~\ref{chp:concl} conclusions and prospects are given.

\section{QCD and chiral Symmetry}
\label{chap:QCD} \setcounter{equation}{0}
\subsection{Effective field theory}\rlabel{Eff}
There is a folklore theorem ascribed to Weinberg which
states~\cite{Weinberg:1978kz}: For a given set of asymptotic
states, perturbation theory with the most general Lagrangian
containing all the terms allowed by the assumed symmetries will
yield the most general S-matrix elements consistent with
analyticity, perturbative unitarity, cluster decomposition and the
assumed symmetries. In other words, regardless of the underlying
theory, when the degrees of freedom and the symmetries relevant to
the energy scale at hand are known, the effective Lagrangian built
based on them will address the same physics as the underlying
theory. If a small parameter, $\lambda$ is also realized in the
effective theory, one can conduct perturbative calculations upon
this parameter. Having this in mind, one would go ahead with
constructing an effective theory of strong interactions in low
energies where the original QCD Lagrangian runs into problems due
to the fact that in the regime $p^2\ll 1$ GeV squared, where the
meson dynamics take place, the QCD coupling constant is large. In
this energy regime, the fundamental particles are hadrons rather
than quarks and gluons. To build a Lagrangian for a process
happening at a scale $p\ll\Lambda$, one can use a expansion in
powers of $p/\Lambda$ where $\Lambda$ is the cut-off energy of the
model. Then, the Lagrangian could be organized as a series of
growing powers of momenta, i.e. of derivatives as
 \be\label{17} {\cal
L}={\cal L}_2+{\cal L}_4+\cdot\cdot\cdot+{\cal
L}_{2n}+\cdot\cdot\cdot\, \ , \ee

where the subscript indicates the number of derivatives. After
building the Lagrangian like this, one should use Weinberg power
counting to realize to which order a given diagram belongs. Based
on the Weinberg power counting scheme, the most important
contribution to a given scattering comes from the tree level
diagram ${\cal L}_2$. The contribution from one loop diagrams is
suppressed with respect to tree level and is the same size as the
level of contribution from the lagrangian ${\cal L}_4$. The
formula determining the power counting quantitatively for a given
diagram is
 \be
D=2+\sum_{n=1}^{\infty}2(n-1)N_{2n}+2N_L\, \ , \ee
 where $N_{2n}$ is
the number of vertices originating from ${\cal L}_{2n}$ and $N_L$
is the number of loops.

\subsection{Linear sigma model}\rlabel{Lin}

 To understand how an effective theory can
describe dynamics correctly, forgetting about the underlying
theory, we digress to the linear sigma model as an example. Let's
start with the linear sigma model Lagrangian
\begin{equation}\label{lagrangi}
L=\frac{1}{2}\partial_\mu\mbox{\boldmath$\phi$}\cdot\partial^\mu\mbox{\boldmath$\phi$}^T-
\mu\left(\mbox{\boldmath$\phi$}\cdot\mbox{\boldmath$\phi$}-a^2\right)^2\, \ ,
\end{equation}
Where the vector field
$\mbox{\boldmath$\phi$}=(\phi_1,\cdot\cdot\cdot,\phi_N)$ is a N-component real
scalar field. The field has a nonzero vacuum expectation value
that is $\phi_0^2=\phi_1^2+\cdot\cdot\cdot+\phi_N=a^2$. We assume that among
an infinite number of ground states that satisfy this condition,
one of them is chosen dynamically so that, the symmetry is
spontaneously broken to the sub group $H\equiv O(N-1)$. This leads
to generation of $N-1$ Goldstone bosons according to the Goldstone
theorem~\cite{Goldstone:1962es}, which are taken to be $\pi^i$.
Taking a simple choice $\mbox{\boldmath$\phi$}_0=(a,0,..,0)$ and
expanding $\mbox{\boldmath$\phi$}$ around
$\mbox{\boldmath$\phi$}_0$, assuming $\sigma\ll a$ as the
parameter of expansion and then integrating out the $\sigma$
field, one finds the corresponding effective Lagrangian of the non
linear sigma model. In Figure~\tref{fig10} the difference between
the case when the sigma particle exists and when it is integrated
out in the limit of $p^2/m_\sigma^2\ll 1$ is depicted.
\begin{figure}
\centering
\setlength{\unitlength}{0.5pt}
\begin{picture}(140,120)(0,-20)
\SetScale{0.5}
\SetWidth{2}
\Line(0,100)(50,50)
\Line(50,50)(0,0)
\Line(50,50)(90,50)
\Line(90,50)(140,0)
\Line(90,50)(140,100)
\Vertex(50,50){5}
\Vertex(90,50){5}
%\Text(75,60)[]{(b)}
\end{picture}
\qquad\qquad \setlength{\unitlength}{0.5pt}
\begin{picture}(100,120)(0,-20)
\SetScale{0.5}
\SetWidth{2}
\Line(0,100)(50,50)
\Line(50,50)(100,100)
\Line(50,50)(0,0)
\Line(50,50)(100,0)
\Vertex(50,50){5}
%\Text(200,45)[]{(a)}
\end{picture}
\caption{
Left is the interaction, mediated via the $\sigma$ particle, right is the same interaction when the mediator has been integrated out.
}
\label{fig10}
\end{figure}

\subsection{Chiral symmetry}\rlabel{Chiral symmetry}
 Returning to QCD, one can observe that the degrees of freedom
 to be dealt with at low energy, namely baryons and mesons are quite
  different from quarks and gluons
 which are the main players at high energies and the method described
  for
 the sigma model to integrate out the heavy field and build up an
 effective theory does not seem to be applicable here. However,
 the fact that there exists an energy gap between the family of
 pseudo scalar mesons and the rest of the hadrons, and that they
 can be accounted for as Goldstone bosons of a broken symmetry,
 encourages us to look for an effective field theory to describe
 their interactions~\cite{Scherer}.
The QCD lagrangian is
\begin{equation}\label{21}
L=\sum_{flavors}
 \bar{\psi}(i\gamma^\mu\partial_\mu+g_sA^\mu\gamma_\mu-m_i)\psi_i-\frac{1}{4}G_{\mu\nu}G^{\mu\nu}\, \ ,
\end{equation}
where $A^\mu$ is the gluon field and $G_{\mu\nu}$ is the gluon
field strength tensor with the definition
\begin{equation}\label{22}
G_{\mu\nu}=\partial_\mu A_\nu-\partial_\nu A_\mu-g_sfA_\mu A_\nu\, \ ,
\end{equation}
with the $f$ coefficients the structure constants of the group
$SU(3)_{colour}$. One can define the left handed and right handed
fields as
\begin{equation}\label{23}
\psi_R=\frac{1}{2}(1+\gamma_5)\psi \qquad
\psi_L=\frac{1}{2}(1-\gamma_5)\psi
\end{equation}
and $\psi=\psi_L+\psi_R$. The QCD Lagrangian when written in terms
of $\psi_L$ and $\psi_R$ is

\begin{eqnarray}\label{24}
L&=&\sum_{flavors}
\bar{\psi}_L(i\gamma^\mu\partial_\mu+g_sA^\mu\gamma_\mu)
\psi_{iL}\cr{}\ &+&
\bar{\psi}_R(i\gamma^\mu\partial_\mu+g_sA^\mu\gamma_\mu)\psi_{iR}-m_i\bar{\psi}_{iL}\psi_{iR}
-m_i\bar{\psi}_{iR}\psi_{iL}-\frac{1}{4}G_{\mu\nu}G^{\mu\nu}\,
\ .
\end{eqnarray}
If one drops the mass terms the Lagrangian is invariant under
following transformations
\begin{equation}\label{25}
\psi_{iL}\rightarrow\exp(-i\alpha_L\cdot\lambda)\psi_{iL}\qquad\psi_{iR}
\rightarrow\exp(-i\alpha_R\cdot\lambda)\psi_{iR}\, \ ,
\end{equation}
where $\lambda^a(a=1,2,\cdot\cdot\cdot,8)$ are the $SU(3)$ Gell-Mann matrices
in the flavor indices. The Lagrangian is said to have an
approximate symmetry $G=SU(3)_L\times SU(3)_R$ or chiral symmetry.
Of course quarks are massive and the chiral symmetry is not
realized fully in nature however, for three lightest quarks $u, d,
s$, it could be assumed to hold approximately. But as this
symmetry is not visible in the spectrum of light
hadrons~\cite{Scherer}, it should be spontaneously broken in
nature due to some spontaneous symmetry breaking (SSB) mechanism.
This leads to the global symmetry $G=SU(3)_L\times SU(3)_R$ to be
reduced to the subgroup $H=SU(3)_V$. This being the case, the
Goldstone theorem dictates that the difference between the
original number of generators and the final ones, should have
turned into Goldstone bosons. In the case at hand the number of
Goldstone bosons is $8$. As the chiral symmetry is also broken
explicitly due to the quark masses in the QCD Lagrangian, the
bosons could be recognized as the pseudo scalar mesons, which have
acquired a small mass due to this explicit symmetry breaking.
\subsection{ChPT}\rlabel{Chpt}
\subsubsection{Lowest order}
 Now
that the ground have been laid, one can go ahead by constructing
an effective QCD theory at low energies. The most general
Lagrangian invariant under Lorentz and chiral transformations in
the lowest order has the form~\cite{Gasser}
\begin{equation}\label{26}
{\cal L}_{2}=\frac{F_0^2}{4}[tr(D_\mu U^\dagger D^\mu U)+2B_0tr(U
M^\dagger+M U^\dagger)]\, \ ,
\end{equation}
where $F_0$ is the pion decay constant in the limits of the
massless pion, $B_0$ is related to the chiral quark condensate and
$U$ can be shown to be the $SU(3)$ matrix, written in terms of the
Goldstone fields as
\begin {equation}\label{U}
U= \exp\left(i{F_0}\phi\right)\, \ ,
\end{equation}
where

\begin{eqnarray}\label{mat}
\phi = \left(
                 \begin{array}{ccc}
                 \frac{1}{\sqrt{2}}\pi^0+\frac{1}{\sqrt{6}}\eta& \pi^+ & K^+\\
                  \pi^- & -\frac{1}{\sqrt{2}}\pi^0+\frac{1}{\sqrt{6}}\eta & K^0\\
                   K^- &  \bar{K}^0  & -\frac{2}{\sqrt{6}}\eta
                 \end{array}
               \right)
\end{eqnarray}

The covariant derivative is \be \label{covariant} D_\mu U =
\partial_\mu U -i l_\mu U + i U r_\mu\, \ , \ee with right and left
external fields reducing to
 \ba\label{rl}
 l_\mu&=& -eQA_\mu\nonumber\\
 r_\mu&=&-eQA_\mu\, \ ,
\ea
for this work with $e$, the electromagnetic coupling and
\[
Q =
{1\over 3}\left(\begin{array}{ccc}
2 & 0  & 0 \\
0 & -1 & 0\\
0 & 0  &-1
\end{array}\right)\, \ .
\]

Now let's see how can one actually calculate with these tools. To
find the amplitude for the scattering $\gamma(q,\varepsilon)
\rightarrow \pi(p)+\pi(p^\prime)$, one has
 \be\label{29}
 r_\mu=l_\mu=-eQ A_\mu
\ee
 and hence
 \ba\label{cova}
D_{\nu}U&=&\partial_\nu U+ieA_\nu[Q,U]\nonumber\\
D_{\nu}U^\dagger&=&\partial_\nu U^\dagger+ieA_\nu[Q,U^\dagger]\, \ . \ea
Then, starting from the lowest order Lagrangian the corresponding
term is

\ba\label{31}
{F^2\over 4}\langle D_\mu U(D^\mu U)^\dagger
\rangle&=&{F^2\over 4}
\langle \partial^\mu U\partial^\mu U^\dagger \rangle\nonumber\\
&-&ieA_\mu{F^2\over 4}\langle Q[(\partial^\mu UU^\dagger-U^\dagger\partial^\mu
 U)] \rangle\nonumber\\
&-& A_\mu A^\mu {F^2\over 4} \langle [Q,U][Q,U^\dagger] \rangle\, \ .
\ea
Putting in from~\rref{mat} and keeping terms only up to second
order of $\phi$, the second term reads
\begin{equation}\label{32}
{\cal L}=-e\frac{i}{2}A_\mu \langle
Q[\partial^\mu\phi,\phi]\rangle\, \ .
\end{equation}
Inserting from~\rref{mat} only for the pion field of $\phi$ one
gets

\begin{eqnarray}\label{33}
Q[\partial^\mu\phi,\phi] = \left(
                 \begin{array}{ccc}
                  2(\partial^\mu\pi^+\pi^--\pi^+\partial^\mu\pi^-) & 0 & 0\\
                  0 & 2(\partial^\mu\pi^-\pi^+\!\!-\!\!\pi^-\partial^\mu\pi^+) & 0\\
                  0  &  0 & 0
                 \end{array}
               \right)
\end{eqnarray}
and after performing the trace one easily finds
\begin{equation}\label{34}
{\cal L}=-eiA_\mu(\partial^\mu\pi^+\pi^--\pi^+\partial^\mu\pi^-)\ .
\end{equation}
Hence, the Feynman rule for the scattering $\gamma(q,\varepsilon)
\rightarrow \pi^+(p)+\pi^-(p^\prime)$, using \be\label{A}
A_\mu(x)=\int{d^3p\over(2\pi)^3}{1\over
\sqrt{2E_\mathbf{p}}}\sum_{r=0}^3\left(a_{\mathbf{p}}^r\varepsilon_\mu^r(p)\exp(-ip\cdot
x)+a_{\mathbf{p}}^{r^\dagger}\varepsilon_\mu^{r\ast}(p)\exp(ip\cdot
x)\right) \ee
 and
\be \label{fi} \phi(x)=\int{d^3p\over(2\pi)^3}{1\over
\sqrt{2\omega_\mathbf{p}}}\Big(a_{\mathbf{p}}^\dagger\exp(ip\cdot
x)+a_{\mathbf{p}}\exp(-ip\cdot x)\Big)\, \ , \ee
 as the photon and pion
field respectively, reads
\begin{equation}\label{35}
{\cal M}= ie\varepsilon\cdot(p+p^\prime)
\end{equation}
and the vertex is proportional to
\be \label{gpp}
ie(p_\mu+p_\mu^\prime)\, \ . \ee
Following the same lines for the scattering
$\gamma(q,\varepsilon)+\gamma(q^\prime,\varepsilon^\prime)\rightarrow
\pi^+(p)+\pi^-(p^\prime)$, the Lagrangian becomes
\begin{equation}\label{36}
{\cal L}=e^2A_\mu A^\mu\pi^+\pi^-
\end{equation}
and the amplitude is
\begin{equation}\label{37}
{\cal M}=2ie^2\varepsilon^{^\prime\star}\cdot\varepsilon\, \ ,
\end{equation}
which leads to the vertex
 \be\label{ggpp}
 2ie^2g_{\mu\nu}\
. \ee
\subsubsection{${L}_9$ and ${L}_{10}$}
\label{chap:L9}
The Lagrangian in order $p^4$ of ChPT has the form~\cite{Gasser}
\ba\label{p4}
 {\cal L}_{4}&=& L_1 \langle D_\mu U^\dagger D^\mu U \rangle^2
+L_2 \langle D_\mu U^\dagger D_\nu U \rangle
     \langle D^\mu U^\dagger D^\nu U \rangle \nonumber\\&&\hspace{-0.5cm}
+L_3 \langle D^\mu U^\dagger D_\mu U D^\nu U^\dagger D_\nu
U\rangle +L_4 \langle D^\mu U^\dagger D_\mu U \rangle \langle
\chi^\dagger U +\chi U^\dagger \rangle \nonumber\\&& +L_5 \langle
D^\mu U^\dagger D_\mu U (\chi^\dagger U+U^\dagger \chi ) \rangle
+L_6 \langle \chi^\dagger U+\chi U^\dagger \rangle^2 \nonumber\\&&
+L_7 \langle \chi^\dagger U-\chi U^\dagger \rangle^2 +L_8 \langle
\chi^\dagger U \chi^\dagger U + \chi U^\dagger \chi U^\dagger
\rangle \nonumber\\&& -i L_9 \langle F^R_{\mu\nu} D^\mu U D^\nu
U^\dagger +
               F^L_{\mu\nu} D^\mu U^\dagger D^\nu U \rangle
\nonumber\\&& +L_{10} \langle U^\dagger  F^R_{\mu\nu} U
F^{L\mu\nu} \rangle\,, \ea where the field strength tensor reads
\be\label{27}
 F_{\mu\nu}^{L(R)} = \partial_\mu l(r)_\nu -\partial_\nu
l(r)_\mu -i \left[ l(r)_\mu , l(r)_\nu \right]\,, \ee
 with
\be\label{28} \chi = 2 B_0\left(s+ip\right)\,, \ee in which $s$,
$p$, $l_\mu$ and $r_\mu$ denote the scalar, pseudo scalar, left
and right handed external fields, respectively~\cite{Gasser}.
 The terms of interest in this Lagrangian for our purpose are those containing ${L}_9$ and ${L}_{10}$.
 These term correspond to the pion charge radius and pion
polarizability.

One can rewrite the term containing ${ L}_9$ as \ba\label{main}
{\cal L}_9&=&-i\langle F_{\mu\nu R}D_\mu UD_\nu U^\dagger+F_{\mu\nu L}D_\mu U^\dagger D_\nu U\rangle\nonumber\\
&=& i\langle D_\mu F_{\mu\nu R}UD_\nu U^\dagger+D_\mu F_{\mu\nu L}U^\dagger D_\nu U\rangle\nonumber\\
&+& i\langle F_{\mu\nu R}UD_\mu D_\nu U^\dagger+F_{\mu\nu
L}U^\dagger D_\mu D_\nu U\rangle={\cal L}_9^1+{\cal L}_9^2\, \ .
\ea
The second term can be written as
\be\label{68}
 {\cal L}_9^2={i\over 2}\langle F_{\mu\nu R}
U[D_\mu,D_\nu]U^\dagger+F_{\mu\nu L}U^\dagger[D_\mu,D_\nu]U\rangle\, \ ,
\ee
which using the equalities
\be\label{69} [D_\mu,D_\nu]U^\dagger= -i(F_{\mu\nu
L}U^\dagger-U^\dagger F_{\mu\nu R})\qquad [D_\mu,D_\nu]U
=-i(F_{\mu\nu R}U-UF_{\mu\nu L})\, \ , \ee
takes the same form as
${\cal L}_{10}$ in the above Lagrangian. Since we are dealing only
with electromagnetic interaction, $F_{\mu\nu L}=F_{\mu\nu R}$ and
${\cal L}_9^1$ becomes 
\be
%\label{1}
 {\cal L}_9^1=\langle D_\mu
F^{\mu\nu}[UD_\nu U^\dagger+U^\dagger D_\nu U]\rangle\, \ , \ee
 Which upon use of the relation~(\ref{cova}) reads
\ba\label{70} {\cal L}_9^1=
\langle D_\mu F_{\mu\nu}\Big[U\partial U^\dagger &+& ieA_\nu U[Q,U^\dagger]\nonumber\\
U^\dagger\partial\nu U &+& ieA_\nu U^\dagger[Q,U]\Big] \rangle\, \
. \ea Then using the definition~(\ref{U}), expanding $U$ and
keeping terms up to the order $\phi^2$ one finds
 \ba\label{71}
U\partial_\nu U^\dagger&=&-\partial_\nu U U^\dagger=-i{\partial_\nu \phi\over F_0}\nonumber\\
U^\dagger\partial_\nu U&=&-\partial_\nu U^\dagger U=
i{\partial_\nu \phi\over F_0} \ea
 and
\ba\label{72}
U[Q,U^\dagger]&\simeq& [Q,{(i\phi)^2\over 2}]+i\phi[Q,-i\phi]\nonumber\\
U^\dagger[Q,U]&\simeq& [Q,{(i\phi)^2\over 2}]-i\phi[Q,i\phi]\, \ .
\ea Using relations~(\tref{mat}) and~(\ref{covariant}), the final
result is \ba\label{73} {\cal L}_9^1=-e\partial_\mu
f^{\mu\nu}\Big[2(\pi^+\partial_\nu\pi^--\pi^-\partial_\nu\pi^+)-2ieA_\nu\pi^+\pi^-\Big]\,
\ . \ea The second part of the Eq.~(\ref{main}) can also be
calculated in the same way to give
 \be\label{74}
 {\cal
L}_9^2=-4e^2f_{\mu\nu}f^{\mu\nu}\pi^+\pi^- \, \ ,
\ee
where
$F_{\mu\nu}=-eQf_{\mu\nu}$
is assumed and
\ba\label{75}
f_{\mu\nu}f^{\mu\nu}&=&(\partial_\mu A_\nu-\partial_\nu A_\mu)(\partial^\mu A^\nu-\partial^\nu A^\mu)\nonumber\\
&=&\partial_\mu A_\nu \partial^\mu A^\nu-\partial_\nu
A_\mu\partial^\nu A^\mu-\partial_\nu A_\mu\partial^\mu
A^\nu+\partial_\nu A_\mu\partial^\nu A^\mu \, \ . \ea
 One can also derive the term
corresponding to ${\cal L}_{10}$ similarly. The only remaining
task is to derive the part related to $F_{\mu\nu}$ and extract the
Feynman rules. For example for the $\gamma\gamma\pi\pi$ process
\be\label{76}
 {\cal M}=\langle kk^\prime|\partial_\mu A_\nu\partial^\mu
A^\nu\pi^+\pi^-|pp^\prime\rangle \, \ , \ee which via using
relation~(\ref{A}) and~(\ref{fi}), leads to the following
invariant amplitude
 \be\label{77} {\cal
M}=4\varepsilon_\mu(p)\varepsilon^\mu(p^\prime)p_\mu
p^{\prime\mu}-4p_{\mu}
p^{\prime\nu}\varepsilon_\nu(p)\varepsilon^\mu(p^\prime)\, \ .
 \ee

To get a better understanding of what $L_9$ and $L_{10}$ actually represent,
the charge radius of the pion is related to its electromagnetic
form factor in the low energy region via the definition~\cite{HLS}

 \be
F^{\pi^\pm}(p^2)=1+{p^2\over 6}\langle
r^{\pi^2}\rangle+\cdot\cdot\cdot\, \ . \ee Comparing this with the
pion form factor in the low energy limit one gets

\be \langle r^{\pi^2}\rangle={3a\over m_\rho^2}\, \ . \ee
Meanwhile, it could be shown that~\cite{Ecker}, $L_9\propto
1/m_\rho^2$ and hence, $L_9$ is proportional to the pion charge
radius. Using similar arguments, $L_{10}$ can be shown to be
related to the pion polarizability.

\section{Hidden local symmetry model}
\label{chap:HLS} \setcounter{equation}{0} As the Hidden Local
Symmetry (HLS) is used extensively in this work, we give a brief
introduction to it in this section. In fact, the HLS considers
vector mesons as its gauge bosons, achieving mass via eating up
the Goldstone bosons appearing as a result of breaking of the
hidden symmetry, which is added to the chiral symmetry of the ChPT
Lagrangian~\cite{HLS1}. Hence this model is a kind of
generalization of perturbation theory. Indeed, ChPT at its tree
level only covers the threshold energy and even after adding the
loop corrcetions, the energy it covers is fully below the chiral
symmetry breaking energy around $1$GeV~\cite{HLS}. As the energy
grows the $\rho$ meson should be inevitably considered. That is
where the HLS takes center stage. There are also some other
compatible models discussed in the literature \cite{HLS}.

As discussed above, symmetry of ChPT is of the type $G_{global}=
SU(N_f)_L\times SU(N_f)_R$, which in the HLS model is extended to
$G_{global}\times H_{local}$ with $H=SU(N_f)_V$. It is interesting
to mention that the HLS model reduces to ChPT in the low energy
region when the vector meson is integrated out. In the HLS model
the variable $U$ of ChPT, introduced in the relation~(\ref{U}) is
divided into two parts
 \be\label{38}
U=\xi_l^\dagger\xi_r \, \ . \ee These new variables can be
parameterized as
 \be \label{39}
\xi_{l,r}=\exp{(i\sigma/F_\sigma)}\exp{(\pm
i\pi/F_\pi)}\qquad\text{with}\qquad [\pi=\pi^aT_a,
\sigma=\sigma^aT_a]\, \ , \ee where $\pi$ denotes the Goldstone
bosons of the global symmetry and have the same definition as
before, while $\sigma$ denotes those of the local symmetry. These
are the Goldstone bosons absorbed by the vector mesons to get
massive. Also, $F_\pi$ and $F_\sigma$ are the corresponding decay
constants respectively and $T_a$ are the generators of the group.
Then, one can introduce the covariant derivative including the
external fields
 \be\label{45}
D_\mu\xi_l=\partial\xi_l-iv_\mu\xi_l+i\xi_l l_\mu\qquad
D_\mu\xi_r=\partial\xi_r-iv_\mu\xi_r+i\xi_r r_\mu\, \ , \ee where
$r$ and $l$ are the same as in~(\ref{rl}), with the gauge fields
of the $H_{local}$ defined as\footnote{This does include an extra
$U(1)$ global but the $\omega$ plays no role in this work.}\
\[
\rho_\mu={v_\mu^a\over g} T_a=
{1\over\sqrt{2}}\left(\begin{array}{ccc}
\frac{1}{\sqrt{2}}\left(\rho_\mu^0+\omega_\mu\right) & \rho_\mu^+ & K_\mu^{\star,+} \\
\rho_\mu^- &-{1\over\sqrt{2}}\left(\rho_\mu^0+\omega_\mu\right)& K_\mu^{\star,0}\\
K_\mu^{\star,-} & \bar{K}_\mu^{\star,0}  & \phi_\mu
\end{array}\right)\, \ ,
\]
satisfying the field strength

\be \label{41} v_{\mu\nu}=\partial_\mu v_\nu-\partial_\nu
v_\mu-i[v_\mu,v_\nu] \, \ . \ee

Then, one can build two independent 1-forms out of the above
variables \be\label{42}
\alpha_\perp^\mu(x)={(D_\mu\xi_r\cdot\xi_r^\dagger-D_\mu\xi_l\cdot\xi_l^\dagger)\over
2i}\qquad
\alpha_\Vert^\mu(x)={(D_\mu\xi_r\cdot\xi_r^\dagger+D_\mu\xi_l\cdot\xi_l^\dagger)\over
2i}\, \ . \ee Using these 1-forms, one can build the lowest order
lagrangian including $\xi_{l,r}$ and $D_\mu\xi_{l,r}$ to the
lowest derivative as
 \be \label{43} {\cal L}={\cal L}_A+a{\cal
L}_V=F_\pi^2
tr[(\hat{\alpha}_\perp^\mu(x))^2]+F_\sigma^2tr[(\hat{\alpha}_\Vert^\mu(x))^2]\,
\ ,
 \ee
where
 \be\label{44} a\equiv {F_\sigma^2/F_\pi^2} \, \ , \ee
 is a constant.

Finally, adding the kinetic term of the gauge bosons, the
Lagrangian, in the unitary gauge $\sigma=0$, takes the form
\ba\label{46}
{\cal L}&=&{\cal L}_A +a{\cal L}_V+{\cal L}_{int}(V_\mu)\nonumber\\
&=& F_\pi^2 tr[(\hat{\alpha}_\perp^\mu(x))^2]+F_\sigma^2tr[(\hat{\alpha}_\Vert^\mu(x))^2]-{1\over 2g^2}tr[V_{\mu\nu}V^{\mu\nu}]\nonumber\\
&=& tr[\partial_\mu\pi\partial^\mu\pi]+ag^2F_\pi^2tr[\rho_\mu\rho^\mu]+2i\left({1\over 2}ag\right)tr\left[\rho^\mu[\partial_\mu,\pi]\right]\nonumber\\
&-&2eagF_\pi^2A^\mu tr[\rho_\mu Q]+2ie\left(1-{a\over 2}\right)A^\mu tr\left[Q[\partial_\mu,\pi]\right]\nonumber\\
&+&ae^2F_\pi^2A_\mu A^\mu tr[QQ]+{4-3a\over
12F_\pi^2}tr\Big[[\partial_\mu,\pi][\partial_\mu,\pi]\Big]+\cdot\cdot\cdot\,
\ , \ea where $g$ is the HLS gauge coupling constant. from this
one can easily observe that the vector meson has acquired mass
equal to $ag^2F_\pi^2$ via the Higgs mechanism. Also other
couplings could be expressed as
 \ba\label{47}
g_{\rho\pi\pi}&=&{1\over 2}ag\nonumber\\
g_\rho &=& agF_\pi^2\nonumber\\
g_{\gamma\pi\pi}&=&\left(1-{a\over 2}\right)e \, \ .\ea

\begin{figure}
\centering

\subfigure[$2eg_{\rho\pi\pi}A^\mu\rho_\mu^0\pi^+\pi^-$]{
   \includegraphics[scale =.15] {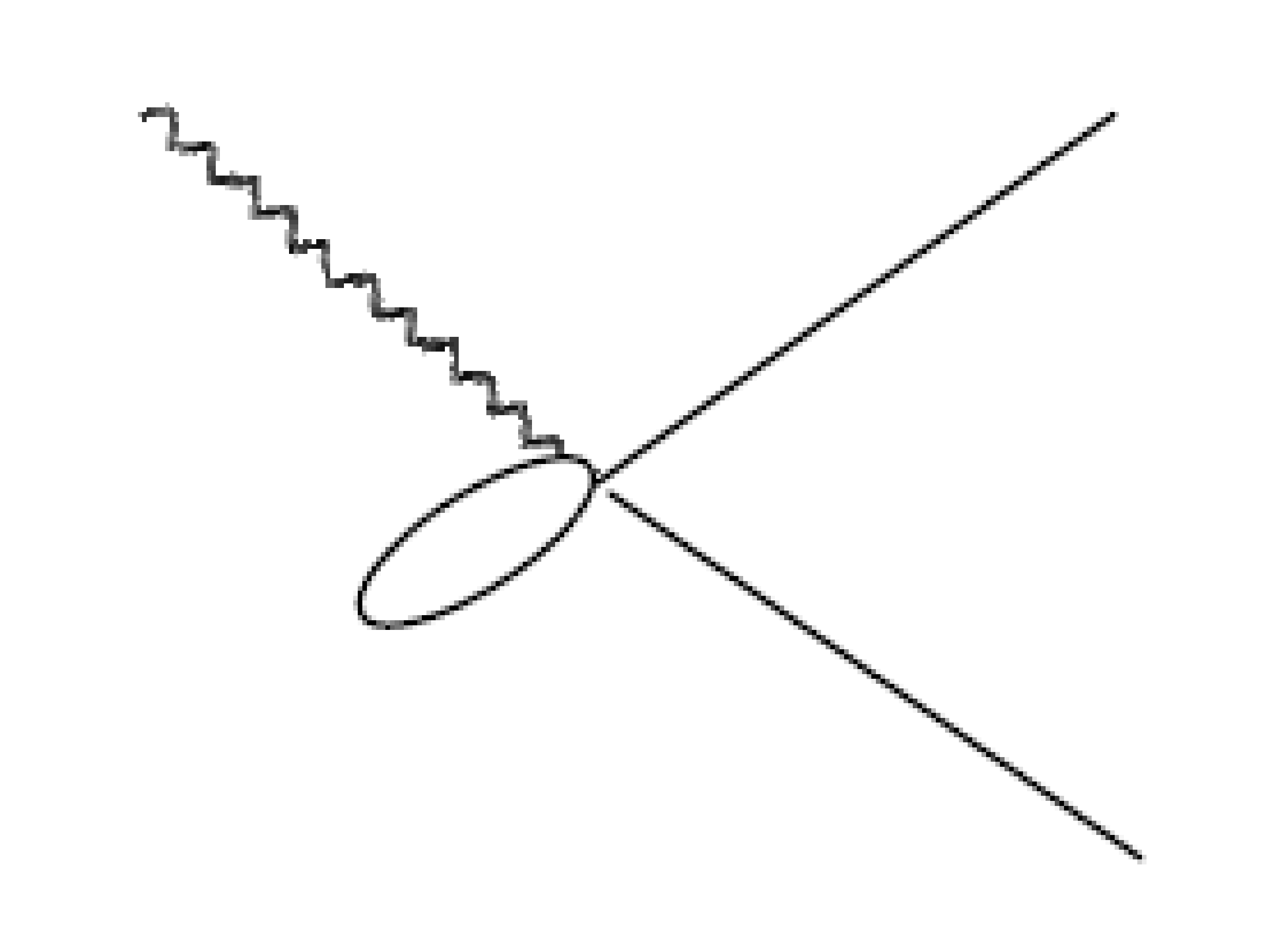}
   %\label{fig:subfig1}\qquad
 }

 \subfigure[$-eg_\rho A^\mu\rho_\mu^0$]{
   \includegraphics[scale =.15] {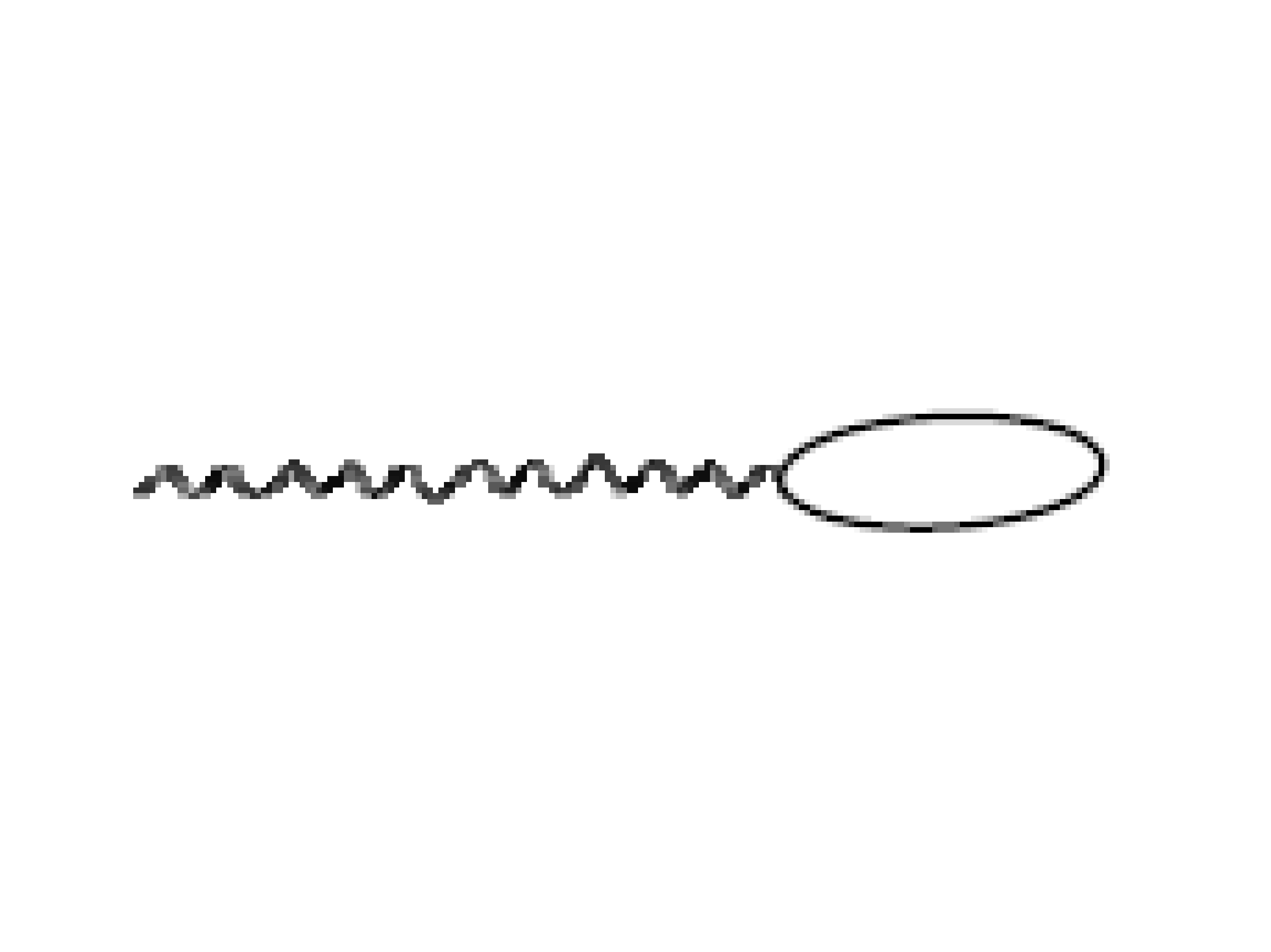}
   %\label{fig:subfig2}\qquad
 }

 \subfigure[$(1-a)e^2A^\mu A_\mu \pi^+\pi^-$]{
   \includegraphics[scale =.15] {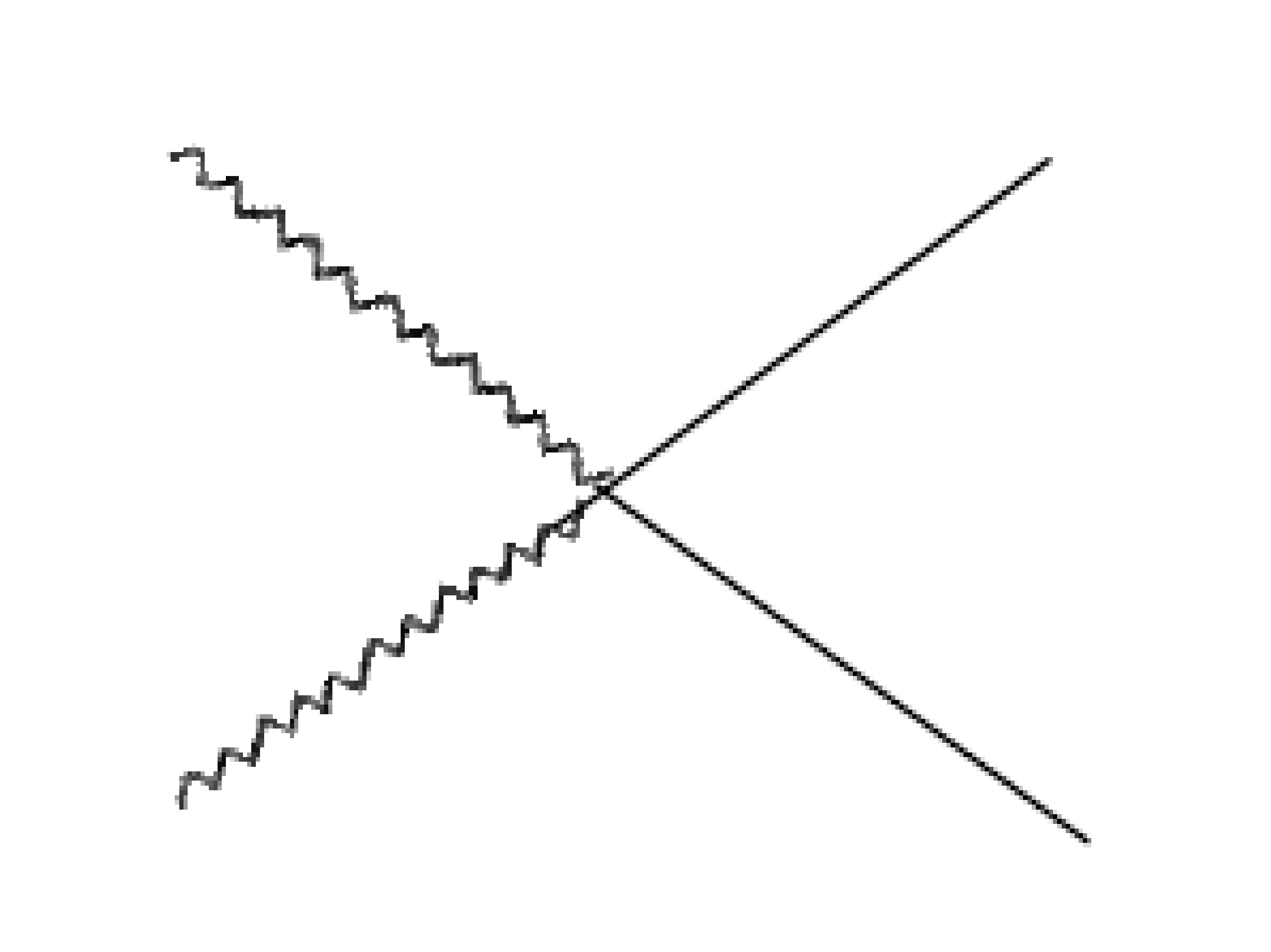}
   %\label{fig:subfig3}\qquad
 }

 \subfigure[$-ig_{\gamma\pi\pi}A_\mu(\pi^+\partial^\mu\pi^--\pi^-\partial^\mu\pi^+)$]{
   \includegraphics[scale =.15] {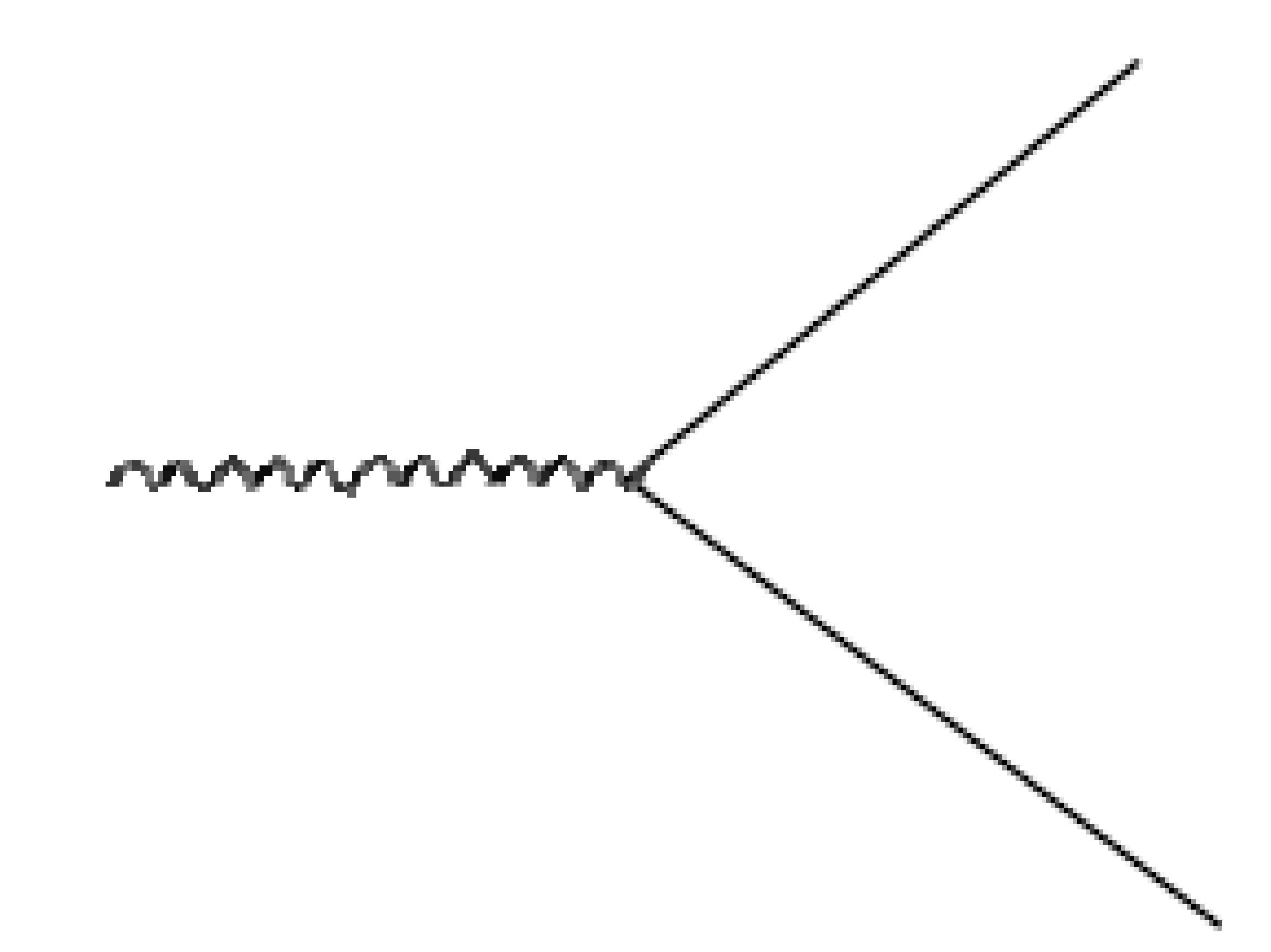}
   %\label{fig:subfig4}\qquad

 }

 \subfigure[$-ig_{\rho\pi\pi}\rho_\mu(\pi^+\partial^\mu\pi^--\pi^-\partial^\mu\pi^+)$]{
   \includegraphics[scale =.15] {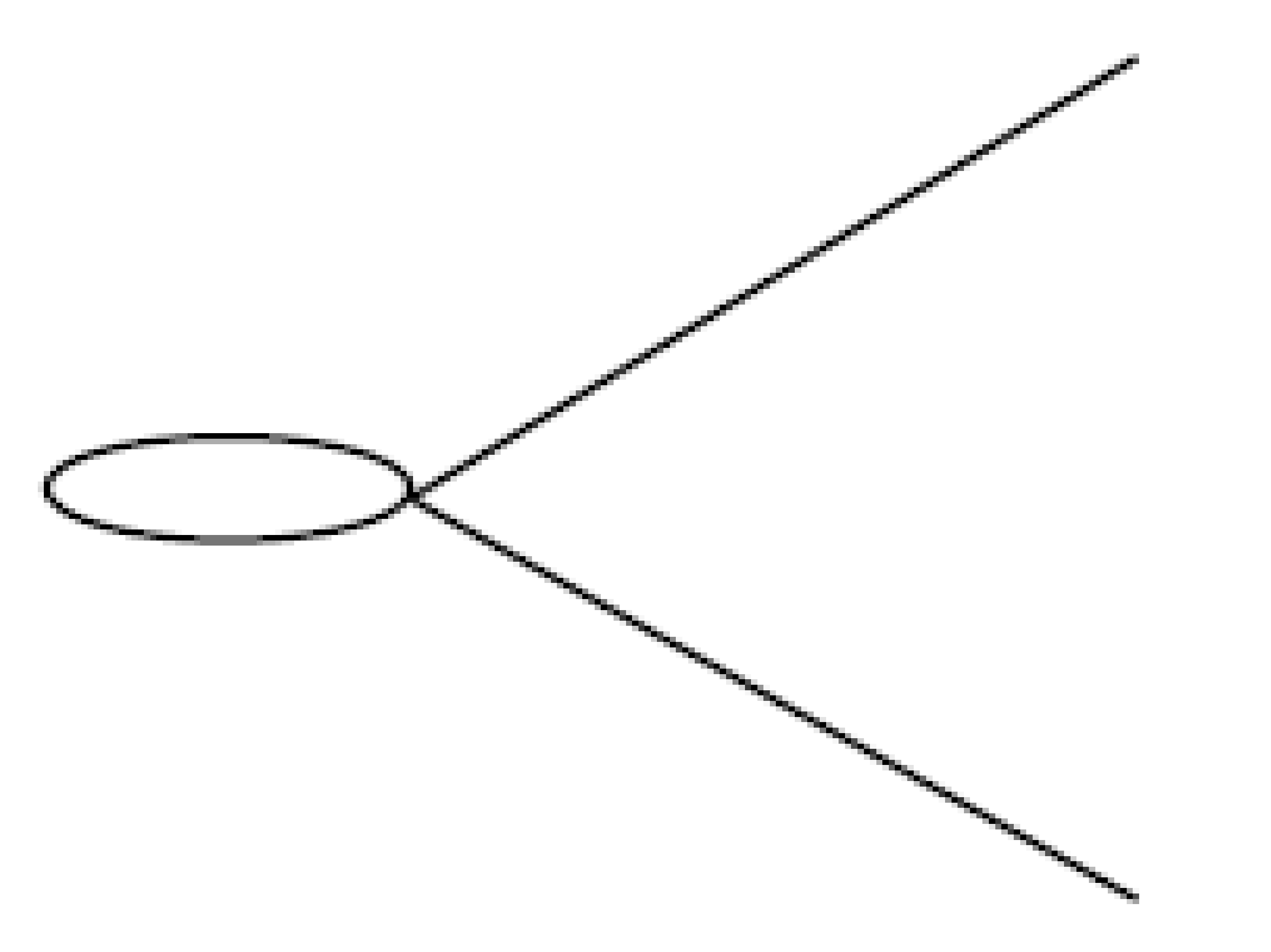}
   %\label{fig:subfig5}\qquad
 }
\caption{Different HLS Lagrangian terms with the corresponding Feynman diagrams.}
\label{gro}
\end{figure}

 The relevant terms of the above Lagrangian
for our purpose in this work, are~\cite{Hayakawa:1996ki}
\ba\label{50} {\cal L}_{int}&=&-eg_\rho
A^\mu\rho_\mu^0-ig_{\rho\pi\pi}\rho_\mu^0\pi^+\overleftrightarrow{\partial}^\mu\pi^--ig_{\gamma\pi\pi}A_\mu\pi^+
\overleftrightarrow{\partial}^\mu\pi^-
\nonumber\\
&+&(1-a)e^2A^\mu A_\mu
\pi^+\pi^-+2eg_{\rho\pi\pi}A^\mu\rho_\mu^0\pi^+\pi^- \, \ .\ea
 It
should be mentioned that the crucial property of this Lagrangian,
regarding our consideration of the HLL scattering is that it does
not have a $\rho^0\rho^0\pi^+\pi^-$ term. Corresponding diagrams
to each of the above terms are depicted in Figure~\tref{gro}.
Another property of the HLS lagrangian is that for the $a=2$ case
it reduces to the VMD for the pion single photon coupling, still
it is different from the full VMD version which includes the
$\rho\rho\pi\pi$ vertex as well.

\section{$\gamma\pi^+\pi^-$ and $\gamma\gamma\pi^+\pi^-$ vertices}
\label{Had} \setcounter{equation}{0}
 Up to this point necessary ingredients of a more
through discussion of the HLL contribution to $a_\mu$ are introduced.
 As discussed in the introduction, to cure the infinities one has to introduce vector
mesons in the calculation of the pion exchange, and this could be done via VMD models
or the HLS model. At this point one is ready to consider what kind
of change happens to the point diagrams of the ChPT Lagrangian
mentioned in the Sec.~\ref{Chpt}, when the HLS or VMD models are taken into
account. In the naive VMD model, one just replaces the photon
propagator with the term below
 \be\label{51} {ig_{\mu\nu}\over
q^2}\rightarrow{ig_{\mu\nu}\over q^2}+{-ig_{\mu\nu}\over
q^2-m_\rho^2}\equiv{ig_{\mu\nu}\over q^2}{m_\rho^2\over
q^2-m_\rho^2}\, \ . \ee
However, this simple model is not
compatible with Ward identities~\cite{Hayakawa:1996ki}.

\begin{figure}
\begin{center}
\includegraphics[width=8cm,height=4cm]{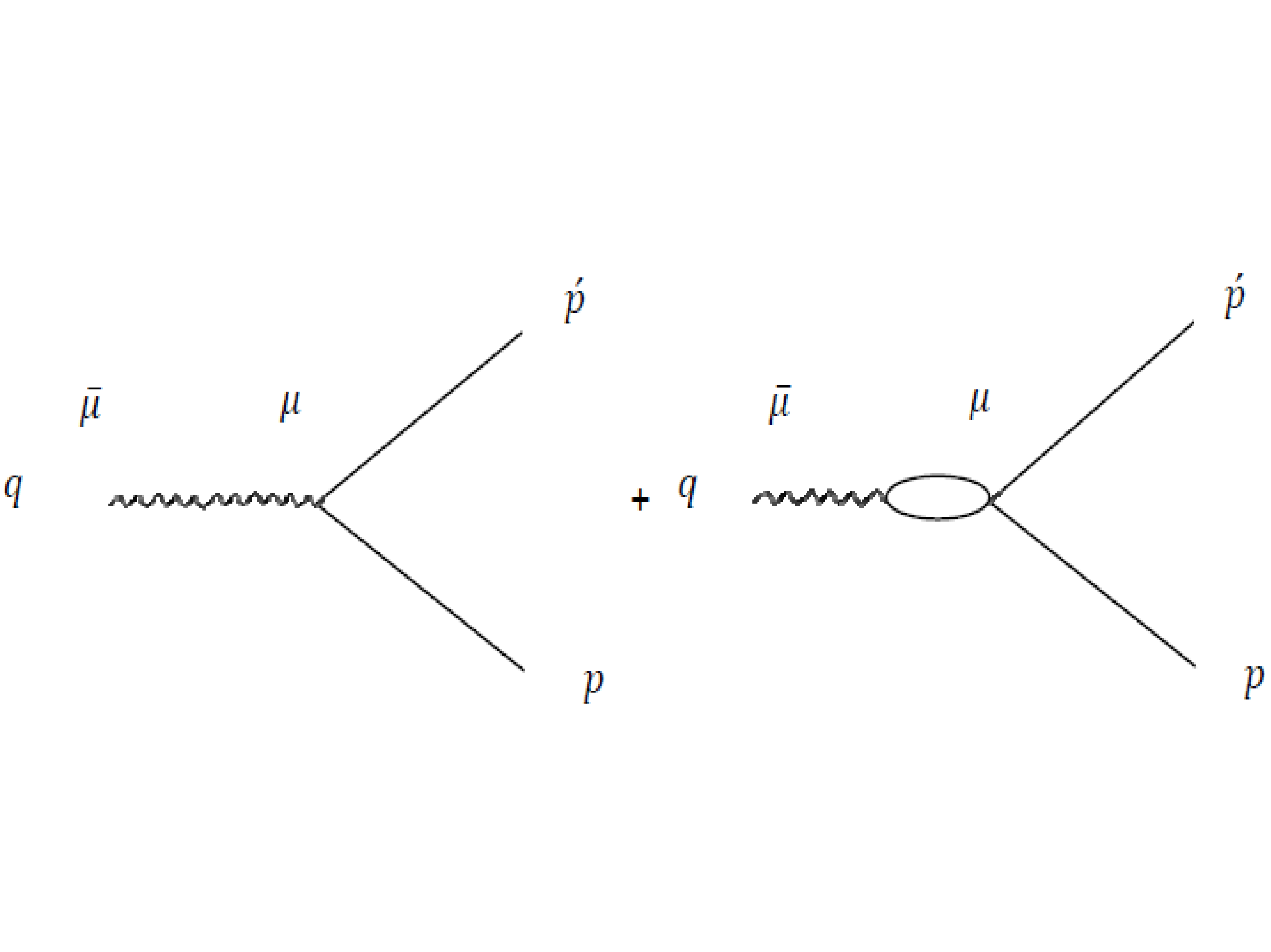}
\end{center}
\caption{The equivalent vertex of the $\gamma\pi\pi$ in the HLS model.}
\label{gppsum}
\end{figure}

\begin{figure}
\begin{center}
\includegraphics[width=10cm,height=6cm]{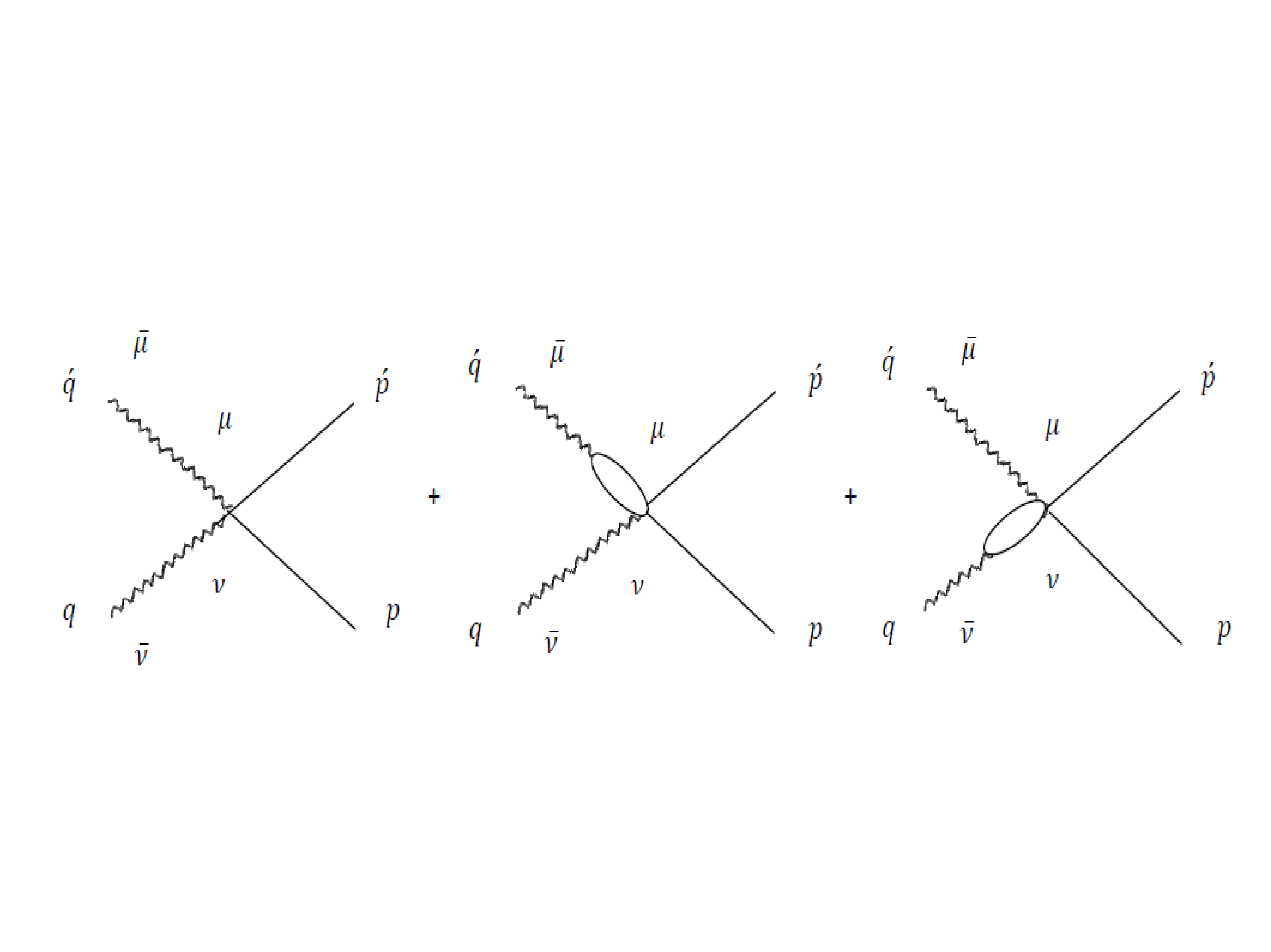}
\end{center}
\caption{The equivalent vertex of $\gamma\gamma\pi\pi$ in the
HLS.} \label{ggppsum2}
\end{figure}

To proceed more systematically, one can note that the amplitude
corresponding to the $\gamma$ to $\rho$ to $\pi\pi$, depicted in
the right of the Figure~\tref{gppsum} is
 \ba\label{52} {\cal M}=
ie\varepsilon\cdot(p+p^\prime){a\over
2}{(-i)(-i)(m_\rho^2g_{\mu\nu}-{q_\mu q_\nu})\over(q^2-m_\rho^2)}
\ea and hence, one only needs to multiply the vertex of
$\gamma\pi\pi$ from~(\ref{gpp}) with
 \ba\label{53}
 (1-{a\over 2})g_{\mu\bar{\mu}}-{a\over2}
{m_\rho^2g_{\mu\bar{\mu}}-q_\mu q_{\bar{\mu}}\over q^2-m_\rho^2
}=g_{\mu\bar{\mu}}-{a\over 2}{q^2g_{\mu\bar{\mu}}-q_\mu
q_{\bar{\mu}}\over q^2-m_\rho^2 }\, \ , \ea
 to find the equivalent vertex including both
diagrams shown in Figure~\tref{gppsum}. Following the same lines,
the amplitude corresponding to Figure~\tref{ggppsum2}, is found by
multiplying the $\gamma\gamma\pi\pi$ vertex~(\ref{ggpp}) with

\ba\label{54}
(1-a)g_{\mu\bar{\mu}}g_{\nu\bar{\nu}}\!\!\!&-&\!\!\!{a\over
2}g_{\nu\bar{\nu}} \Big({m_\rho^2g_{\mu\bar{\mu}}-q_\mu
q_{\bar{\mu}}\over q^2-m_\rho^2 }\Big) -{a\over 2}g_{\mu\bar{\mu}}
\Big({m_\rho^2g_{\nu\bar{\nu}}-q_\nu q_{\bar{\nu}}\over q^2-m_\rho^2 }\Big)\nonumber\\
\!\!\!&=&\!\!\!\Big[g_{\mu\bar{\mu}}g_{\nu\bar{\nu}}
+g_{\mu\bar{\mu}}{a\over2}{p^2g_{\nu\bar{\nu}}-p_{\bar{\nu}}p_{\nu}\over
m_\rho^2-p^2}
+g_{\nu\bar{\nu}}{a\over2}{q^2g_{\mu\bar{\mu}}-q_{\bar{\mu}}q_{\mu}\over
m_\rho^2-q^2}\Big]\, \ . \ea

In the full VMD version, one multiplies the point like
$\gamma\pi\pi$ vertex with \be\label{fullvmd1}
{m_\rho^2g_{\mu\nu}-m_\rho^2{q_\mu q_\nu}\over m_\rho^2-q^2} \ee

and the
$\gamma\gamma\pi\pi$ vertex of~(\ref{ggpp}) with the term
 \be\label{fullvmd2}
{m_\rho^2g_{\nu\bar{\nu}}-p_{\bar{\nu}}p_{\nu}\over
m_\rho^2-p^2}{m_\rho^2g_{\mu\bar{\mu}}-q_{\bar{\mu}}q_{\mu}\over
m_\rho^2-q^2}\, \ .
\ee

These new vertices are fully gauged and chiral invariant as
mentioned in Ref.~\cite{Bijnens:1995xf}.
 One can also follow the same procedure to retrieve the
desired Feynmen rules for the $L_9$ and $L_{10}$. The extension of
the point vertex $\gamma\pi\pi$ is achieved when multiplied with
 \ba\label{78}
 g_{\mu\bar{\mu}}+{L_9}
\left(q^2g_{\mu\bar{\mu}}-q_\mu q_{\bar{\mu}}\right)\ea and the
amplitude corresponding to the $\gamma\gamma\pi\pi$ vertex,
including the $p^4$ corrections, should be multiplied with

\ba\label{79} g_{\mu\bar{\mu}}g_{\nu\bar{\nu}}
&+&g_{\mu\bar{\mu}}{L_9}\left({p^2g_{\nu\bar{\nu}}-p_{\bar{\nu}}p_{\nu}}\right)
+g_{\nu\bar{\nu}}{L_9}\left({q^2g_{\mu\bar{\mu}}-q_{\bar{\mu}}q_{\mu}}\right)\nonumber\\
&+&\left(L_9+L_{10}\right)\Big(q\cdot
pg_{\mu\bar{\mu}}g_{\nu\bar{\nu}}-g_{\mu\nu}p_{\bar{\mu}}q_{\bar{\nu}}\Big)\,
\ , \ea where $L_{9}$ and $L_{10}$ are constants involved in the
Lagrangian~(\tref{p4}).

\subsection{High energy limit}
In this section, the high energy limits and the matching with low
energy limits are considered.
\begin{figure}
\centering
\subfigure[]{
   \includegraphics[scale =.12] {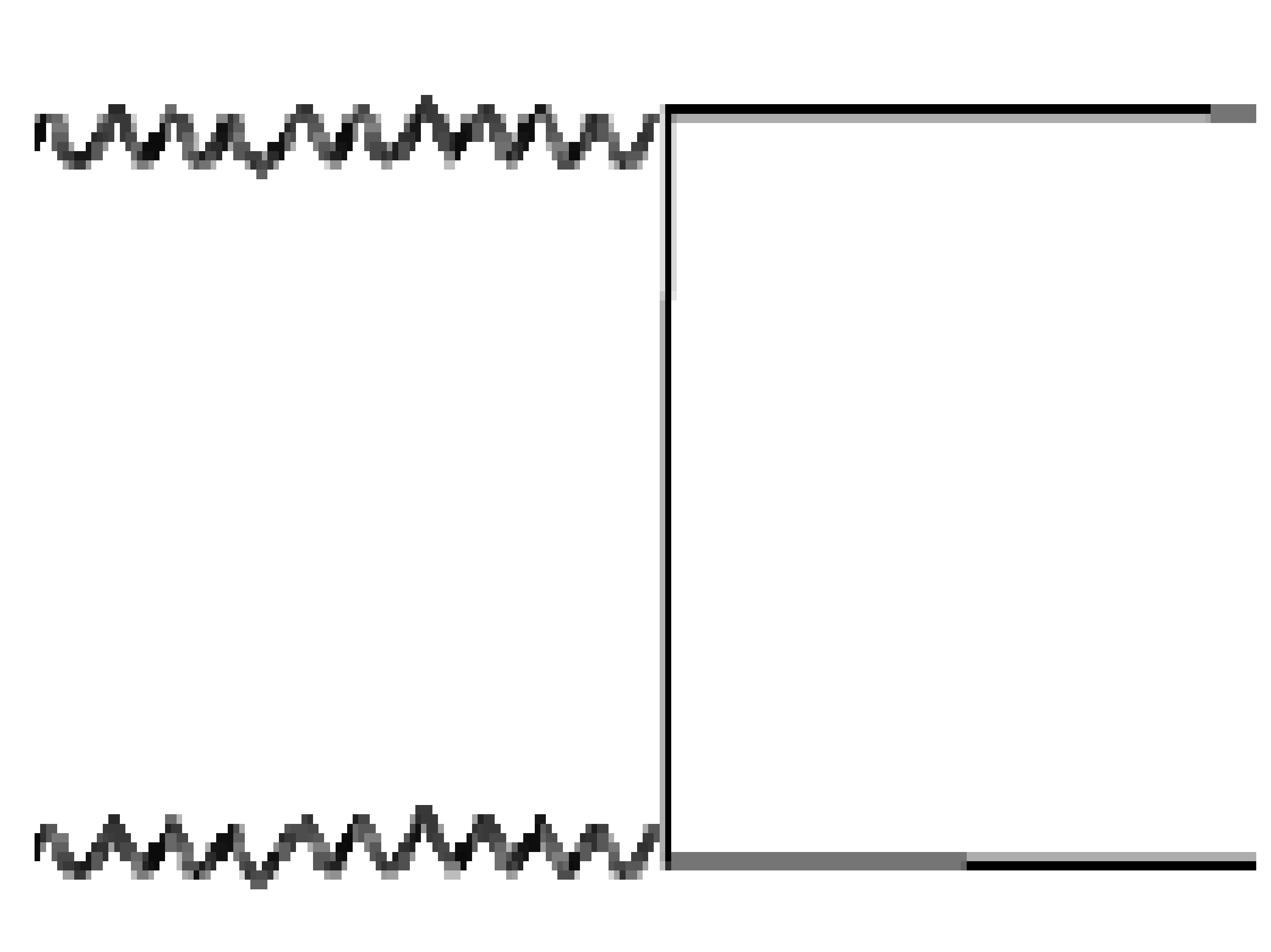}
   \label{fig:subfig1}\qquad
 }
 \subfigure[]{
   \includegraphics[scale =.15] {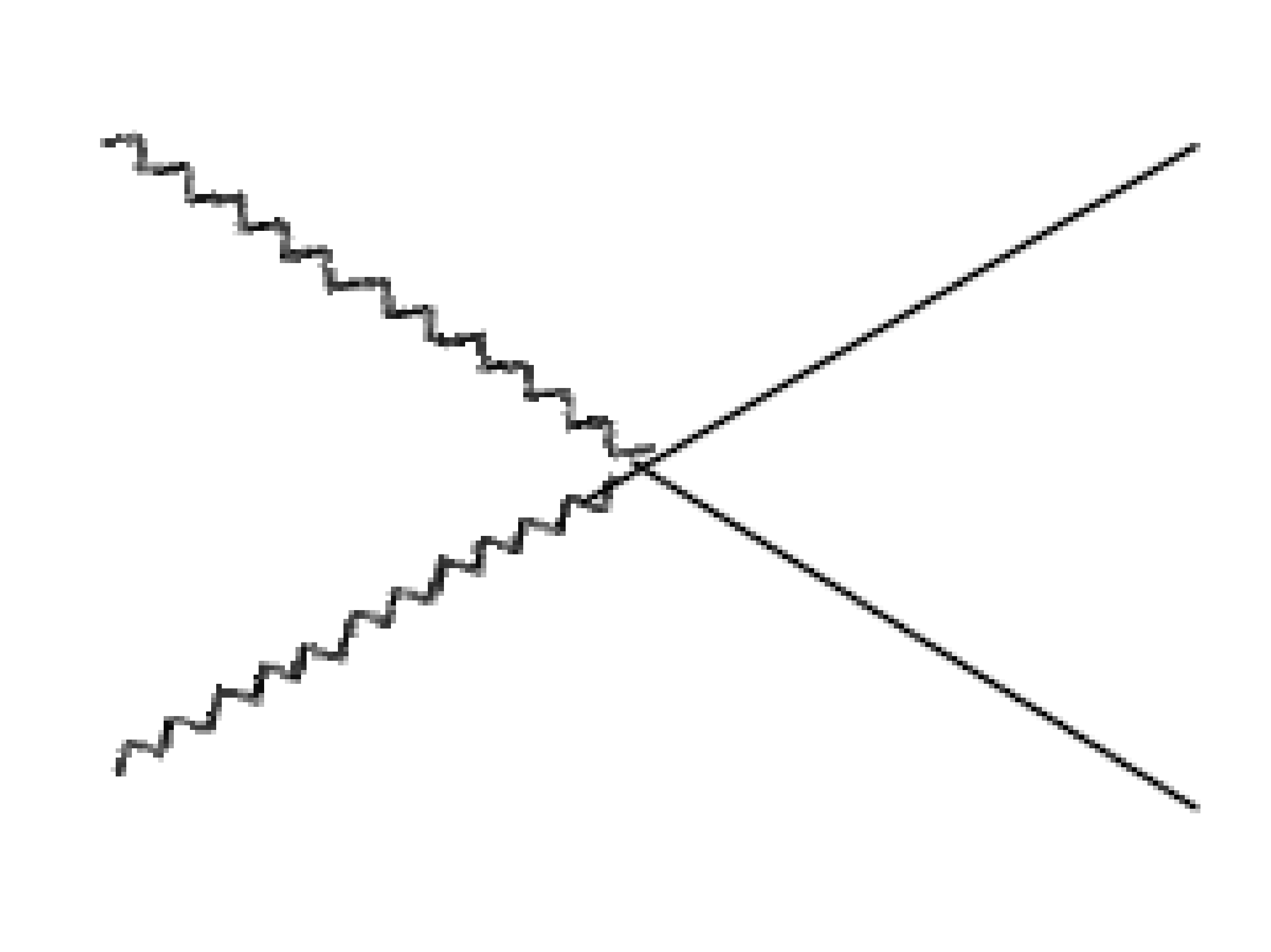}
   \label{fig:subfig2}\qquad
 }
\caption{The $\gamma\gamma\pi\pi$ vertex.}
\label{OPE}
\end{figure}

It can be shown that the $\gamma\gamma\pi\pi$ amplitude for two
high energy photons with momenta $P_1\simeq -P_2\simeq P$ is
proportional to $1/P^2$. This is done by using the operator
product expansion for two vector currents and showing that the
matrix element of the leading part which is proportional to an
axial current vanishes. Hence, when $P\rightarrow \infty$ the
amplitude vanishes. The amplitude corresponding to the
diagram~\tref{OPE} is

\ba\label{OPE1}
{\cal M}=ie^2 \Big\{{\left(P_{1\mu}-2K_{1\mu}\right)\left(P_{2\nu}-2K_{2\nu}\right)\over \left(P_2-K_2\right)^2-m_\pi^2}
        +{\left(P_{2\nu}-2K_{1\nu}\right)\left(P_{1\mu}-2K_{2\mu}\right)\over \left(P_1-K_2\right)^2-m_\pi^2}
        - 2g_{\mu\nu}
 \Big\}\, \ .
\ea For $P_1\simeq -P_2\simeq P\rightarrow \infty$ this reads \be
{\cal M}=\left(2g_{\mu\nu}-2{P_\mu P_\nu\over P^2}\right)\, \ ,
\ee which does not vanish fast enough. For the VMD case the above
amplitude should be multiplied with~(\ref{fullvmd2}). The
resulting amplitude vanishes at order $P^0$. Now let us examine
the HLS case. Then, the first and second term of~(\ref{OPE1}) are
multiplied with~(\ref{53}) and the third is multiplied
with~(\ref{54}). In the high energy limit the leading term is \ba
{\cal M}=2\left( g_{\mu\nu}-{P_\mu P_\nu\over
P^2}\right)\left(1-a\right)\, \ . \ea This is only satisfied for
$a=1$. However, the case HLS with $a=2$ does not uphold this
condition. Hence, one can infer that something must be wrong with
it.
\section{Muon magnetic anomaly from light by light amplitude}
\label{chap:Muon}\setcounter{equation}{0}
\subsection{General}\rlabel{general}
 The response of a muon carrying momentum $p$ to an external electromagnetic
field $A_\mu$ with momentum transferred $p_3\equiv p-p^\prime$ is
described by the matrix element
\begin{equation}\label{56}
{\cal M}\equiv-\mid e\mid
A_\rho\bar{u}(p^\prime)\Gamma^\rho(\acute{p},p)u(p)\, \ ,
\end{equation}
with
\begin{equation}\label{57}
\Gamma^\rho(p^\prime,p)=F_1(p_3^2)\gamma^\rho-\frac{i}{2m_l}F_2(p_3^2)\sigma^{\rho\nu}p_{3\nu}
-F_3(p_3^2)\gamma_5\sigma^{\rho\nu}p_{3\nu}+F_4(p_3^2)[p_3^2\gamma^\rho-2m_lp_3^\rho]\gamma_5\,
\ .
\end{equation}
The two first form factors are known as the Dirac and the Pauli
form factor, respectively. In fact~\cite{Bijnens:1995xf}, the
magnetic moment of the fermion in magnetons is $\mu\equiv
2(F_1(0)+F_2(0))$ and in analogy with the classical limit,
described in the introduction, one can define the gyromagnetic
ratio as $g\equiv 2\mu$ and the anomalous magnetic moment as
$a\equiv(g-2)/2=F_2(0)$~\cite{Bijnens:1995xf}. The form factor
$F_3(p_3^2)$ can be different from zero provided parity and time
reversal invariance are broken and for $F_4(p_3^2)$ to be nonzero,
parity invariance should be broken. Therefore, both are absent in
our survey. Since the task of computation of $\Gamma_\rho(
p^\prime,p)$ is very involved especially for higher order
corrections, one can project out the form factor of interest,
$F_2(p_3^2)$ in our case, and then the general form of the
contribution can be shown to be~\cite{Knecht:2003kc}
\be\label{Damu} a_\mu^{\rm light-by-light} = -{1\over 48 m} \tr
[(\slashed{p} +m )\Gamma^{\lambda\beta}(0) \, (\slashed{p} +m )
[\gamma_\lambda,\gamma_\beta ] ] \, \ . \ee
 Defining the four
point function $\Pi_{\rho\nu\alpha\lambda}$ as

\ba\label{fourp} \Pi^{\rho\nu\alpha\lambda}(p_1,p_2,p_3)&=&i^3\int
d^4x_1\int d^4x_2\int d^4x_3\exp i\left(p_1\cdot x_1+p_2\cdot
x_2+p_3\cdot x_3\right)\nonumber\\&\times&\langle 0\mid
T{j_\rho(0)j_\nu(x_1)j_\alpha(x_2)j_\lambda(x_3)}\mid 0\rangle \ea
and using the Ward identity to rewrite it in the form

\be \Pi^{\rho\nu\alpha\lambda}(p_1,p_2,p_3)=
-p_{3\beta} {\delta \Pi^{\rho\nu\alpha\beta}(p_1,p_2,p_3) \over
\delta p_{3\lambda}}\, \ ,\ee
 the $\Gamma^{\lambda\beta}(0)$ for the Figure~\tref{HLL} writes
 \ba\label{MLB} \Gamma^{\lambda
\beta}(p_3)&=& \vert e\vert^6 \int {{\rm d}^4 p_1 \over (2\pi )^4}
\int {{\rm d}^4p_2\over (2\pi )^4} \, \, {1\over q^2\, p_1^2 \,
p_2^2 (p_4^2-m^2) \,
(p_5^2 - m^2)}\nonumber\\
&\times& \left[ {\delta \Pi^{\rho\nu\alpha\beta} (p_1,p_2,p_3)
\over \delta p_{3\lambda}} \right] \gamma_\alpha (\slashed{p}_4 +m
)\gamma_\nu (\slashed{p}_5 +m ) \gamma_\rho  \, \ .\ea

with $p_4=p^\prime-p_2$, $p_5=p-q$.
 The most formidable task ahead is then to build the relevant
four point functions and to calculate the integral~(\ref{MLB}).
 One should note that this four-point function can be decomposed by using
Lorentz covariance as follows
\ba\label{58} \fourp &\equiv&
\Pi^{1}(p_1,p_2,p_3)
 g^{\rho\nu} g^{\alpha\beta} +
\Pi^{2}(p_1,p_2,p_3) g^{\rho\alpha} g^{\nu\beta}
\nonumber\\
&+&\Pi^{3} (p_1,p_2,p_3)
g^{\rho\beta} g^{\nu\alpha} \nonumber \\
&+&\Pi^{1jk}(p_1,p_2,p_3)
 g^{\rho\nu} p_j^\alpha p_k^\beta +
\Pi^{2jk}(p_1,p_2,p_3)
 g^{\rho\alpha} p_j^\nu p_k^\beta \nonumber \\
&+& \Pi^{3jk}(p_1,p_2,p_3)
 g^{\rho\beta} p_j^\nu p_k^\alpha +
\Pi^{4jk}(p_1,p_2,p_3)
 g^{\nu\alpha} p_j^\rho p_k^\beta \nonumber \\
&+& \Pi^{5jk}(p_1,p_2,p_3)
 g^{\nu\beta} p_j^\rho p_k^\alpha +
\Pi^{6jk}(p_1,p_2,p_3)
 g^{\alpha\beta} p_j^\rho p_k^\nu \nonumber \\
&+& \Pi^{ijkm}(p_1,p_2,p_3)
 p_i^\rho p_j^\nu p_k^\beta p_m^\alpha \, ,
\ea
 where $i,j,k,m =$ 1, 2 or 3 and repeated indices are summed.
There are in total 138 $\Pi$-functions. However, due to Ward
identities \ba \rlabel{gauge}
p_{1\nu}\fourp = p_{2\alpha}\fourp =\nonumber \\
p_{3\beta}\fourp = q{_\rho}\fourp = 0\, \ ,
 \ea
all of these functions are not independent and using these
identities frequently, the overall number could be reduced to $43$
independent $\Pi^{ijkm}(p_1,p_2,p_3)$ functions, $32$ of which
contribute to $a_\mu$, Ref.~\cite{Bijnens:1995xf}. When the
functions are found, one should add them up, derivate them with
respect to $p_3$, then set $p_3=0$ and put them into the
integral~(\ref{MLB}).

\subsection{Integration}\rlabel{int}
In this work the main focus is to calculate the
contribution of the charged pion loop light by light scattering to
$a_\mu$. However, to get familiar with the overall idea behind the
mathematical approach, it would be illuminating to start with the
simpler calculation for the pion exchange, shown in Figure~\tref{pionpol}.

\subsubsection{Pion exchange}\rlabel{exchange}

\begin{figure}
\begin{center}
\includegraphics[width=6cm,height=4cm]{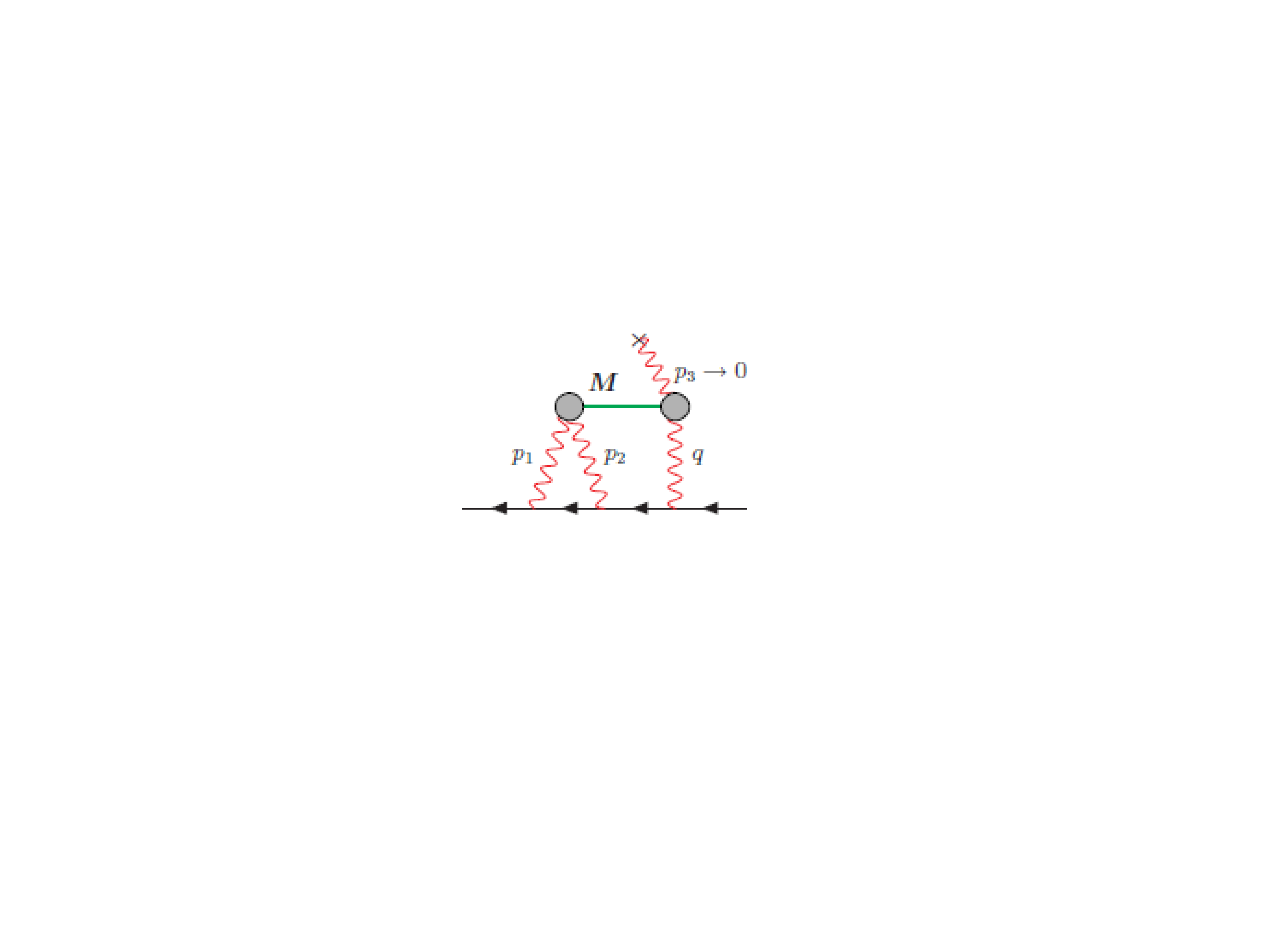}
\end{center}
\caption{The pion exchange HLL contribution to
$a_\mu$. Figure from~\cite{Bijnens:2007pz}.}
\label{pionpol}
\end{figure}
The key object that is used for this case is the
$\pi\gamma^\star\gamma^\star$ amplitude, which can be calculated
via~\cite{Jegerlehner:2009ry}

\ba \int\!\!\!&d^4&\!\!\!x\exp i\left(p\cdot x\right)\times\langle
0\mid
T\{j_\mu(x_1)j_\nu(x_2)\}\mid \pi^0(p)\rangle \nonumber\\
&=&\varepsilon_{\mu\nu\alpha\beta}p_1^\alpha p_2^\beta{\cal
F}_{\pi^0\gamma\gamma}(m_\pi^2,p_1^2,p_2^2)\, \ , \ea
 where,
${\cal F}_{\pi^0\gamma\gamma}$ is the form factor function and
$p_1$ and  $p_2$ are the photon momenta involved in the
$\pi^0\gamma\gamma$ vertex. Calculating this amplitude for each of
the vertices of the Figure~\tref{pionpol}, constructing the whole amplitude,
 taking derivative respect to $p_3$,
putting $p_3=0$ and plugging into the relation~(\ref{Damu}) for
three different permutations of Figure~\tref{pionpol}, one finds

\ba \label{MLB1}
{a_\mu^{LbL}} &=& -e^6\int{ d^4p_1\over(2\pi)^4}
\int{d^4p_2\over(2\pi)^4}{1\over
{p_1^2p_2^2(p_1+p_2)^2[(p+p_1)-m_\mu^2][(p-p_2)-m_\mu^2]}}\cr{}\cr
&\times& \Big[{{\cal F}_{\pi^{0\star}\gamma^\star\gamma^\star}(p_2^2,p_1^2,q^2)
{\cal F}_{\pi^{0\star}\gamma^\star\gamma}(p_2^2,p_2^2,0)\over {p_2^2-m_\pi^2}}T_2(p_1,p_2;q)\nonumber\\
&+& {{\cal
F}_{\pi^{0\star}\gamma^\star\gamma^\star}(q^2,p_1^2,p_2^2){\cal
F}_{\pi^{0\star}\gamma^\star\gamma}(q^2,q^2,0)\over
{q^2-m_\pi^2}}T_1(p_1,p_2;q) \big]\, \ , \ea with \ba\label{59}
T_1(p_1,p_2;q)&=&{16\over 3}(p\cdot p_1)(p\cdot
p_2)(p_1\cdot p_2)-{16\over 3}(p\cdot p_2)^2p_1^2\nonumber\\
&-&{8\over3}(p\cdot
p_1)(p_1\cdot p_2)p_2^2 + 8(p\cdot p_2)p_1^2p_2^2\nonumber\\
&-&{16\over 3}(p\cdot p_2)(p_1\cdot p_2)^2 +{16\over 3}m_\mu^2
p_1^2p_2^2\nonumber\\ &-&{16\over 3} m_\mu^2(p_1\cdot p_2)^2 \ea
\ba\label{60}
 T_2(p_1,p_2;q)&=&{16\over 3}(p\cdot p_1)(p\cdot p_2)(p_1\cdot
p_2)-{16\over 3}(p\cdot p_1)^2p_2^2\nonumber\\&+&{8\over3}(p\cdot
p_1)(p_1\cdot p_2)p_2^2 + {8\over 3}(p\cdot p_1)(p_1\cdot
p_2)p_2^2\nonumber\\&+&{8\over 3}(p\cdot p_1)(p_1^2 p_2^2)
+{8\over 3}m_\mu^2 p_1^2p_2^2\nonumber\\&-&{8\over 3}
m_\mu^2(p_1\cdot p_2)^2\, \ , \ea where $q=-(p_1+p_2)$ has been
used in the limit that $p_3$ vanishes. Also, $p$ is the muon
momentum. This is an eight dimensional integral to be done. In
general three of the integrations can be done analytically and one
is left with a five dimensional integral consisting of three
angles and two moduli. Then, the angles could be reduced to one,
using the Gegenbauer polynomials
technique~\cite{Jegerlehner:2009ry}. Using this technique, the
${a_\mu^{LbL}}$ can be averaged over the direction of the muon in
space such that
 \be \label{61}
<\cdot\cdot\cdot>={1\over{2\pi^2}}\int d\Omega(\hat{p})\, \ . \ee
To do so, one defines $(4)\equiv(P+P_1)^2+m_\mu^2$ and
$(5)\equiv(P-P_2)^2+m_\mu^2$ with $P^2=-m_\mu^2$, to
find~\cite{Jegerlehner:2009ry} \ba\label{62}
<{1\over(4)}{1\over(5)}>&=&{1\over m_\mu^2 R_{12}}\arctan\Big({zx\over{1-zt}}\Big)\nonumber\\
<(P.P_1){1\over (5)}>&=& (P_1\cdot P_2){(1-R_{m2})^2\over{8m_\mu^2}}\nonumber\\
<(P.P_2){1\over (4)}>&=& (P_1\cdot P_2){(1-R_{m1})^2\over{8m_\mu^2}}\nonumber\\
<{1\over (4)}>&=&-{(1-R_{m1})\over 2m_\mu^2}\nonumber\\
<{1\over (5)}>&=&-{(1-R_{m2})\over 2m_\mu^2}\, \ , \ea
 where
 \be \label{63} R_{mi}=\sqrt{1+{4m_\mu^2\over Q_i^2}} \ee
  and
\be\label{64} z={P_1P_2\over 4m_\mu^2}(1-R_{m1})(1-R_{m2})\, \ . \ee
Also,
$t=\cos\theta$ and $\theta$ is the azimuthal angle between the momenta $P_1$
and $P_2$.

 The integral~(\ref{MLB1}) reduces to a three dimensional
integral
\ba\label{65}
 {a_\mu^{LbL}} = -{2\alpha^3\over3\pi^2}\int_0^\infty
dP_1dP_2\int_{-1}^{+1}dt\sqrt{1-t^2}P_1^3P_2^3
&\Bigg[&{F_1I_1(P_1,P_2,t)\over(P_2^2+m_\pi^2)}\nonumber\\
&+&{F_2I_2(P_1,P_2,t)\over (Q^2+m_\pi^2)}\Bigg] \, \ , \ea where
\ba\label{66}
I_1(P_1,P_2,t)&=&1/(P_1^2P_2^2Q^2)\Bigg[X(P_1,P_2,t)\Big(8Q^2(P_1\cdot
P_2)-2P_2^2(P_2^4/m_\mu^2-2P_2^2)\nonumber\\ &-&
2P_2^2Q^2(2-P_2^2/m_\mu^2+2(P_1\cdot P_2)/m_\mu^
2)+4P_1^4\nonumber\\
&-&4P_1^2Q^2-2P_1^2P_2^2(4+P_1^2/m_\mu^2-2P_2^2/m_\mu^2)+2/m_\mu^2\Big)\nonumber\\
&-& 2Q^2(1+(1-R_{m1})(P_1\cdot P_2)/m_\mu^2)
+P_2^2(2-(1-R_{m1})P_2^2/m_\mu^2)\nonumber\\
&+&P_2^2Q^2(1-R_{m1})/m_\mu^2+P_1^2(2+(1-R_{m1})^2(P_1\cdot
P_2)/m_\mu^2)\nonumber\\ &+& 3P_1^2P_2^2(1-R_{m1})/m_\mu^2\Bigg]\,
\ ,\ea \ba\label{67}
I_2(P_1,P_2,t)&=&1/(P_1^2P_2^2Q^2)\Bigg[X(P_1,P_2,t)\Big(4Q^2(P_1\cdot
P_2)+2P_2^4-2P_2^2Q^2+2P_1^4\nonumber\\ &-&
2P_1^2Q^2-4P_1^2P_2^2-4/m_\mu^2\Big)
-2Q^2-3P_2^2Q^2(1-R_{m2})/(2m_\mu^2)\nonumber\\&-&
3P_1^2Q^2(1-R_{m1})/(2m_\mu^2)+P_2^2(2+3(1-R_{m2})P_2^2/(2m_\mu^2))\nonumber\\&+&(1-R_{m2})^2(P_1\cdot
P_2)/(2m_\mu^2) +P_1^2(2+3(1-R_{m1})P_1^2/(2m_\mu^2))\nonumber\\
&+&(1-R_{m1})^2(P_1\cdot P_2)/(2m_\mu^2) -
P_1^2P_2^2(2-R_{m1}-R_{m2})/(2m_\mu^2)\Bigg]\, \ . \ea The
auxiliary function is 
\be
%\label{1}
 X(P_1,p_2,t)={1\over
p_1p_2x}\arctan\left({zx \over 1-zt}\right)\, \ . \ee So, using
this technique, one manages to reduce the eight dimensional
integral~(\ref{MLB1}) to a three dimensional one. Instead of $t$
one can also use $Q^2=P_1^2+P_2^2+2P_1P_2$ as variable. Results
for the different calculations for the pion exchange are shown in
Table 2. These correspond to different choices of ${\cal
F}_{\pi^{0\star}\gamma^\star\gamma}(q^2,p_1^2,p_2^2)$.

\begin{table}
\begin{center}
\begin{tabular}{|c|c|c|}
\hline
pion exchange contribution &$a_\mu$ $\times$  $10^{10}$  \\
\hline
Bijnens, Pallante and Prades~\cite{Bijnens:1995cc} &5.6\\
Hayakawa and Kinoshita~\cite{Hayakawa:1995ps} &5.7\\
Knecht and Nyffeler~\cite{Nyffeler} &5.8\\
Melnikov and Vainshtein~\cite{Melnikov:2003xd} &7.65\\
\hline
\end{tabular}
\end{center}
\caption{Results of different calculations for the pion exchange
contribution to $a_\mu$.} \label{table3}
\end{table}

\subsubsection{Bare pion loop}\rlabel{bare}

What was described above for the pion exchange case we now extend
to the charged pion loop case which is the topic of this work.
This is the contribution of scalar QED which is renormalizable.
For the pion loop after evaluating the four--point function
in~(\ref{fourp}) all functions in (\ref{58}) are nonzero.

\begin{figure}
\begin{center}
\includegraphics[width=9cm,height=6cm]{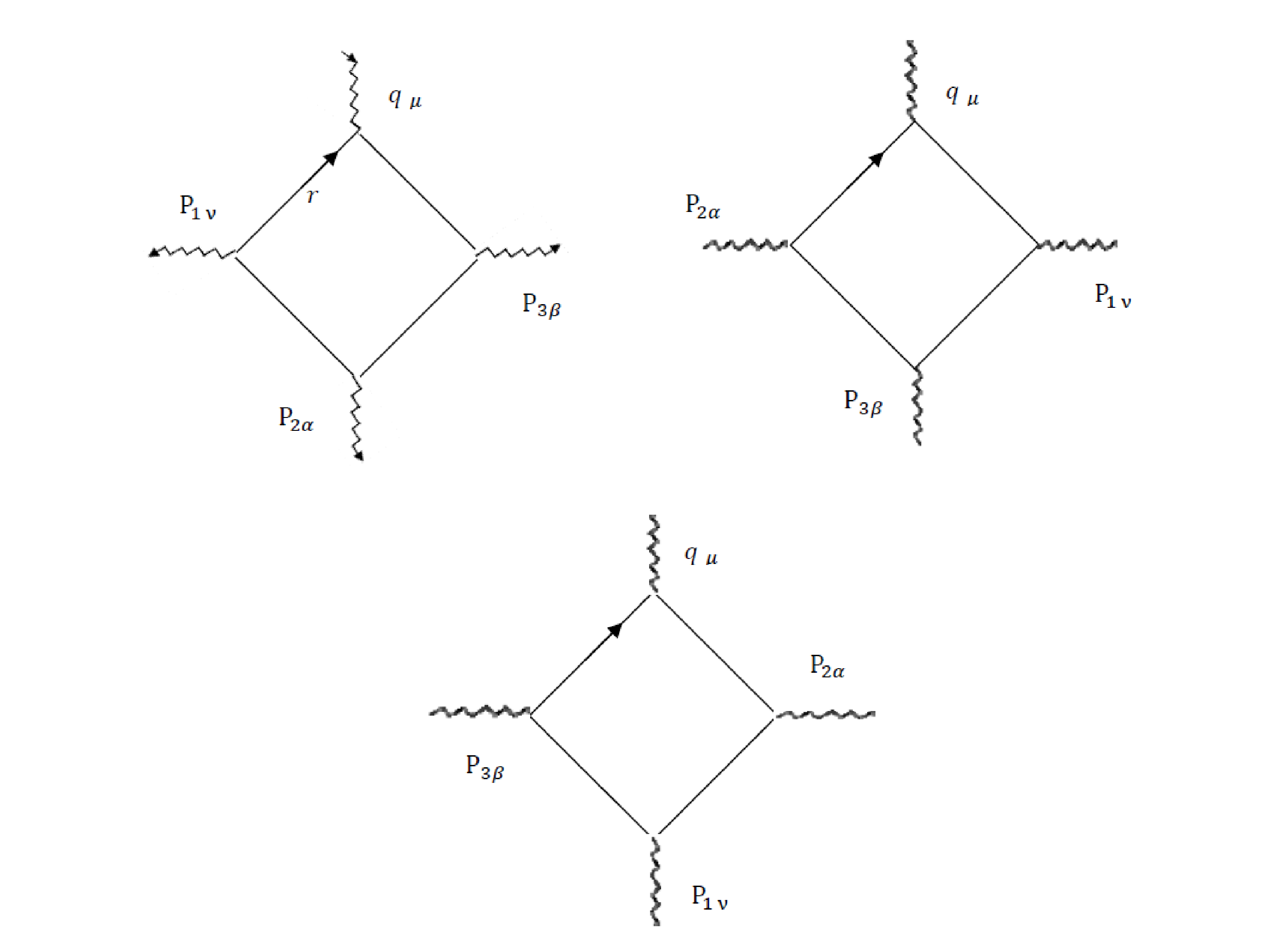}
\end{center}
\caption{The box diagram contribution to the $a_\mu$ HLL.}
\label{box}
\end{figure}

\begin{figure}
\begin{center}
\includegraphics[width=9cm,height=6cm]{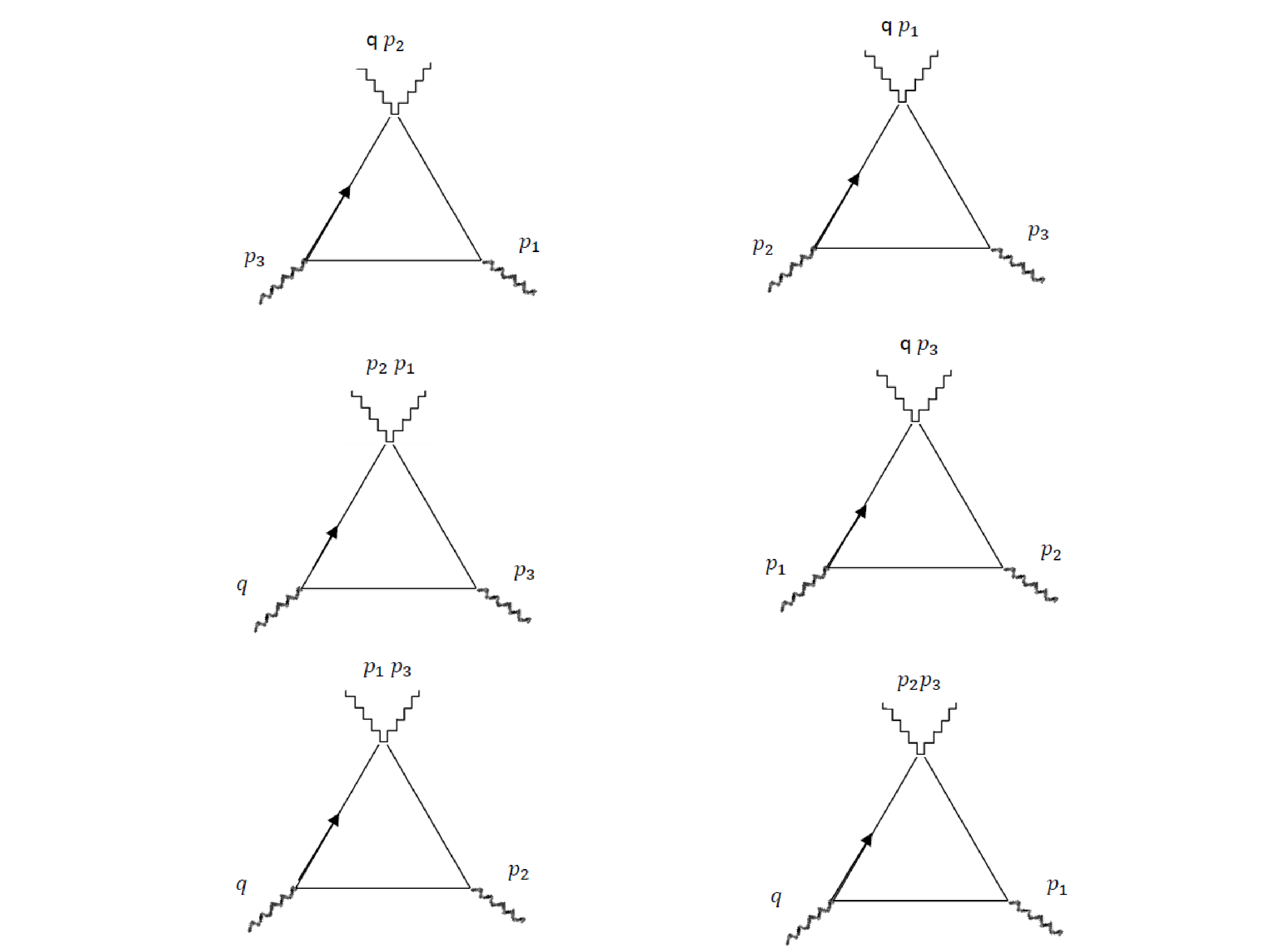}
\end{center}
\caption{The triangle diagram contribution to the $a_\mu$ HLL.}
\label{triangle}
\end{figure}

\begin{figure}
\begin{center}
\includegraphics[width=9cm,height=6cm]{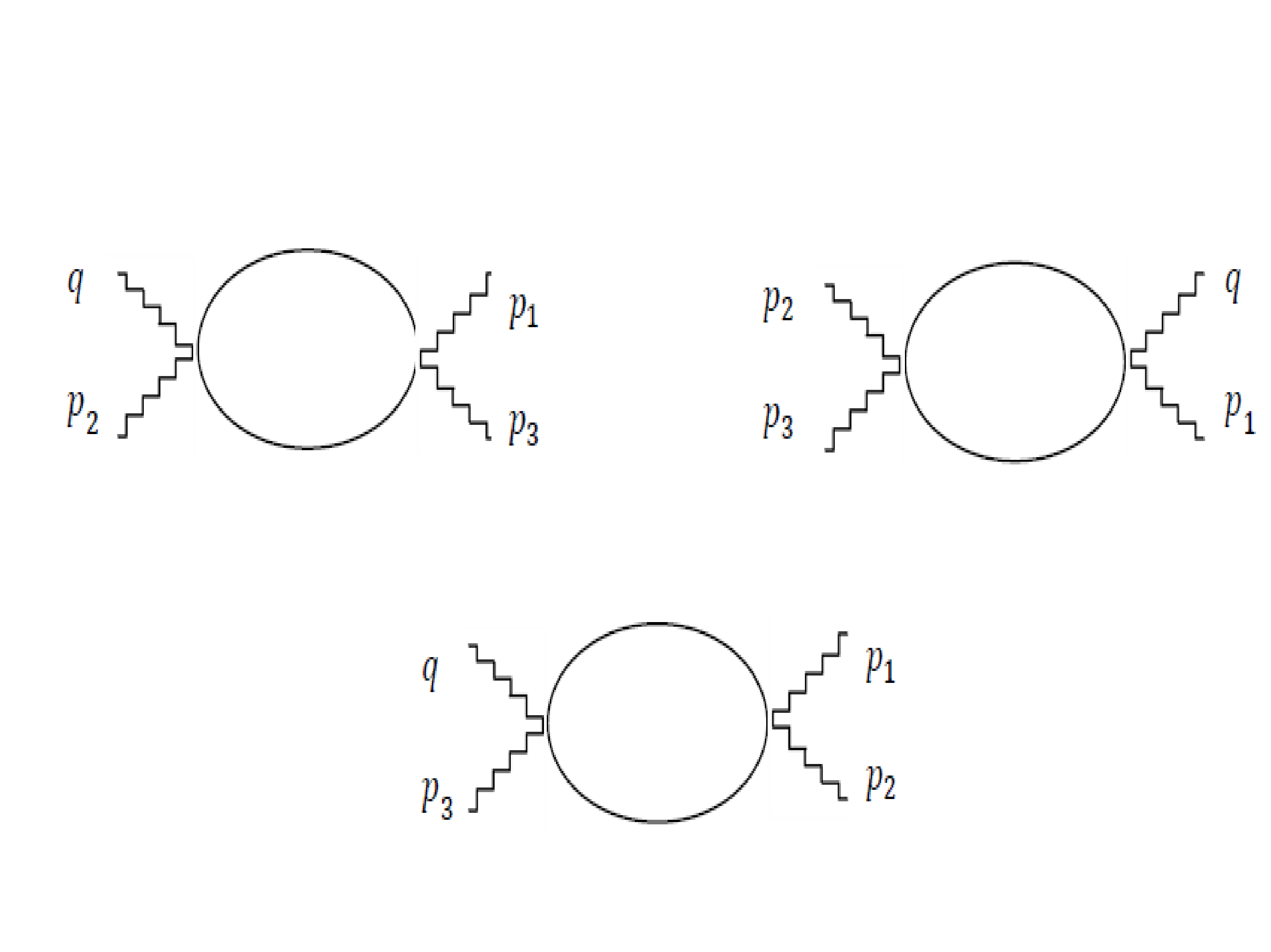}
\end{center}
\caption{The bulb diagram contribution to the $a_\mu$ HLL.}
\label{bulb}
\end{figure}

The contribution of the hadronic light by light scattering arises
at the order of $(\alpha/\pi)^3$. This includes box
 diagrams shown in Figure~\tref{box}, triangle diagrams in Figure~\tref{triangle} and bulb diagrams,
 shown in Figure~\tref{bulb}. These diagrams, along with their charge conjugates, add
up to 21.

 Among the contributions to the four point function denoted in the relation~(\ref{58}),
the $ \Pi^{ijkm}(p_1,p_2,p_3)$ ones originate only from the box
diagrams of Figure~\tref{box}, the $\Pi^{ijk}(p_1,p_2,p_3)$ ones
can originate both from the box and triangle diagrams and the
$\Pi^{i} (p_1,p_2,p_3)$ functions come only from the bulb
diagrams. One needs to find these functions, take their derivative
with respect to $p_3$, set $p_3=0$ and plug into the
relation~(\ref{Damu}). To do so, we have used the code FORM.

Let us first illustrate the procedure with the corresponding four
point function for the first diagram of the Figure~\tref{box}.
This diagram gives
 \ba\label{a1}
\!\!\!\!&\Pi&\!\!\!\!_{\mu\nu\alpha\beta}=\nonumber\\\!\!\!\!&{1\over
i}&\!\!\!\!\int{d^dr\over(2\pi)^d}{i^4\times
i^4\over(r^2-m^2)((r+p_1)^2-m^2)((r+p_1+p_2)^2-m^2)((r+p_1+p_2+p_3)^2-m^2)}\nonumber\\\times(2r\!\!\!\!&+&\!\!\!\!
p_1+p_2+p_3)_\mu(2r+p_1)_\nu(2r+2p_1+p_2)_\alpha(2r+2p_1+2p_2+p_3)_\beta\,
\ . \ea Using the Feynman parametrization method \be\label{a2}
{1\over
abcd}=6\int_0^1dx\int_0^{1-x}dy\int_0^{1-x-y}dz{1\over[a(1-x-y-z)+bx+cy+dz]^4}\,
\ , \ee
 one obtains
\ba\label{a3}
 \Pi_{\mu\nu\alpha\beta}={6\over
i}\int{d^4\tilde{r}\over(2\pi)^4}\int_0^1dx\int_0^{1-x}\!\!\!&dy&\!\!\!\int_0^{1-x-y}dz
{1\over\tilde{r}^2-\tilde{m}^2}\nonumber\\(2r+p_1+p_2+p_3)_\mu(2r+p_1)_\nu(2r\!\!\!\!&+&\!\!\!\!2p_1+p_2)_\alpha(2r+2p_1+2p_2+p_3)_\beta\,
\ , \ea
 with
 \ba\label{a4}
\tilde{m}^2&=&m^2+\Big((x+y+z)p_1+(y+z)p_2+zp_3\Big)^2-xp_1^2-y(p_1+p_2)^2-z(p_1+p_2+p_3)^2\nonumber\\
\tilde{r}&=&r+(x+y+z)p_1+(y+z)p_2+zp_3 \, \ . \ea To deal with the
integrals we have used the relations \ba\label{a5}
\int{d^dr\over(2\pi)^d}\tilde{r}_\mu\tilde{r}_\nu\tilde{r}_\alpha\tilde{r}_\beta
f(\tilde{r}^2)&=&\int{d^dr\over(2\pi)^d}{1\over
d(d+2)}\tilde{r}^4(g_{\mu\nu}g_{\alpha\beta}+g_{\mu\beta}g_{\alpha\nu}+g_{\nu\beta}g_{\alpha\mu})f(\tilde{r}^2)\nonumber\\
\int{d^dr\over(2\pi)^d}\tilde{r}_\mu\tilde{r}_\nu
f(\tilde{r}^2)&=&\int{d^dr\over(2\pi)^d}{1\over
d}\tilde{r}^2g_{\mu\nu}f(\tilde{r}^2)\, \ , \ea and that integrals
with odd powers in the numerator vanish. The remaining integrals
to be done are \ba\label{int1} {1\over d(d+2)}{1\over
i}\int{d^d\tilde{r}\over(2\pi)^d}{\tilde{r}^4\over(\tilde{r}^2-\tilde{m}^2)^4}&=&{1\over
d(d+2)i}\int{dr^d\over(2\pi)^d}\nonumber\\ \Bigg[
{1\over(\tilde{r}^2-\tilde{m}^2)^2}&+&{2\tilde{m}^2\over(\tilde{r}^2-\tilde{m}^2)^3}+{\tilde{m}^4\over(\tilde{r}^2-\tilde{m}^2)^4}
\Bigg]\nonumber\\
&=&{1\over24}{1\over16\pi^2}[{1\over\tilde{\varepsilon}}+{5\over6}-1-ln{\tilde{m}^2\over\mu^2}-1+{1\over6}]\nonumber\\
&=&{1\over24}{1\over16\pi^2}[{1\over\tilde{\varepsilon}}-1-ln{\tilde{m}^2\over\mu^2}]\,
\ , \ea

\ba\label{int2} {1\over d}{1\over
i}\int{d^d\tilde{r}\over(2\pi)^d}{\tilde{r}^2\over(\tilde{r}^2-\tilde{m}^2)^4}&=&{1\over
di}\int{dr^d\over(2\pi)^d}\nonumber\\ \Bigg[
{1\over(\tilde{r}^2-\tilde{m}^2)^3}&+&{\tilde{m}^2\over(\tilde{r}^2-\tilde{m}^2)^4}\Bigg]\nonumber\\={1\over
4}{1\over 16\pi^2}\left({-1\over 2\tilde{m}^2}+{1\over
6\tilde{m}^2}\right)&=&{1\over 16\pi^2}{-1\over12\tilde{m}^2} \ea
and \be\label{int3} {1\over
i}\int{d^d\tilde{r}\over(2\pi)^d}{1\over(\tilde{r}^2-\tilde{m}^2)^4}={1\over
16\pi^2}{1\over6}{1\over\tilde{m}^4}\, \ . \ee where $\mu$ is the
subtraction scale of the problem at hand, $d=4-2\varepsilon$,
${1/\tilde{\varepsilon}}={1/(\varepsilon-ln4\pi+\gamma+1)}$,
 $\gamma$ is the Euler-Mascheroni constant and $\varepsilon$ creeps in from the dimensional
 regularization. The expressions for the integral
\be
\int{d^d\tilde{r}\over(2\pi)^d}{1\over(\tilde{r}^2-\tilde{m}^2)^4}
\ee
 can be found in~\cite{Scherer}. Then, one evaluates the derivative
 $\partial\Pi_{\mu\nu\alpha\beta}/\partial p_3$, which means deriving $\tilde{m}^2$
 or occurrences of $p_3$ in the numerator, and put $p_3=0$
 to find the relevant function to be integrated over. A similar
 procedure is done for the other box diagrams.

For the first triangle diagram shown in Figure~\tref{triangle} the four point
function reads
 \ba\label{a6}
 \Pi_{\mu\nu\alpha\beta}&=&\nonumber\\&{1\over
i}&\int{d^4r\over(2\pi)^4}{i^3\times
i^3\over(r^2-m^2)((r+p_2)^2-m^2)((r+p_2+p_3)^2-m^2)}\nonumber\\\times\!\!\!&g_{\mu\nu}&\!\!\!(2r+p_2)_\alpha(2r+2p_2+p_3)_\beta\,
\ , \ea
which can be parameterized using
 \be\label{a7}
 {1\over
abc}=6\int_0^1dx\int_0^{1-x}dy{1\over[a(1-x-y)+bx+cy]^3}\, \ , \ee
to give
 \be\label{a8}
 \Pi_{\mu\nu\alpha\beta}={1\over
i}\int{d^4\tilde{r}\over(2\pi)^4}{i^3\times
i^3\over\tilde{r}^2-\tilde{m}^2}\nonumber\\g_{\mu\nu}(2r+p_2)_\alpha(2r+2p_2+p_3)_\beta\,
\ , \ee
 with
\ba\label{a9}
\tilde{m}^2&=&m^2+\Big((x+y)p_2+yp_3\Big)^2-xp_2^2-y(p_2+p_3)^2\nonumber\\
\tilde{r}&=&r+(x+y)p_2+yp_3\, \ . \ea
Then using again~(\ref{a5})
and
 \ba\label{a11} {1\over d}{1\over
i}\int{d^d\tilde{r}\over(2\pi)^d}{\tilde{r}^2\over(\tilde{r}^2-\tilde{m}^2)^3}&=&\nonumber\\
&=&{1\over 4}{1\over 16\pi^2}\left[{1\over\tilde{\varepsilon}}+{1\over2}-1-ln{\tilde{m}^2\over\mu^2}-{1\over2}\right]\nonumber\\
&=&{1\over 4}{1\over
16\pi^2}\left[{1\over\tilde{\varepsilon}}-1-ln{\tilde{m}^2\over\mu^2}\right]\
, \ea one gets \ba\label{a12}
 {1\over
i}\int{d^d\tilde{r}\over(2\pi)^d}{1\over(\tilde{r}^2-\tilde{m}^2)^3}=
{1\over 4}{1\over 16\pi^2}{-1\over 2\tilde{m}^2} \, \ . \ea
 And
finally, for the first bulb diagram of Figure~\tref{bulb} one has
\ba\label{a13} \Pi_{\mu\nu\alpha\beta}&=&\nonumber\\&{1\over
i}&\int{d^4 r\over(2\pi)^4}{i^2\times
i^2\over(r^2-m^2)((r+p_2+p_3)^2-m^2)}\nonumber\\\times
2\!\!\!&g_{\mu\nu}&\!\!\!2g_{\alpha\beta}={1\over
i}\int{d^4\tilde{r}\over(2\pi)^4}\int_0^1d
x{1\over(\tilde{r}^2-\tilde{m}^2)^2}\times
4g_{\mu\nu}g_{\alpha\beta}\nonumber\\&=&4{1\over
16\pi^2}\left[{1\over\tilde{\varepsilon}}-1-ln{\tilde{m}^2\over\mu}\right]g_{\mu\nu}g_{\alpha\beta}\,
\ , \ea
 with
 \be\label{a14}
 \tilde{r}=r+x(p_2+p_3) \ee
and
 \be\label{a15}
 \tilde{m}^2=m_p^2+x(x-1)(p_2+p_3)\, \ .
 \ee
 It is
important to notice that the divergent parts of each diagrams,
when added up, cancel each other and the remaining part which
contributes to the $a_\mu$ is finite. As a consequence, all
$ln\mu^2$ dependent terms above disappear miraculously.

After doing the above integrations, depending on the type of the
diagram, we apply the Gegenbauer polynomial method to perform five
of the integrations in~(\ref{MLB}) similar to the steps that led
to~(\ref{65}). The final formula is rather long and is not
presented here. Besides the $P_1, P_2, t$ or $P_1, P_2, Q$
integration one is always left with one, two or three Feynman
parameters that should also be integrated over. One can always
shift the parameters in the case of box and triangle diagrams so
that, the denominator is independent of one of them, and it could
be analytically integrated out, reducing the size of the
expressions considerably. It turns out that the different
$\tilde{m}^2$ for the different box diagrams can all be brought in
the same form as well reducing the size of the expressions
considerably. The final integral to be done is
 a five or four dimensional integral, which we have dealt with using the Monte Carlo
routine VEGAS.

\subsubsection{HLS}\rlabel{hls1}
When trying the same procedure for the HLS case, we can reuse a
lot of the previous subsection since the vertices are related by
(\ref{53}) and (\ref{54}). One should be careful since, terms
including $p.\tilde{r}$ will also appear in the numerator. For
example, for the Figure~\tref{HLL}, the four
point function of the HLS is
 \ba\label{a16}
\!\!\!\!&\Pi&\!\!\!\!_{\mu\nu\alpha\beta}=\nonumber\\\!\!\!\!&{1\over
i}&\!\!\!\!\int{d^4r\over(2\pi)^4}{i^4\times
i^4\over(r^2-m^2)((r+p_1)^2-m^2)((r+p_1+p_2)^2-m^2)((r+p_1+p_2+p_3)^2-m^2)}\nonumber\\
\times(2r\!\!\!\!&+&\!\!\!\!p_1+p_2+p_3)_\mu\left(g_{\mu\bar{\mu}}-{q^2g_{\mu\bar{\mu}}-q_\mu
q_{\bar{\mu}}\over
q^2-m_\rho^2 }\right)\nonumber\\
(2r\!\!\!\!&+&\!\!\!\!p_1)_\nu\left(g_{\nu\bar{\nu}}-{p_1^2g_{\nu\bar{\nu}}-p_{1\nu}
p_{1\bar{\nu}}\over
p_1^2-m_\rho^2 }\right)\nonumber\\
(2r\!\!\!\!&+&\!\!\!\!2p_1+p_2)_\alpha\left(g_{\alpha\bar{\alpha}}-{p_2^2g_{\alpha\bar{\alpha}}-p_{2\alpha}
p_{2\bar{\alpha}}\over
p_2^2-m_\rho^2 }\right)\nonumber\\
(2r\!\!\!\!&+&\!\!\!\!2p_1+2p_2+p_3)_\beta\left(g_{\beta\bar{\beta}}-{p_3^2g_{\beta\bar{\beta}}-p_{3\beta}
p_{3\bar{\beta}}\over p_3^2-m_\rho^2 }\right) \, \ . \ea The four
point function of the first diagram of Figure~\tref{triangle}
becomes
 \ba\label{HLS1}
 \Pi_{\mu\nu\alpha\beta}&=&\nonumber\\\!\!\!\!&{1\over
i}&\!\!\!\!\int{d^4r\over(2\pi)^4}{i^3\times
i^3\over(r^2-m^2)((r+p_2)^2-m^2)((r+p_2+p_3)^2-m^2)}\nonumber\\\times\!\!\!&g_{\mu\nu}&\!\!\!\!2\Big(g_{\mu\bar{\mu}}g_{\nu\bar{\nu}}
+g_{\mu\bar{\mu}}{a\over2}{p_1^2g_{\nu\bar{\nu}}-p_{1\bar{\nu}}p_{1\nu}\over
m_\rho^2-p_1^2}
+g_{\nu\bar{\nu}}{a\over2}{q^2g_{\mu\bar{\mu}}-q_{\bar{\mu}}q_{\mu}\over
m_\rho^2-q^2}\Big)\!\!\!\nonumber\\(2r\!\!\!\!&+&\!\!\!\!p_2)_\alpha\Big(g_{\alpha\bar{\alpha}}-{p_2^2g_{\alpha\bar{\alpha}}-p_{2\alpha}
p_{2\bar{\alpha}}\over p_2^2-m_\rho^2
}\Big)\nonumber\\(2r\!\!\!\!&+&\!\!\!\!2p_2+p_3)_\beta\Big(g_{\beta\bar{\beta}}-{p_3^2g_{\beta\bar{\beta}}-p_{3\beta}
p_{3\bar{\beta}}\over p_3^2-m_\rho^2 }\Big)\, \, \ea and the four
point function corresponding to the first bulb of
Figure~\tref{bulb} is \ba\label{HLS2}
\Pi_{\mu\nu\alpha\beta}&=&\nonumber\\&{1\over i}&\int{d^4
r\over(2\pi)^4}{i^2\times
i^2\over(r^2-m^2)((r+p_2+p_3)^2-m^2)}\nonumber\\\times
2g_{\mu\nu}2g_{\alpha\beta} \Big(g_{\mu\bar{\mu}}g_{\nu\bar{\nu}}
\!\!\!&+&\!\!\!g_{\mu\bar{\mu}}{a\over2}{p_1^2g_{\nu\bar{\nu}}-p_{1\bar{\nu}}p_{1\nu}\over
m_\rho^2-p_1^2}
+g_{\nu\bar{\nu}}{a\over2}{q^2g_{\mu\bar{\mu}}-q_{\bar{\mu}}q_{\mu}\over
m_\rho^2-q^2}\Big)\nonumber\\
\Big(g_{\beta\bar{\beta}}g_{\alpha\bar{\alpha}}
\!\!\!&+&\!\!\!g_{\beta\bar{\beta}}{a\over2}{p2^2g_{\alpha\bar{\alpha}}-p_{2\bar{\alpha}}p_{2\alpha}\over
m_\rho^2-p_2^2}
+g_{\alpha\bar{\alpha}}{a\over2}{p_3^2g_{\beta\bar{\beta}}-p_{3\bar{\beta}}p_{3\beta}\over
m_\rho^2-p_3^2}\Big)\,
\ . \ea
 To deal with terms like $p\cdot\tilde{r}$ one can use the relations~(\ref{int1}),~(\ref{int2})
and~(\ref{int3}), but some of the $\tilde{r}$s couple to $p$.

 Furthermore, as for the infinities in the HLS approach, they only
 cancel out after taking the derivative with respect to $p_3$ and
 setting $p_3=0$. The HLS is not a renormalizable
 theory,and thus the result could have been divergent but surprisingly, the contribution to $a_\mu$ is finite.

\subsubsection{Full VMD}\rlabel{VMD}
As it was described in Sec.~\ref{Had} in the full VMD, $\gamma\pi\pi$ vertex is multiplied with
(\ref{fullvmd1}). However, how it deals with the need of being chiral and gauge invariant, that
the naive VMD model does not meet. In fact, it could be shown that covering the photon legs
with vector mesons when using~(\ref{fullvmd1}), because of the Ward identities in~\ref{gauge}, always
the second term in this expression cancels and the result is just like multiplying each photon leg with
${m_\rho^2/ (m_\rho^2- q^2)}$, and this is fully invariant under above mentioned symmetries.

Now let us illustrate how the four point function changes in this model. Again using~(\ref{fullvmd1}),~(\ref{fullvmd1})
and~(\ref{a1}) one finds
\ba\label{VMDf}
\!\!\!\!&\Pi&\!\!\!\!_{\mu\nu\alpha\beta}=\nonumber\\\!\!\!\!&{1\over
i}&\!\!\!\!\int{d^4r\over(2\pi)^4}{i^4\times
i^4\over(r^2-m^2)((r+p_1)^2-m^2)((r+p_1+p_2)^2-m^2)((r+p_1+p_2+p_3)^2-m^2)}\nonumber\\
\times(2r\!\!\!\!&+&\!\!\!\!p_1+p_2+p_3)_\mu\left({g_{\mu\bar{\mu}}m_\rho^2-{q_\mu q_{\bar{\mu}} }\over m_\rho^2-q^2}\right)\nonumber\\
(2r\!\!\!\!&+&\!\!\!\!p_1)_\nu\left({g_{\bar{\nu}\nu}m_\rho^2-{p_{1\nu} p_{1\bar{\nu}}}\over m_\rho^2-p_1^2}\right)\nonumber\\
(2r\!\!\!\!&+&\!\!\!\!2p_1+p_2)_\alpha\left({g_{\alpha\bar{\alpha}}m_\rho^2-{P_{2\alpha} p_{2\bar{\alpha}} }\over m_\rho^2-p_2^2}\right)\nonumber\\
(2r\!\!\!\!&+&\!\!\!\!2p_1+2p_2+p_3)_\beta\left({g_{\beta\bar{\beta}}m_\rho^2-{p_{3\beta}p_{3\bar{\beta}}
}\over m_\rho^2-p_3^2}\right) \, \ . \ea For the first triangle
diagram of Figure~\tref{triangle} the four--point function writes
\ba\label{VMDf1}
 \Pi_{\mu\nu\alpha\beta}&=&\nonumber\\\!\!\!\!&{1\over
i}&\!\!\!\!\int{d^4r\over(2\pi)^4}{i^3\times
i^3\over(r^2-m^2)((r+p_2)^2-m^2)((r+p_2+p_3)^2-m^2)}\nonumber\\\times\!\!\!&g_{\mu\nu}&\!\!\!\!2\Big({m_\rho^2g_{\nu\bar{\nu}}-p_{1\bar{\nu}}p_{1\nu}\over
m_\rho^2-p_1^2}{m_\rho^2g_{\mu\bar{\mu}}-q_{\bar{\mu}}q_{\mu}\over
m_\rho^2-q^2}\Big)\!\!\!\nonumber\\(2r\!\!\!\!&+&\!\!\!\!p_2)_\alpha\Big({g_{\alpha\bar{\alpha}}m_\rho^2-{p_{2\alpha}
p_{2\bar{\alpha}} }\over
m_\rho^2-p_2^2}\Big)\nonumber\\(2r\!\!\!\!&+&\!\!\!\!2p_2+p_3)_
\beta\Big({g_{\beta\bar{\beta}}m_\rho^2-{p_{3\beta}p_{3\bar{\beta}}
}\over m_\rho^2-p_3^2}\Big)\, \, \ea and for the first bulb
diagram of Figure~\tref{bulb} one has \ba\label{VMDf2}
\Pi_{\mu\nu\alpha\beta}&=&\nonumber\\&{1\over i}&\int{d^4
r\over(2\pi)^4}{i^2\times
i^2\over(r^2-m^2)((r+p_2+p_3)^2-m^2)}\nonumber\\\!\!\!\!&\times&\!\!\!\!
2g_{\mu\nu}2g_{\alpha\beta}
\Big({m_\rho^2g_{\nu\bar{\nu}}-p_{1\bar{\nu}}p_{1\nu}\over
m_\rho^2-p_1^2}{m_\rho^2g_{\mu\bar{\mu}}-q_{\bar{\mu}}q_{\mu}\over
m_\rho^2-q^2}\Big)\nonumber\\
\!\!\!\!&\times&\!\!\!\!\Big({m_\rho^2g_{\alpha\bar{\alpha}}-p_{2\bar{\alpha}}p_{2\alpha}\over
m_\rho^2-p_2^2}{m_\rho^2g_{\beta\bar{\beta}}-p_{3\bar{\beta}}p_{3\beta}\over
m_\rho^2-p_3^2}\Big)\,
\ . \ea

\subsubsection{$L_9$ and $L_{10}$}\rlabel{l9l10}
Since Ref.~\cite{Ramsey-Musolf} argued that order $p^4$ effects
might be important, we have also calculated the contributions of
$L_9$ and $L_{10}$ to this order. Using the previous results, the
four point function of the Figure~\tref{HLL}, taking into account
the $L_9$ and $L_{10}$ corrections, takes the form
 \ba\label{L9L10}
\!\!\!\!&\Pi&\!\!\!\!_{\mu\nu\alpha\beta}=\nonumber\\\!\!\!\!&{1\over
i}&\!\!\!\!\int{d^4r\over(2\pi)^4}{i^4\times
i^4\over(r^2-m^2)((r+p_1)^2-m^2)((r+p_1+p_2)^2-m^2)((r+p_1+p_2+p_3)^2-m^2)}\nonumber\\
\times(2r\!\!\!\!&+&\!\!\!\!p_1+p_2+p_3)_\mu\Big(g_{\mu\bar{\mu}}+{L_9}
(q^2g_{\mu\bar{\mu}}-q_\mu q_{\bar{\mu}})\Big)\nonumber\\
(2r\!\!\!\!&+&\!\!\!\!p_1)_\nu\Big(g_{\nu\bar{\nu}}+{L_9}
(p_1^2g_{\nu\bar{\nu}}-p_{1\nu} p_{1\bar{\nu}})\Big)\nonumber\\
(2r\!\!\!\!&+&\!\!\!\!2p_1+p_2)_\alpha\Big(g_{\alpha\bar{\alpha}}+{L_9}
(p_2^2g_{\alpha\bar{\alpha}}-p_{2\alpha} p_{2\bar{\alpha}})\Big)\nonumber\\
(2r\!\!\!\!&+&\!\!\!\!2p_1+2p_2+p_3)_\beta\Big(g_{\beta\bar{\beta}}+{L_9}
(p_3^2g_{\beta\bar{\beta}}-p_{3\beta} p_{3\bar{\beta}})\Big)
\, \ .
\ea
The same function for the first triangle diagram of Figure~\tref{triangle} writes
 \ba\label{L9L101}
 \Pi_{\mu\nu\alpha\beta}&=&\nonumber\\\!\!\!\!&{1\over
i}&\!\!\!\!\int{d^4r\over(2\pi)^4}{i^3\times
i^3\over(r^2-m^2)((r+p_2)^2-m^2)((r+p_2+p_3)^2-m^2)}\nonumber\\\times\!\!\!&g_{\mu\nu}&\!\!\!\!\Big(g_{\mu\bar{\mu}}g_{\nu\bar{\nu}}
+g_{\mu\bar{\mu}}{L_9}\left({p_1^2g_{\nu\bar{\nu}}-p_{1\bar{\nu}}p_{1\nu}} \right)
+g_{\nu\bar{\nu}}{L_9}\left({q^2g_{\mu\bar{\mu}}-q_{\bar{\mu}}q_{\mu}}\right)\nonumber\\
\!\!\!\!&+&\!\!\!\!(L_9+L_{10})\left(q\cdot
p_1g_{\mu\bar{\mu}}g_{\nu\bar{\nu}}-g_{\mu\nu}p_{1\bar{\mu}}q_{\bar{\nu}}\right)\Big)\!\!\!\nonumber\\(2r\!\!\!\!&+&\!\!\!\!p_2)_\alpha
\Big(g_{\alpha\bar{\alpha}}+{L_9}
(p_2^2g_{\alpha\bar{\alpha}}-p_{2\alpha} p_{2\bar{\alpha}})\Big)\nonumber\\(2r\!\!\!\!&+&\!\!\!\!2p_2+p_3)_\beta\Big(g_{\beta\bar{\beta}}+{L_9}
(p_3^2g_{\beta\bar{\beta}}-p_{3\beta} p_{3\bar{\beta}})\Big)\,
\,
\ea
and the four point function of the first bulb diagram of the Figure~\tref{bulb} is
\ba\label{L9L102}
\Pi_{\mu\nu\alpha\beta}&=&\nonumber\\&{1\over
i}&\int{d^4 r\over(2\pi)^4}{i^2\times
i^2\over(r^2-m^2)((r+p_2+p_3)^2-m^2)}\nonumber\\\times
2g_{\mu\nu}2g_{\alpha\beta}
\Big(g_{\mu\bar{\mu}}g_{\nu\bar{\nu}}
\!\!\!\!&+&\!\!\!\!g_{\mu\bar{\mu}}{L_9}\left({p_1^2g_{\nu\bar{\nu}}-p_{1\bar{\nu}}p_{1\nu}} \right)
+g_{\nu\bar{\nu}}{L_9}\left({q^2g_{\mu\bar{\mu}}-q_{\bar{\mu}}q_{\mu}}\right)\nonumber\\
\!\!\!\!&+&\!\!\!\!(L_9+L_{10})\left(q\cdot
p_1g_{\mu\bar{\mu}}g_{\nu\bar{\nu}}-g_{\mu\nu}p_{1\bar{\mu}}q_{\bar{\nu}}\right)\Big)\nonumber\\
\Big(g_{\alpha\bar{\alpha}}g_{\beta\bar{\beta}}
\!\!\!\!&+&\!\!\!\!g_{\alpha\bar{\alpha}}{L_9}\left({p_3^2g_{\beta\bar{\beta}}-p_{3\bar{\beta}}p_{3\beta}} \right)
+g_{\beta\bar{\beta}}{L_9}\left({p_2^2g_{\alpha\bar{\alpha}}-p_{2\bar{\alpha}}p_{2\alpha}}\right)\nonumber\\
\!\!\!\!&+&\!\!\!\!(L_9+L_{10})\left(p_2\cdot
p_3g_{\alpha\bar{\alpha}}g_{\beta\bar{\beta}}-g_{\alpha\beta}p_{3\bar{\alpha}}p_{2\bar{\beta}}\right)\Big)\,
\ . \ea

Methods for doing the integrals are the same as of the HLS part.
Furthermore, in this case the $\Pi^{\mu\nu\alpha\beta}$ are finite
whereas, contribution to $a_\mu$ are not finite.

\section{Relevant Momentum Regions for the pion Loop Contribution.}
\label{chp:momentum} \setcounter{equation}{0}
\subsection{Dependence on the photon cut--off $\Lambda$}
Up to this point, the process needed to be taken to calculate the
$a_\mu$ for each approach has been described. It is now relevant
to consider the way $a_\mu$ behaves under change of momenta. In
fact, one expects that since the HLS model and the ChPT in higher
orders are non renormalizable, it should be somehow visible
through the $a_\mu$ as well. Furthermore, after all one of the
main goals of the present work has been to deal with differences
of the VMD and HLS model. One can also expect that the differences
should somehow reveal themselves via these considerations. To see
how, we have calculated different values of $a_\mu$ for different
cut--offs so that $P_1<\Lambda,P_2<\Lambda,Q<\Lambda$ and results
are shown in Table\tref{table4}.

\begin{table}
\begin{center}
\begin{tabular}{|c|c|c|c|c|c|}
\hline
Cut-off &\multicolumn{5}{c|}{ $10^{10} a_\mu$}\\
\hline
GeV &  bare&  VMD   & HLS $a=2$& $L_9, L_{10}$& HLS $a=1$\\
\hline
0.5 &$-$1.71(7) & $-$1.16(3)  &$-$1.05(1)&$-$1.64(1)&$-$1.35(1) \\
0.6 &$-$2.03(8) & $-$1.41(4)  &$-$1.15(1)&$-$1.80(1)&$-$1.59(1) \\
0.7 &$-$2.41(9) & $-$1.46(4)  &$-$1.17(1)&$-$1.85(1)&$-$1.76(1) \\
0.8 &$-$2.64(9) & $-$1.57(6)  &$-$1.16(1)&$-$1.79(1)&$-$1.88(1) \\
1.0 &$-$2.97(12)& $-$1.59(15) &$-$1.07(1)&$-$1.53(1)&$-$2.03(1) \\
2.0 &$-$3.82(18)& $-$1.70(7)  &$-$0.68(1)&$+$1.15(1)&$-$2.16(1) \\
4.0 &$-$4.12(18)& $-$1.66(6)  &$-$0.50(1)&$+$6.18(1)&$-$2.14(1) \\
\hline
\end{tabular}
\end{center}
\caption{Results of different calculations for the pion loop.}
\label{table4}
\end{table}
 As can be seen, the HLS and the VMD have the same behavior in low momentum region however,
for higher momenta, HLS starts to behave oddly. Meanwhile, the $L_{9}, L_{10}$ behavior
for higher momenta is as expected due to the non--renormalizability of ChPT in order $p^4$.

\subsection{Anatomy of the relevant momentum regions for the pion Loop
Contribution.}
\begin{figure}\label{VMD1}
\begin{center}
\includegraphics[width=10cm,height=6cm]{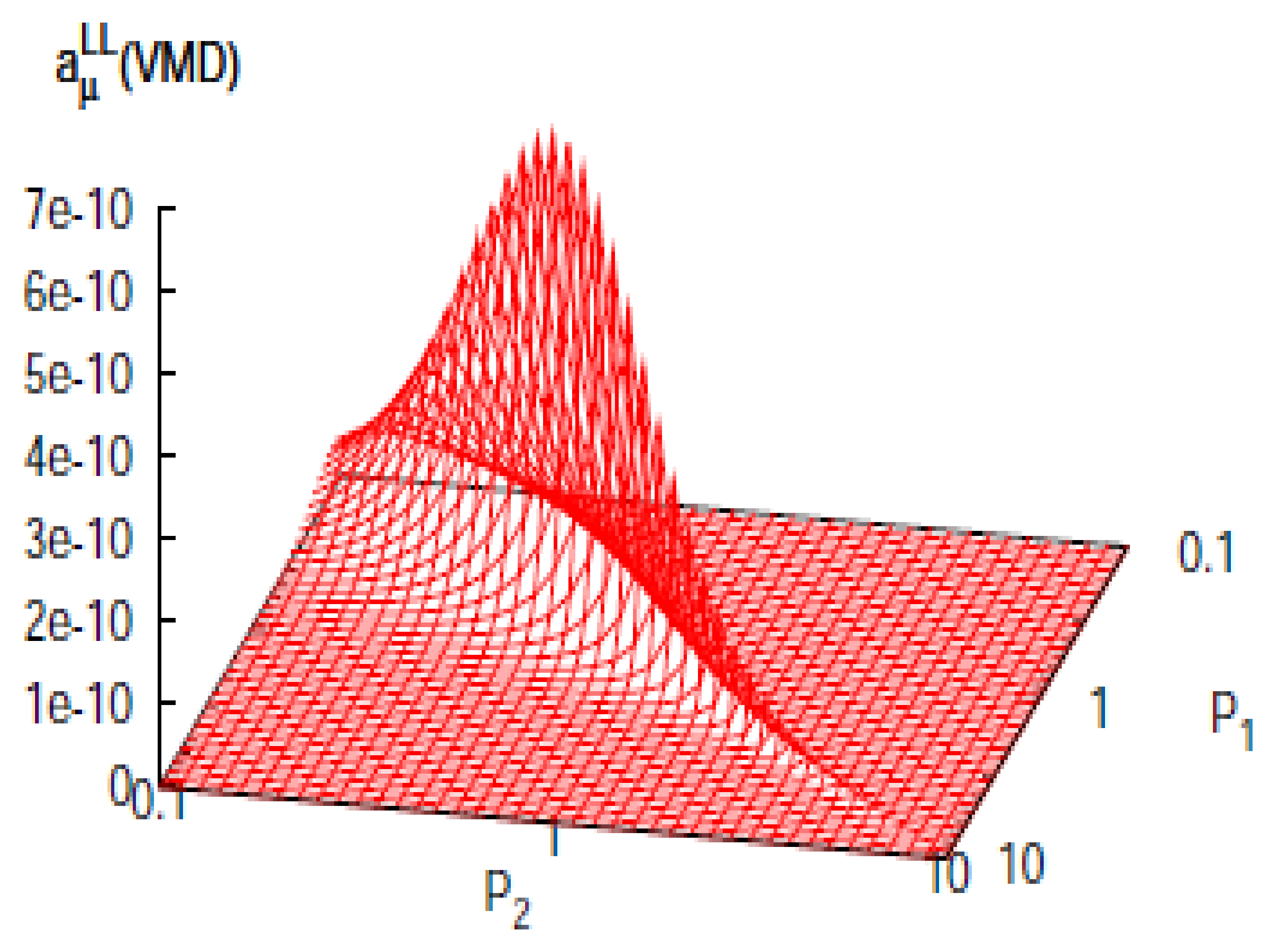}
\end{center}
\caption{$a_\mu^{LL}$ of the Eq.~(\ref{amup1p2}) as a function of $P_1$ and $P_2$
 for the VMD choice for the pion exchange. $a_\mu$ is directly related
to the volume under the surface. Figure from~\cite{Bijnens:2007pz}.}
\label{amup1p2}
\end{figure}

 But,
as it has been discussed in the Ref.~\cite{Bijnens:2007pz}, one
can also investigate how different regions of momentum, contribute
differently to the HLL contribution to the $a_\mu$. The technique
is the same with calculating the whole value of $a_\mu$ but this
time, instead of taking the integral over all variables, one
leaves $P_1^2=-p_1^2$, $P_2^2=-p_2^2$ and $Q^2=-q^2$ unintegrated.
 Mathematically we
write
 \ba\label{amup1p2int}
a_\mu&=&\int dl_1 dl_2 a_\mu^{LL}(l_1,l_2)\nonumber\\
&=&\int dl_1 dl_2dl_q a_\mu^{LLQ}(l_1,l_2,l_q)\, \ , \ea with
$l_1=log(P_1/GeV)$, $l_2=log(P_2/GeV)$ and $l_q=log(Q/GeV)$. To
see exactly how the momentum region above the $1$ GeV contribute
to the $a_\mu$, it is better to use logarithmic scale as has been
discussed in~\cite{Bijnens:2007pz}. Also, the total amount of
$a_\mu$ is proportional to the volume under the surface of each
diagram. For example, as is shown in the Figure~\tref{amup1p2},
the most important contribution of the pion exchange to the
$a_\mu$, via the VMD model, is coming from the low region of
momenta, which is expected since, because of the usage of vector
meson legs, the large momenta are strongly suppressed.
Furthermore, the concentration is around the equal values of
momenta $P_1$ and $P_2$. It should also be noticed that, the whole
value of the $a_\mu$ is proportional to the volume under the
surface, after integrating over the whole region of momenta. Also,
as it is easier to deal with plots with positive values,
$-a_\mu$ is drawn in the figures for the pion loop.

\begin{figure}
\begin{center}
\includegraphics[width=10cm,height=6cm]{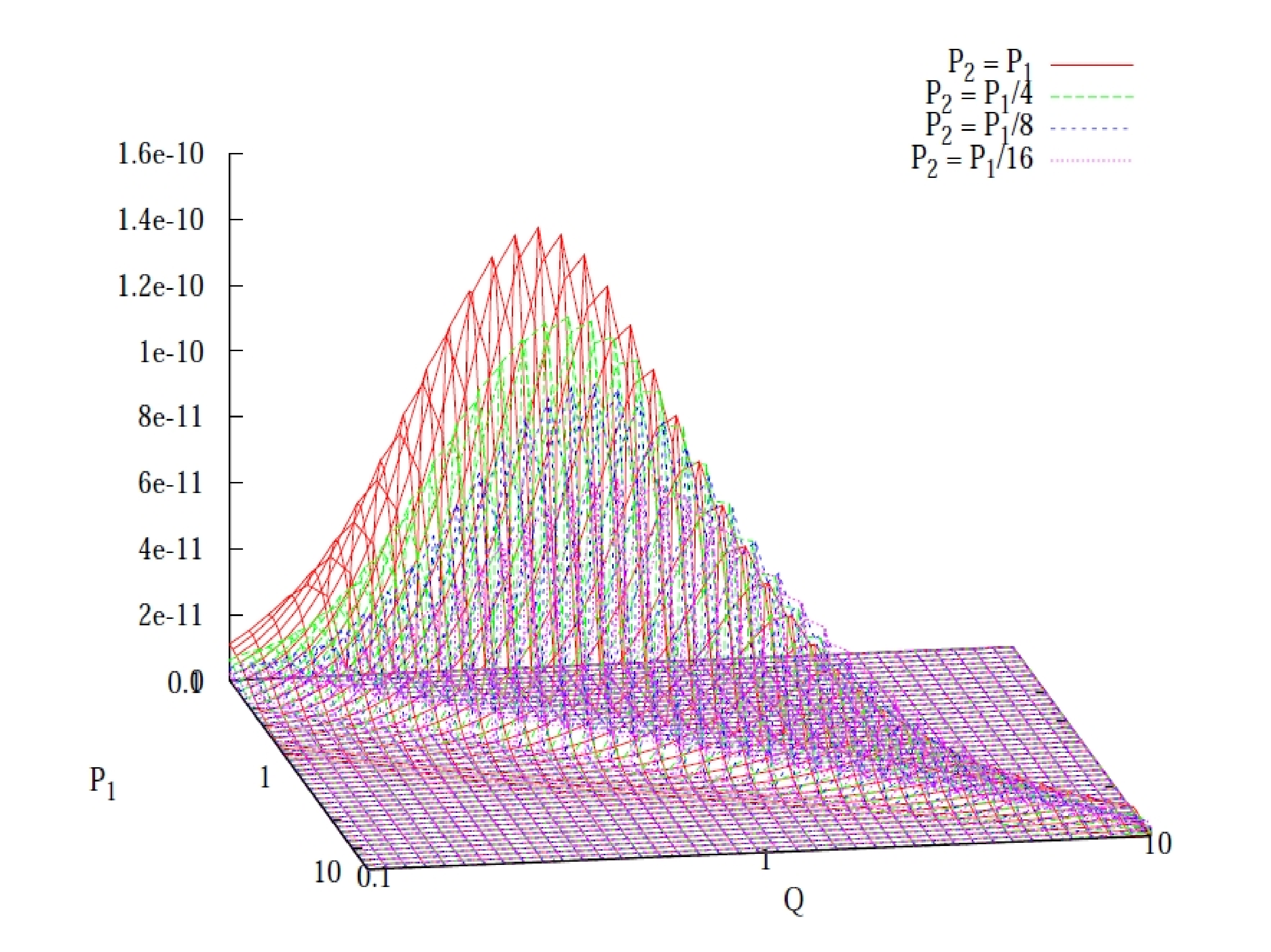}
\end{center}
\caption{$-a_\mu^{LL}$ as a function of different ratios of $P_1$ and $P_2$ versus $Q$ 
for the bare pion loop choice for. $-a_\mu$ is directly related
to the volume under the surface.}
\label{comparep1p2bare}
\end{figure}

\begin{figure}
\begin{center}
\includegraphics[width=10cm,height=6cm]{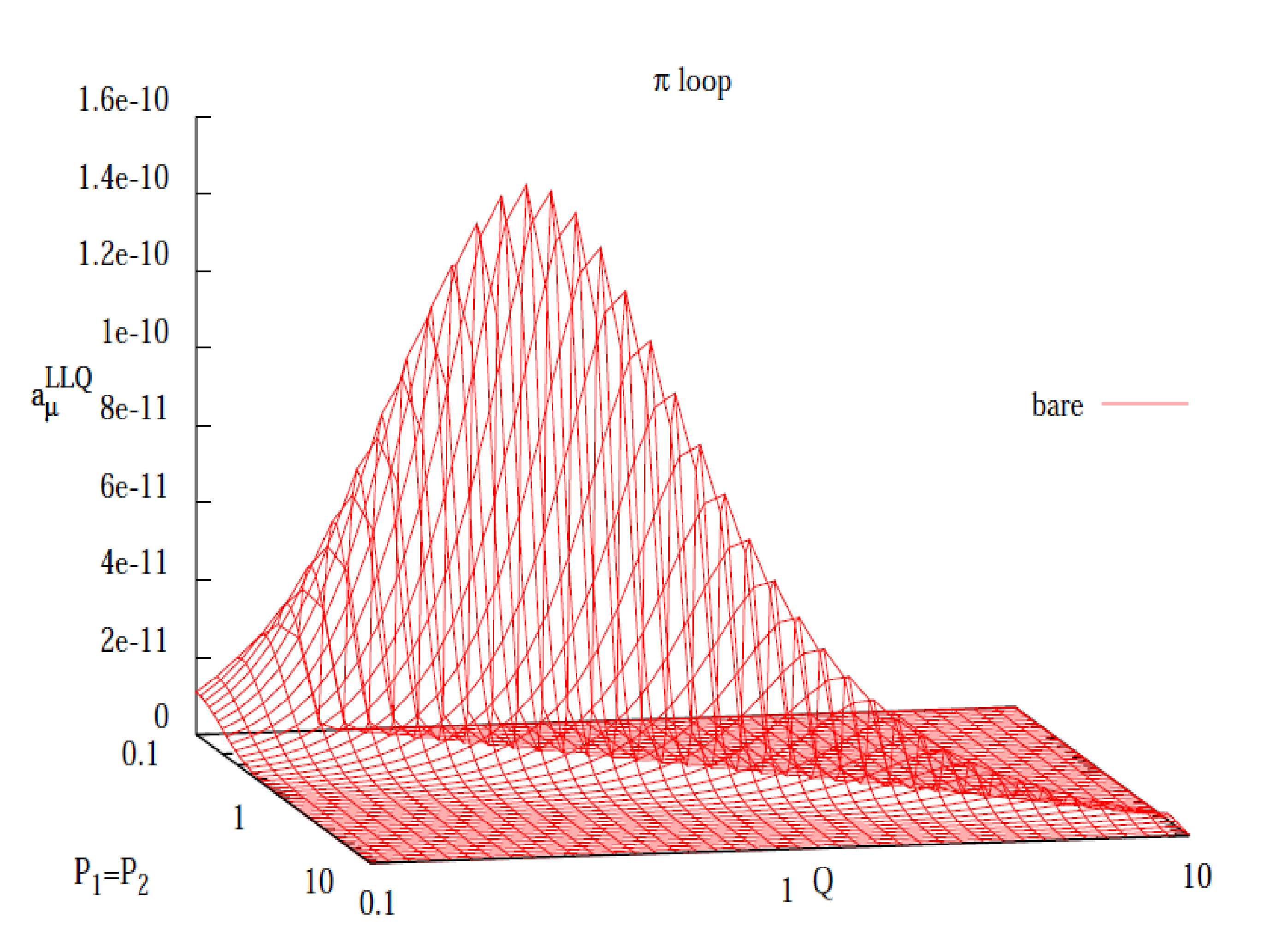}
\end{center}
\caption{$-a_\mu^{LLQ}$ as a function of $P_1=P_2$ and $Q$ for the bare pion loop choice. $-a_\mu$ is directly related
to the volume under the surface.}
\label{barep1Q}
\end{figure}

As can be seen from Figure~\tref{amup1p2}, the momentum is
concentrated along the line $P_1=P_2$. Following the same lines as
in~\cite{Bijnens:2007pz}, we have done the same calculation for
the Bare, VMD, HLS and $L_9, L_{10}$ approaches to the charged
pion loop contribution to the $a_\mu$ and compared them. For each
case we have shown the distribution of $a_\mu^{LLQ}$ versus $P_1,
Q$ and $ Q$. Figures~\tref{comparep1p2bare} and~\tref{barep1Q}
belong to the bare pion loop case, where the peak is in the low
momentum region but, a large part comes from the region above $1$
GeV. I Figure~\tref{comparep1p2bare} we also show the cases for
$P_1\neq P_2$. It is clear that the parts with $P_1$ significantly
different from $P_2$ contribute less. In Figure~\tref{barep1Q} we
show the case $P_1=P_2$ alone.

\begin{figure}
\begin{center}
\includegraphics[width=10cm,height=6cm]{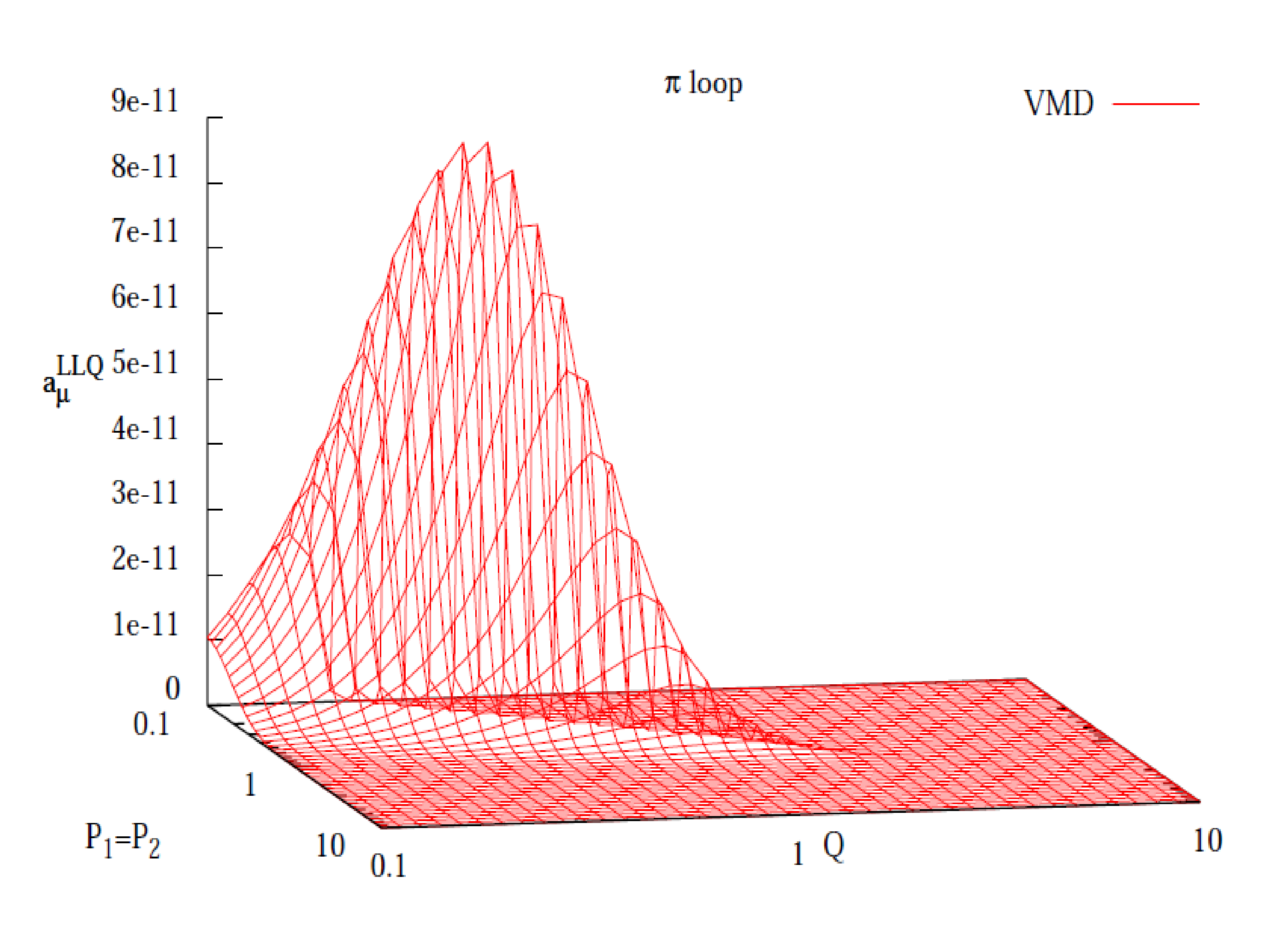}
\end{center}
\caption{$-a_\mu^{LLQ}$ as a function of $P_1=P_2$ and $Q$ for the VMD choice. $-a_\mu$ is directly related
to the volume under the surface.}
\label{VMDp1Q}
\end{figure}

\begin{figure}
\begin{center}
\includegraphics[width=10cm,height=6cm]{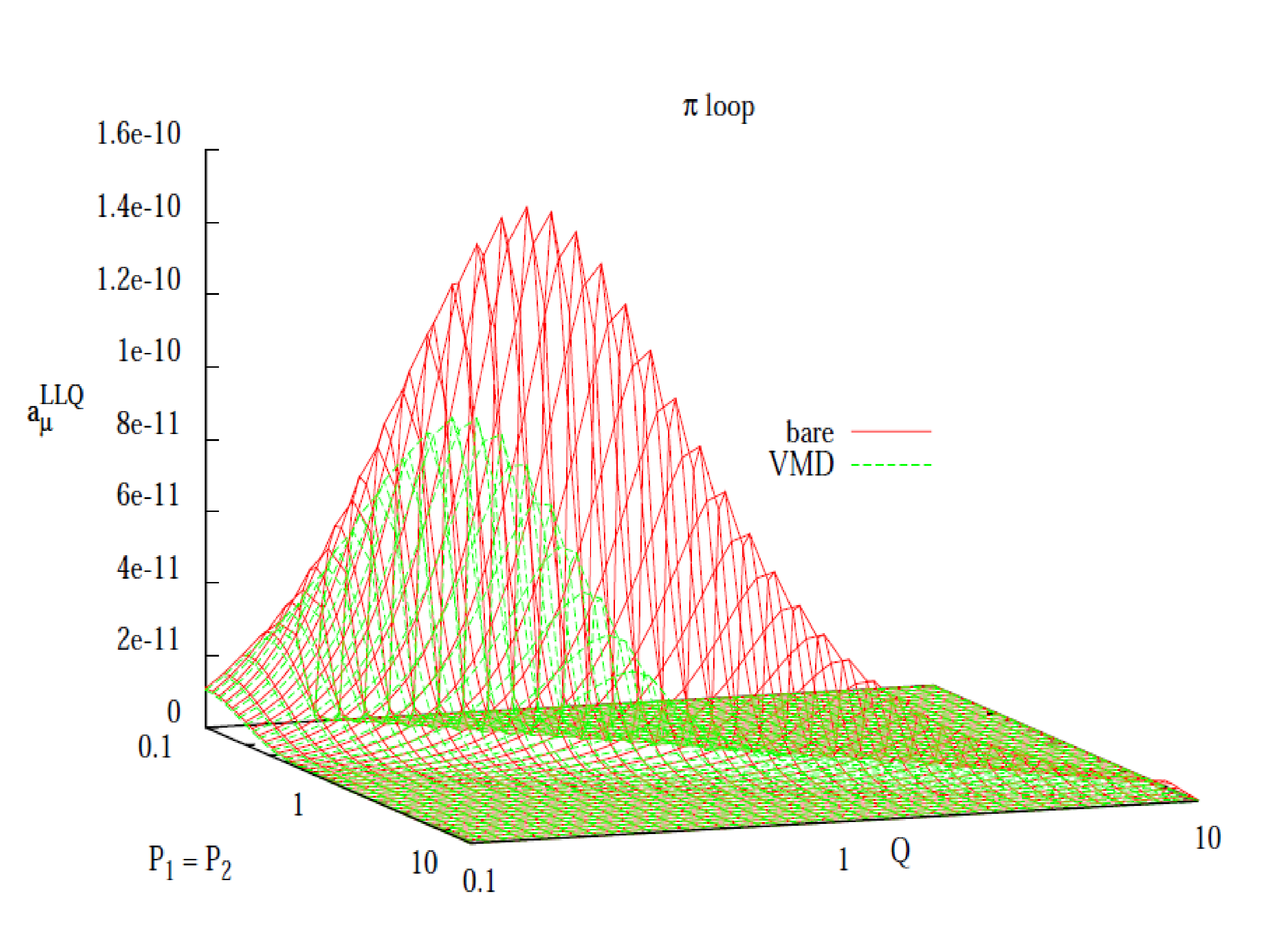}
\end{center}
\caption{$-a_\mu^{LLQ}$ as a function of $P_1=P_2$ and $Q$ for the VMD and the bare pionloop choice. $-a_\mu$ is directly related
to the volume under the surface.}
\label{VMDbare}
\end{figure}

Figures~\tref{VMDp1Q} shows the VMD case, which is obviously
suppressed respect to the bare pion loop case, while the pick
still lies in the low momentum region. Figure~\tref{VMDbare}
compares the bare and the VMD cases.

\begin{figure}
\begin{center}
\includegraphics[width=10cm,height=6cm]{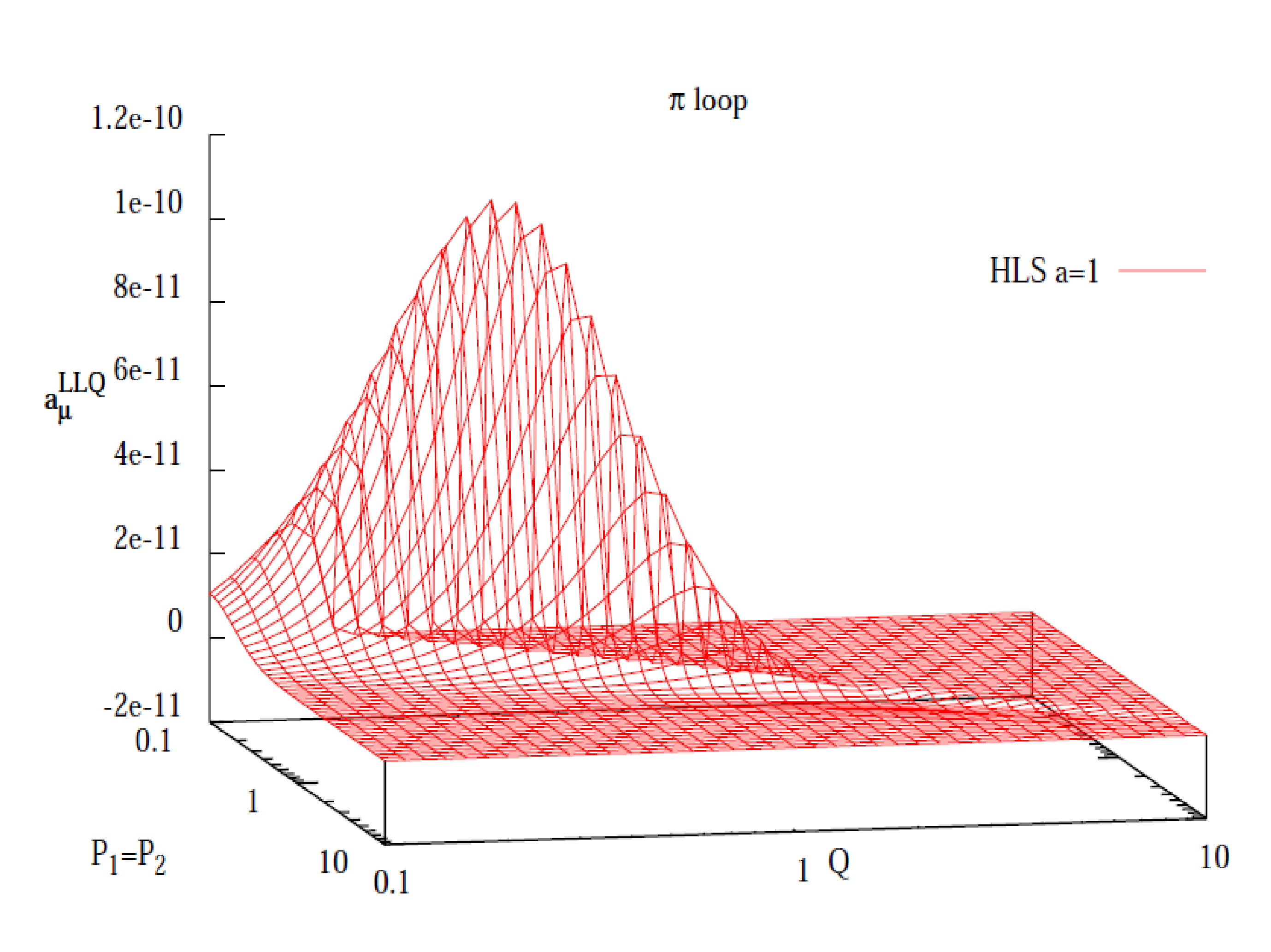}
\end{center}
\caption{$-a_\mu^{LLQ}$ as a function of $P_1=P_2$ and $Q$ for the HLS, $a=1$ choice. $-a_\mu$ is directly related
to the volume under the surface.}
\label{HLSp1Qa=1}
\end{figure}

\begin{figure}
\begin{center}
\includegraphics[width=10cm,height=6cm]{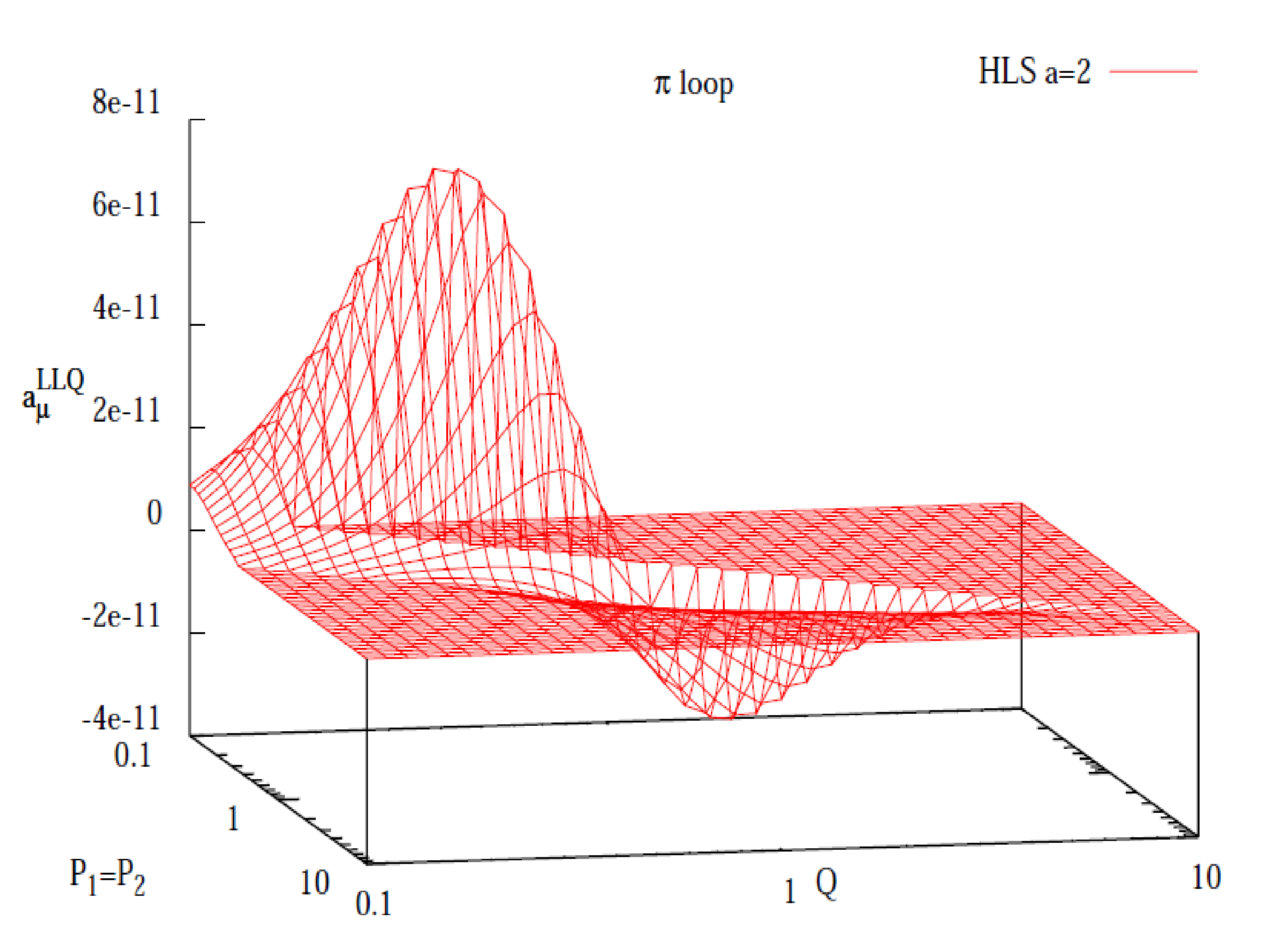}
\end{center}
\caption{$-a_\mu^{LLQ}$ as a function of $P_1=P_2$ and $Q$ for the HLS, $a=2$ choice. $-a_\mu$ is directly related
to the volume under the surface.}
\label{HLSp1Qa=2}
\end{figure}

\begin{figure}
\begin{center}
\includegraphics[width=10cm,height=6cm]{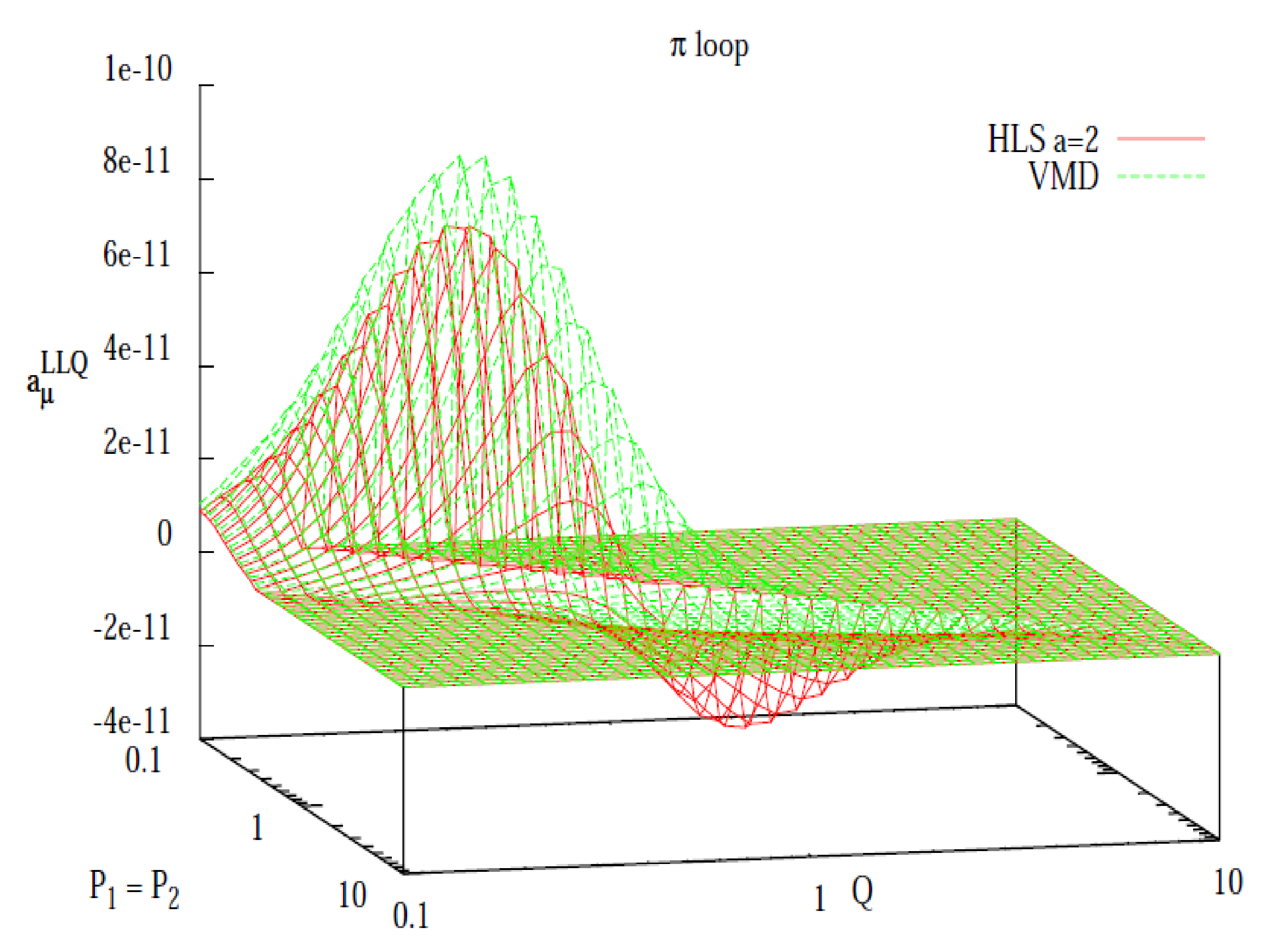}
\end{center}
\caption{$-a_\mu^{LLQ}$ of the Eq.~(\ref{amup1p2int}) as a function of $P_1=P_2$ and $Q$ for the VMD and the HLS choices.}
\label{gnuhlsvmd}
\end{figure}
\begin{figure}

\begin{center}
\includegraphics[width=10cm,height=6cm]{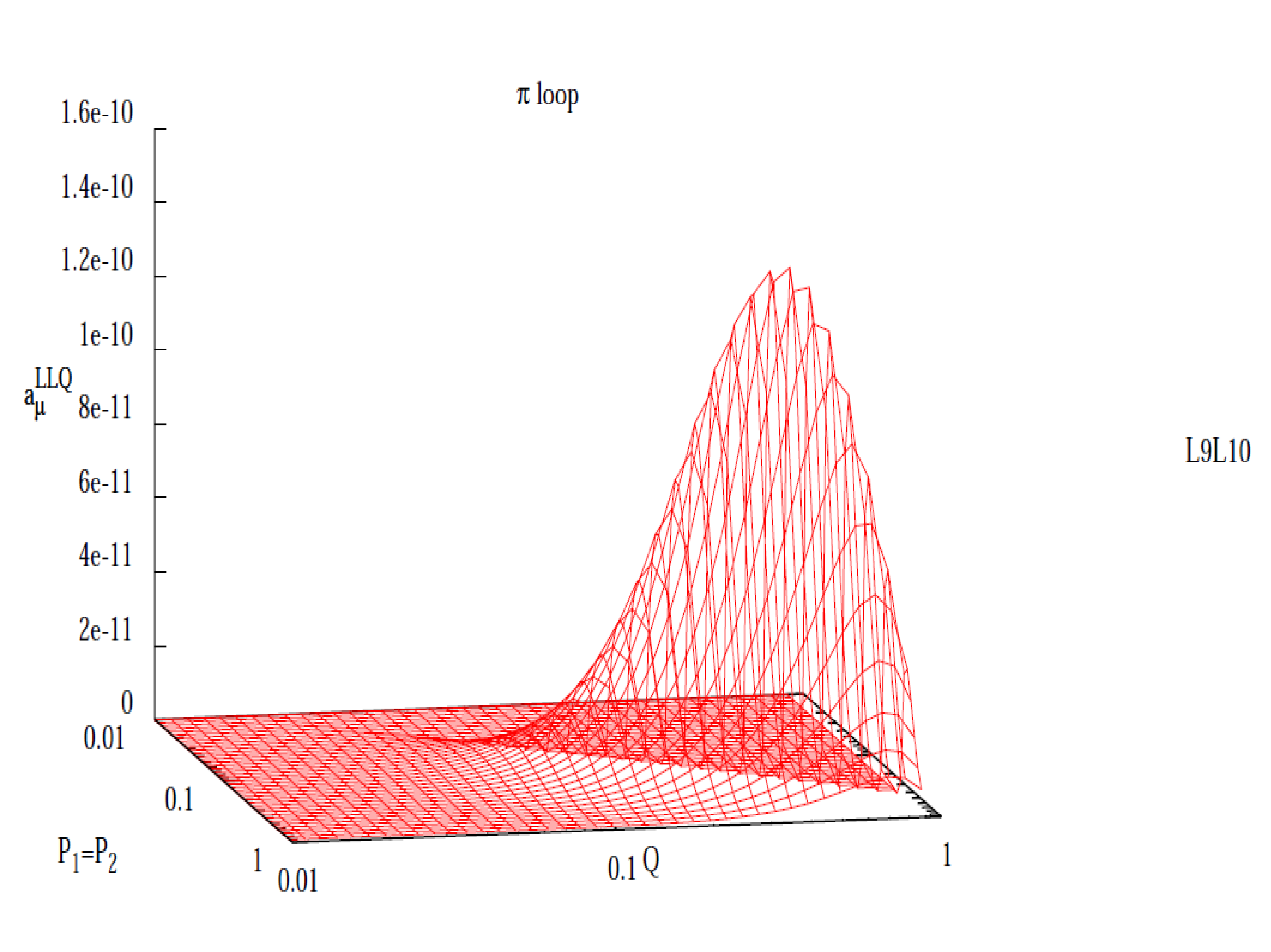}
\end{center}
\caption{$-a_\mu^{LLQ}$ as a function of $P_1=P_2$ and $Q$ for the $L_9$, $L_{10}$ choice. $-a_\mu$ is directly related
to the volume under the surface.}
\label{L9L10p1Q}
\end{figure}

\begin{figure}
\begin{center}
\includegraphics[width=10cm,height=6cm]{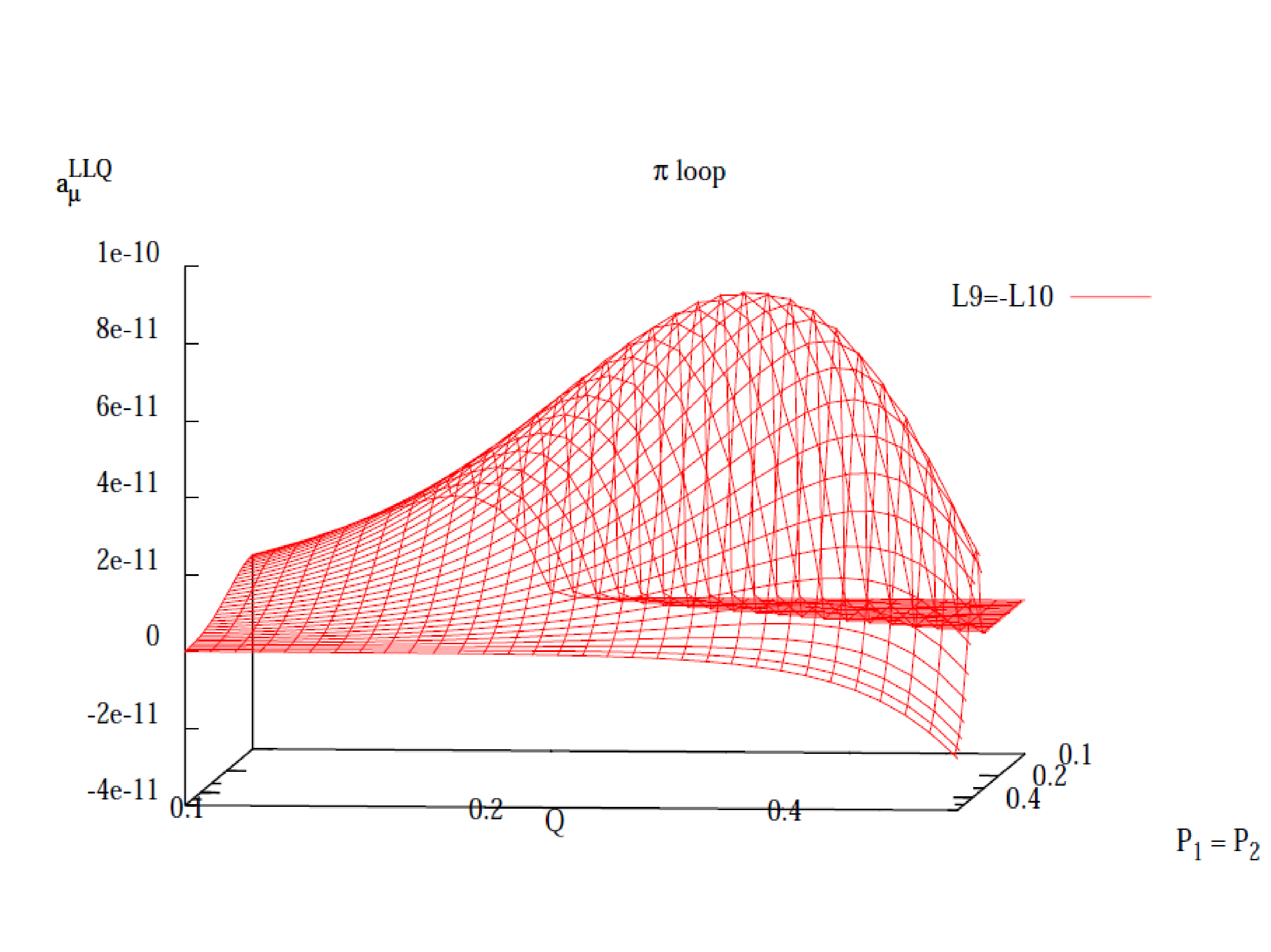}
\end{center}
\caption{$-a_\mu^{LLQ}$ of the Eq.~(\ref{amup1p2int}) as a function of $P_1=P_2$ and $Q$ for the $L_9$ choice. $-a_\mu$ is directly related
to the volume under the surface.}
\label{L9p1Q}
\end{figure}

\begin{figure}
\begin{center}
\includegraphics[width=10cm,height=6cm]{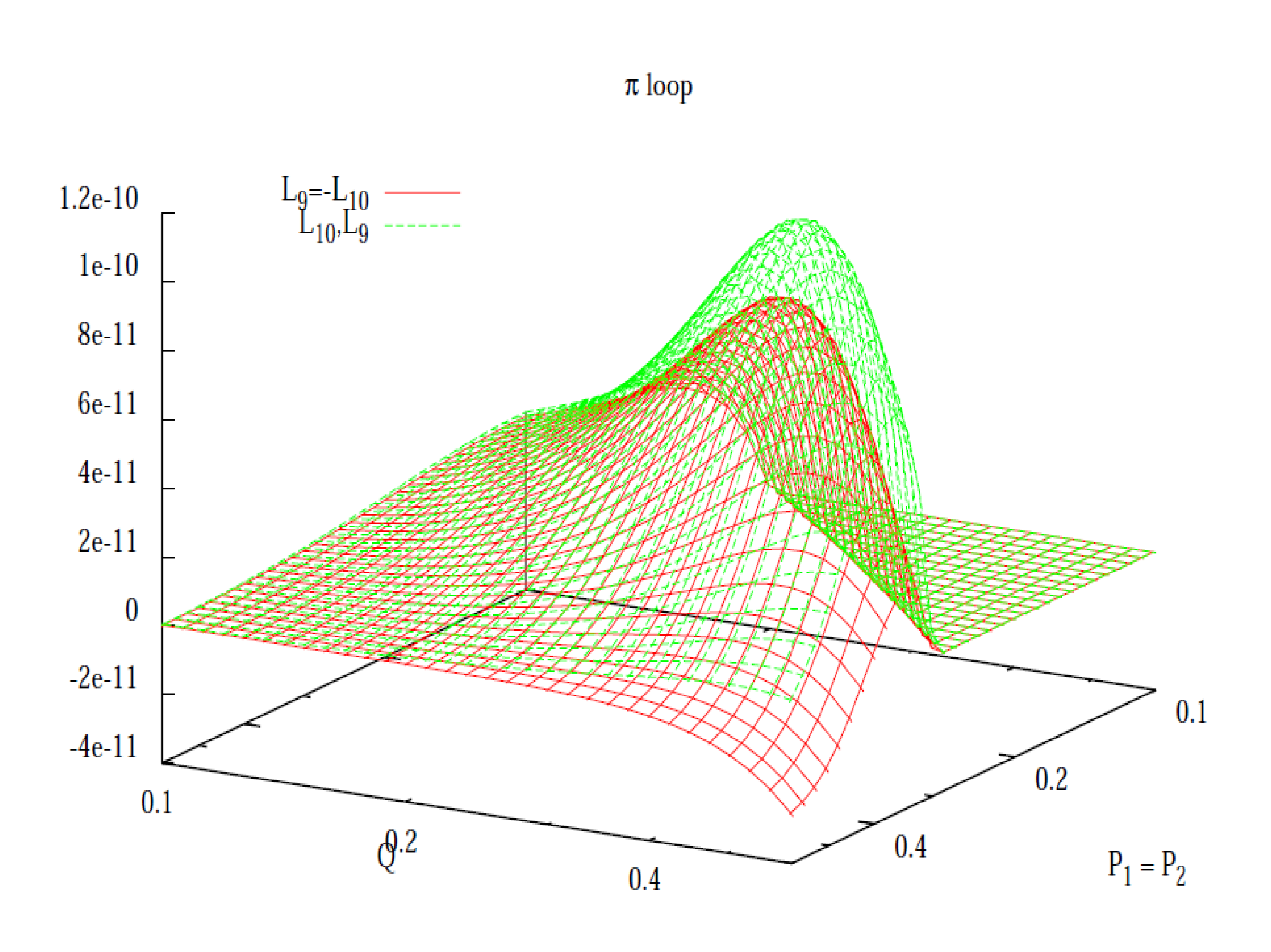}
\end{center}
\caption{$-a_\mu^{LLQ}$ as a function of $P_1=P_2$ and $Q$ for the $L_9$ and $L_9, L_{10}$ choice. $-a_\mu$ is directly related
to the volume under the surface.}
\label{L9-L9L10}
\end{figure}

\begin{figure}
\begin{center}
\includegraphics[width=10cm,height=6cm]{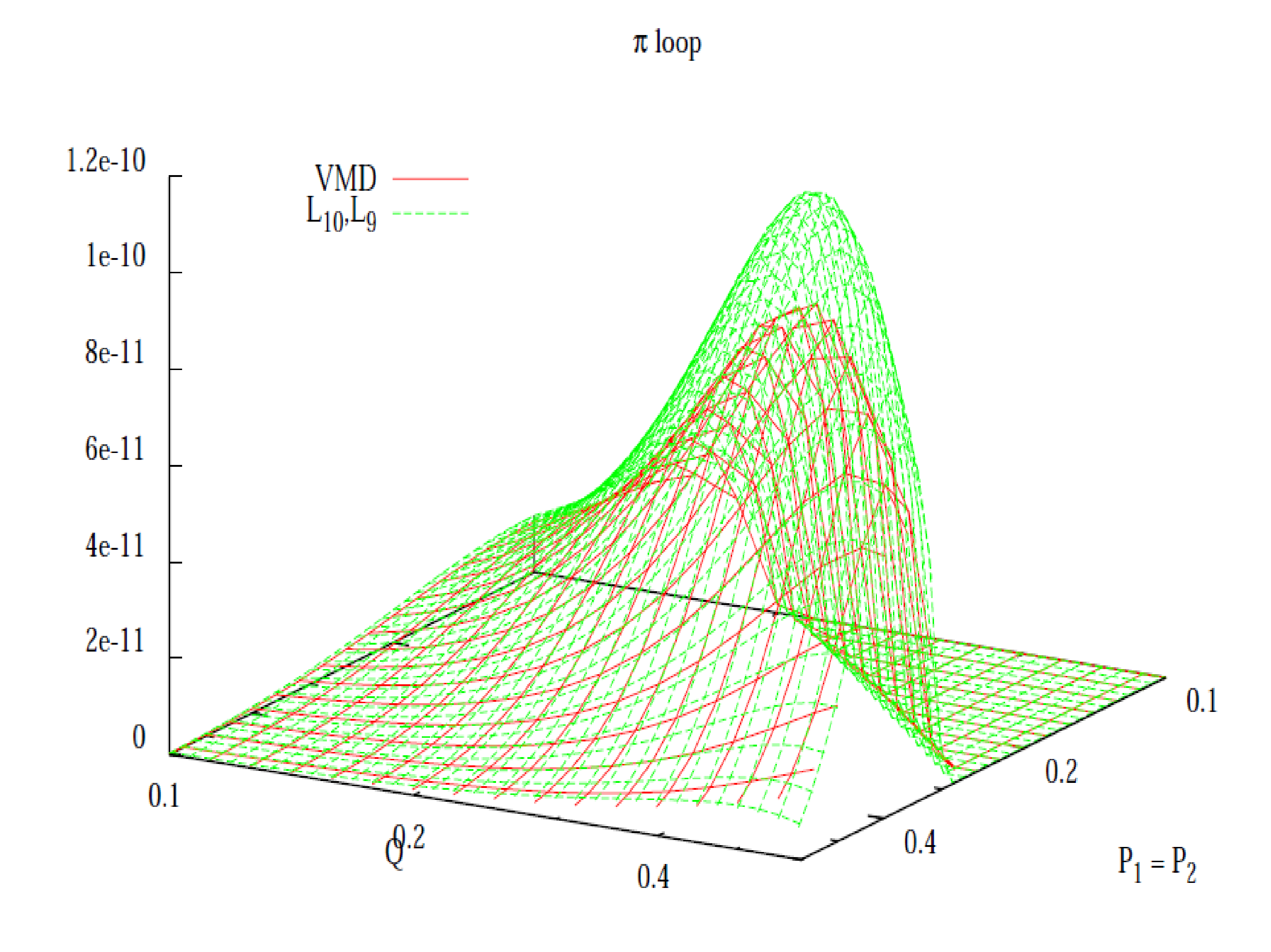}
\end{center}
\caption{$-a_\mu^{LLQ}$ as a function of $P_1=P_2$ and $Q$ for the VMD and $L_9, L_{10}$ choice. $-a_\mu$ is directly related
to the volume under the surface.}
\label{VMDL9L10}
\end{figure}

\begin{figure}
\begin{center}
\includegraphics[width=10cm,height=6cm]{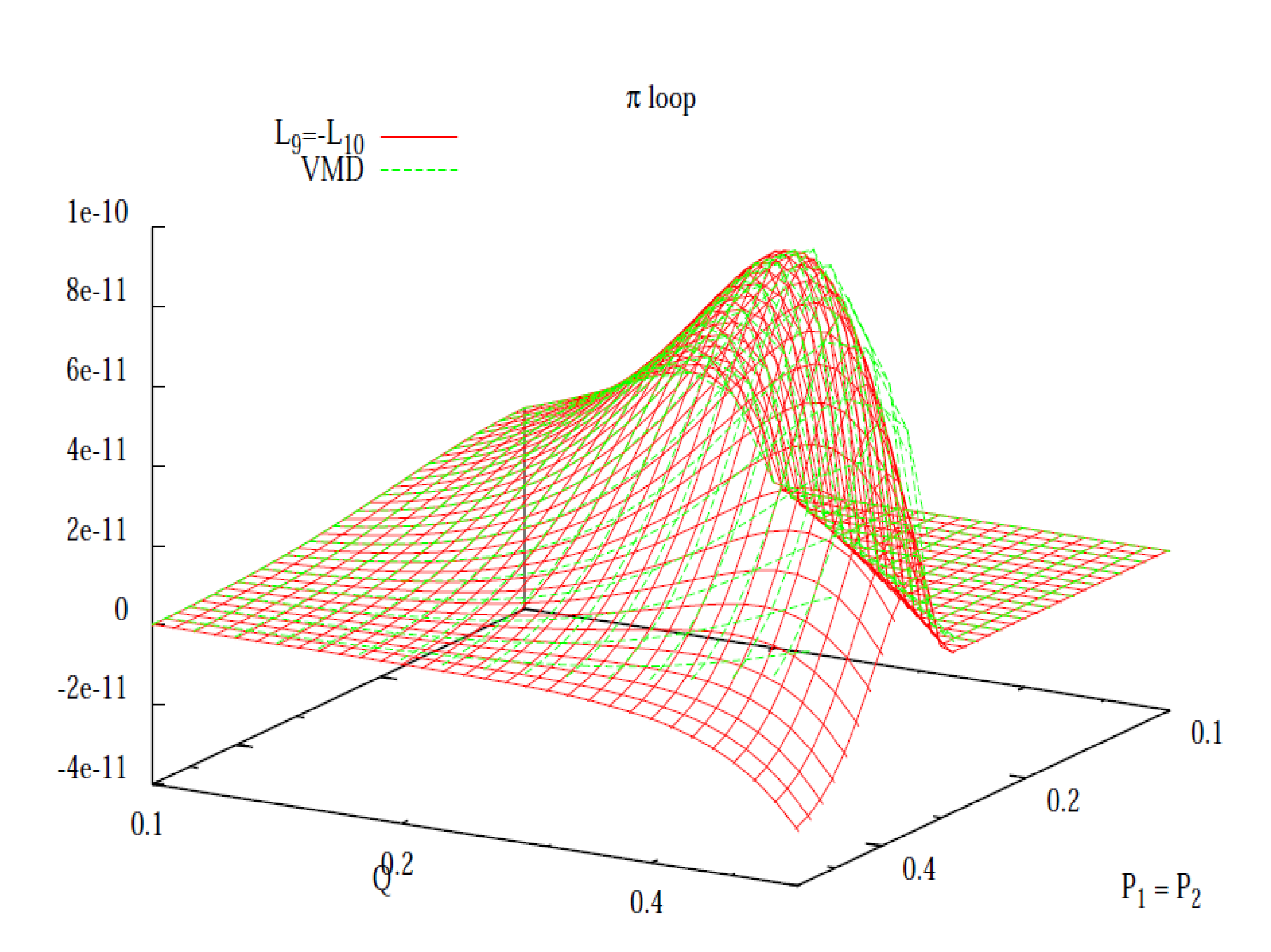}
\end{center}
\caption{$-a_\mu^{LLQ}$ as a function of $P_1=P_2$ and $Q$ for the VMD and $L_9$ choice. $-a_\mu$ is directly related
to the volume under the surface.}
\label{VMDL9}
\end{figure}

 Figures~\tref{HLSp1Qa=1}
and~\tref{HLSp1Qa=2} belong to the HLS case for $a=1$ and $a=2$ respectively, where the second one
 should reproduce
the VMD results. Here is where the surprise comes in and, as can
be seen, the low momentum region peak follows with a dip at the
high momentum region. The resulting graphes of HLS and VMD are
co--plotted in Figure~\tref{gnuhlsvmd}, in terms of $P_1=P_2$ and
$Q$. Conclusion to be drawn from this diagram is, in both
approaches the main body of the contribution comes from the low
momentum region and
 in the VMD case the large Q tail is larger. But, the large negative
contribution to $a_\mu$ in the HLS side needs some justification.
This finding gives a better insight into the nature of the
difference which has led to such a dramatic variation between the
full VMD and the HLS results. It should be mentioned that, as has
been already noticed in the Ref.~\cite{Hayakawa:1996ki}, the
difference would stem from the lack of the $\rho\rho\pi\pi$ vertex
in the HLS lagrangian. The case $a=1$ in the HLS, which has a
better higher energy behavior, makes the dip of the $a=2$ case
vanish.

Figures~\tref{L9L10p1Q} and~\tref{L9p1Q} show results of
calculation for the $L_9, L_{10}$ and $L_9=- L_{10}$ cases and
Figure~\tref{L9-L9L10} compares them. It should be mentioned that,
as ChPT in the order $p^4$ is nonrenormalizable, the overall value
of the $a_\mu$ in these cases are cutoff dependent. There is one
specific property of the $L_9$ and $L_{10}$ terms of the $p^4$
Lagrangian which we would like to emphasis that is, when one sets
$L_{10}=-L_{9}$, the $L_9$ part contribution behaves
 like the HLS with $a=2$ and the VMD part to order $p^4$. This could be seen in the
Figures~\tref{VMDL9L10}
 and~\tref{VMDL9} and could be justified via relation~(\ref{79}) for the
 $\gamma\gamma\pi\pi$ vertex, when
 compared
to the relation~(\ref{54}). Since, resorting to the fact that
$L_9\propto 1/m_\rho^2$~\cite{Ecker}, relation~(\ref{79}) plays a similar role
as the relation~(\ref{54}) does in the limit $L_{10}=-L_{9}$.

\section{Conclusions and Prospects}
In this work we have recalculated the previous results for the HLL
pion loop contribution to the muon magnetic anomaly via the sQED,
the VMD model and the HLS model and all results are in good
agreement with the previous ones. To do so, we have extended the
Gegenbauer polynomial technique to the pion loop case to calculate
 the integrals involved.

We have also added the next to leading order ChPT corrections to the lowest order results and have shown that,
in the corresponding energy region, results are in good agreement with predictions of other models namely, the VMD and the
HLS with $a=2$, as expected.

By investigating the momentum regions that each model predicts to have a part in the $a_\mu^{LbL}$, we have found out why
the HLS prediction for the pion loop contribution is so different with that of the VMD.

Also, using the OPE approach to the $\gamma\gamma\pi\pi$
amplitude, it has been shown that the VMD lives up to the
expectation but, HLS with $a=2$ does not and hence, the HLS can be
ruled out as a valid model to consider the pion loop contribution
to the muon anomalous magnetic moment. The $a=1$ case which has a
better higher energy behavior has final results similar to VMD.

\label{chp:concl} \setcounter{equation}{0}

\section*{Acknowledgments}
I would like to thank my supervisor for teaching me nearly all I
know about this subject. I also would like to thank Stefan Lanz
for his endless patience with my questions, my family in Iran and
my friends Behruz Bozorg, Roland Katz, Pouria Jaberi, 
Donya Mashallah Poor, Reza Jafari Jam and Sajjad Sahbaei for their support.

\end{document}